\documentclass[aps,prd,nofootinbib,longbibliography]{revtex4-1}
\usepackage{amsmath}
\usepackage[T1]{fontenc}
\usepackage{amsfonts}
\usepackage{amssymb}
\usepackage{dsfont}
\usepackage{mathrsfs}
\usepackage[mathscr]{euscript}
\usepackage{yhmath}
\usepackage{graphicx}
\usepackage{epstopdf}
\usepackage{caption}
\usepackage{tensor}
\usepackage{enumerate}
\usepackage{enumitem}
\usepackage{color}
\usepackage[all]{xy}
\usepackage{hyperref}
\hypersetup{
    colorlinks=true, 
    linktoc=page,    
    linkcolor=blue,  
    urlcolor=magenta
}

\newcommand{\tr}{\mathrm{Tr}}
\newcommand{\Tr}[1]{\mathrm{Tr}\left(#1 \right)}

\newcommand{\va}{\scriptscriptstyle}
\allowdisplaybreaks

\newcommand{\be}{\nopagebreak[3]\begin{equation}}
\newcommand{\ee}{\end{equation}}

\newcommand{\bee}{\nopagebreak[3]\begin{equation*}}
\newcommand{\eee}{\end{equation*}}

\newcommand{\ba}{\nopagebreak[3]\begin{eqnarray}}
\newcommand{\ea}{\end{eqnarray}}

\newcommand{\baa}{\nopagebreak[3]\begin{eqnarray*}}
\newcommand{\eaa}{\end{eqnarray*}}

\newcommand{\la}{\label}
\newcommand{\n}{\nonumber}

\newcommand{\C}{\mathbb{C}}

\newcommand{\R}{\mathbb{R}}

\newcommand{\I}{\mathbb{I}}

\begin{document}

\title{Quantum evolution of black hole initial data sets: Foundations}
\date{\today}
\author{Emanuele Alesci$^1$}
\email{eza69@psu.edu}
\author{Sina Bahrami$^1$}
\email{ sina.bahrami@psu.edu}
\author{Daniele Pranzetti$^2$}
\email{dpranzetti@perimeterinstitute.ca}
\affiliation{$^1$Institute for Gravitation and the Cosmos, Pennsylvania State University, University Park, Pennsylvania 16802, USA}
\affiliation{$^2$Perimeter Institute for Theoretical Physics,  Waterloo, Ontario N2L 2Y5, Canada}
\begin{abstract}
We construct a formalism for evolving spherically symmetric black hole initial data sets within a canonical  approach to quantum gravity. This problem can be formulated precisely in quantum reduced loop gravity, a framework which has been successfully applied to give a full theory derivation of loop quantum cosmology. We extend this setting by implementing a particular choice of partial gauge which is  then used to select a kinematical Hilbert space where the symmetry reduction is imposed through semiclassical states. The main result of this investigation is an effective Hamiltonian that can be used to solve for quantum black hole geometries by evolving classical black hole initial data sets.
\end{abstract}

\maketitle
\tableofcontents

\section{Introduction}

{\it Singularities are generic predictions of general relativity}; 
this was the conclusion of the first singularity theorem that
was discovered by Roger Penrose in 1964 \cite{Penrose:1964wq}. Prior
to this, it was widely suspected that singularities are aspects of algebraically special spacetimes or end results of highly symmetric processes \cite{lif-khal}, as in the gravitational collapse of a spherically symmetric compact object \cite{opp-sny}. 
Penrose's arguments instead attributed the occurrence of singularities to the formation of trapped surfaces in geometries where the Ricci curvature tensor satisfies $R_{ab} k^a k^b \geq 0$ for all null vectors $\vec{k}$. \footnote{Penrose's theorem requires global hyperbolicity for the spacetime.} Further progress concerning the formation of singularities in general relativity was subsequently made by the landmark theorems of Hawking and Penrose \cite{hawk-sing-1, hawk-pen}. \footnote{See \cite{hawk-ellis} for an extensive discussion of these results.}

Despite their robustness, singularity theorems are reliable only in the regime where spacetime geometry is classical. This, however, runs 
contrary to what they set out to accomplish. In fact, one expects quantum  corrections to classical geometry to become relevant on scales where $|R_{abcd} R^{abcd}| \gtrsim l_p ^{-4}$, where $R_{abcd}$ is the Riemann curvature tensor and $l_p \sim 10^{-33} \text{cm}$ is the Planck length. For the most elementary examples of singularity in general relativity, $|R_{abcd} R^{abcd}|$ blows up as one approaches the singularity. Therefore, whether singularities generically form in nature as predicted by the singularity theorems hinges on how spacetime behaves in the quantum domain. 

The degree to which quantum effects modify classical singularities  has long been a subject of speculation. Penrose argued in \cite{penrose-time} that understanding the quantum structure of the initial spacetime singularity is the key to resolving one of the long-standing puzzles in theoretical physics, namely the second law of thermodynamics and the origin of the observed (albeit minute) time asymmetry in nature. \footnote{A microscopic example of this time-asymmetry is the $CP$-violating decay of $K^0$ meson \cite{cp-1,cp-2,cp-3}.} Should quantum effects resolve this
initial spacetime singularity by replacing it with a bounce, 
it may be difficult to concoct a compelling explanation for 
the second law of thermodynamics \cite{pen-hawk-deb}. \footnote{We refer the interested reader to \cite{penrose-time} where the {\it vanishing Weyl curvature hypothesis} is explained.} \footnote{Steinhardt and Turok have argued in \cite{StT1, StT2}
that in ekpyrotic models the second law of thermodynamics is respected; the total entropy increases from cycle to cycle while the entropy density undergoes periodic behaviour. Nonetheless, ekpyrotic models do not provide any explanations for the origin of the second law.}
However, in the context of quantum geometry, one expects  the notion of entropy and how it evolves to be scale dependent. Therefore, the fate of the second law in quantum gravity is tied to understanding the quantum structure of the Universe near the cosmological singularities as well
as the exact mechanism by which the classical and continuous spacetime manifold emerges from them.  

 Aside from cosmological singularities,  one can also ponder upon consequences of black hole singularities being removed by quantum effects. In that case, a resolution for a number of outstanding riddles and open questions may be within reach. Of significant importance are the cosmic censorship hypothesis \cite{csc} and the information loss paradox of black hole evaporation \cite{bh-evap}. For the latter, if the semiclassical approximation scheme is correct, the unitarity principle  of quantum mechanics is violated by pure states evolving into mixed states during black hole evaporation (see \cite{preskill} for a review of the basic arguments). 
Several ideas have been proposed to resolve this paradox with some being more radical than others. On the more exotic side, there are ideas proposing black hole event horizons turning into ``firewalls'' \cite{2013JHEP...02..062A} or ``fuzzballs'' \cite{Mathur:2005zp}. In these speculative scenarios, phenomena in gross violation of the equivalence principle are required to occur on a black hole event horizon in order to shut down correlations between the infalling and outgoing pair of created particles. On the more conservative end of the spectrum, there are proposals such as 
Planck-sized remnants of the evaporation process \cite{Aharonov:1987tp, PhysRevD.46.1347}, 
and singularity resolution by quantum gravity effects that lead to an extension of the spacetime diagram for evaporating black holes \cite{ash-bojo}. However, a clear-cut resolution to the aforementioned paradox has not yet emerged  
in these latter proposals either. Indeed the idea of black hole remnants has been criticized for a lack of viability as it requires a Planck-sized object to have an enormous entropy, roughly on the order of $M^2$ where $M$ is the black hole mass, which in turn leads to  the infinite pair production problem, although objections have been raised against this critique in the literature (see, e.g.,  \cite{preskill,  Banks:1992is, Giddings:1993vj, Hossenfelder:2009xq} for a discussion). Similarly in the case where the black hole singularity is resolved and a classical spacetime  emerges to the causal future of  the would-be singularity, a description for a concrete mechanism that purifies the early radiation degrees of freedom is lacking (see, however, \cite{Perez:2014xca} for an interesting proposal).
Beyond all speculations, a definitive fate of a classical singularity is only predicted by a detailed full quantum gravity calculation.
Given the intimate relation between the last stages of black hole evaporation and Planck-scale physics, this paradox will likely be resolved by a better understanding of a quantum gravity description.

 Presently,
string theory and loop quantum gravity (LQG) have been 
sufficiently developed to allow for problems of this kind to be
explored within their respective frameworks. 
In the past two decades some evidence has emerged in string theory in favor of quantum singularity resolution. One of the earliest 
results is by Horowitz {\it et al.} \cite{hor-ross} where AdS/CFT 
duality was used to argue that the interior of a black hole in anti-de Sitter space is described by a singularity-free supersymmetric field theory on the boundary. Similar results have been obtained for cosmological singularities using the AdS/CFT duality \cite{turok1,turok2,das,hertog-hor}. Aside from AdS/CFT, other well-known proposals include models with tachyon condensation \cite{eva1,eva2}, matrix models \cite{mmodel, Ishino:2006nx}, and models that use orientifolds \cite{orient}. Nevertheless, no full theory calculation has been produced that can provide a definitive answer.    

LQG  \cite{Thiemann:2007zz} embodies the most studied nonperturbative quantization program of the gravitational field. Questions such as the quantum fate of classical spacetime singularities can in principle be formulated and investigated within this framework \cite{DiazPolo:2011np, Perez:2017cmj}. On the specific subject of black hole singularities, a number of very important studies \cite{Modesto:2004xx, Modesto:2005zm, Ashtekar:2005qt}\cite{Bohmer:2007wi,Campiglia:2007pb,Chiou:2008nm}\cite{Corichi:2015xia, Cortez:2017alh, Olmedo:2017lvt} have been conducted where the primary focus is on the geometry interior to the event horizon of a Schwarzschild black hole. The aforementioned geometry is homogeneous and can be described by the Kantowski-Sachs type metric. This particular geometry can be treated as a minisuperspace for which the techniques developed in the cosmological context by loop quantum cosmology (LQC) \cite{Ashtekar:2003hd, Ashtekar:2008zu, Ashtekar:2011ni} are available and can be readily used. The results of these investigations point to a singularity resolution of the bouncing cosmological type \cite{Ashtekar:2006uz, Ashtekar:2006wn, Singh:2009mz}.  Further evidence of singularity resolution  due to the implementation of the LQG dynamics was provided also in \cite{Pranzetti:2012pd, Pranzetti:2012dd}.

The study of the complete phase space in the symmetry reduced case started with the work of Kuchar \cite{Kuchar:1994zk} in metric variables and the work of Thiemann and and Kastrup \cite{Thiemann:1992jj} in complex Ashtekar variables which was then revised using LQG techniques \cite{Campiglia:2007}. In \cite{Bojowald:1999eh, Bojowald:2004ag, Bojowald:2005cb}  the Ashtekar-Barbero connection was used to obtain a kinematical description along with the Hamiltonian constraint. In \cite{ Gambini:2013ooa, Gambini:2013hna} the extension of the previous results within a classical modification of the Dirac algebra that transforms the symmetry reduced case in a Lie algebra allowed one to define the physical Hilbert space and observables corresponding to a metric that was shown to be free of singularity.  

A parallel line of investigation was conducted within the framework of covariant LQG, namely the spinfoam models \cite{Perez:2012wv}. The idea is that if the singularity is removed as in the homogeneous LQC, it is reasonable to expect a black hole-white hole transition named a Planck star \cite{Rovelli:2014cta} that can be modeled with a nonsingular metric \cite{Haggard:2014rza,DeLorenzo:2014pta} of the Hayward type \cite{Hayward:2005gi}.  The tunneling can then be studied in terms of transition amplitudes between coherent states representing  classical spacetimes \cite{Christodoulou:2016vny, Christodoulou:2018ryl}. Phenomenological consequences were discussed in \cite{Barrau:2014hda, Barrau:2014yka, Bianchi:2018mml}.

An alternative approach to model semiclassical and continuous spherically symmetric geometry has recently been pursued within the framework of group field theory (GFT) \cite{Oriti:2013aqa} in its operatorial formulation, providing a second quantized version of LQG. The main idea behind this approach is to describe homogeneous continuum geometries in terms of GFT condensate states encoding the information in a condensate wave function depending on a few collective variables \cite{Oriti:2015qva}. This allows one to model a black hole geometry by starting from the full theory and implementing the symmetry reduction at the quantum level \cite{Oriti:2015rwa, Oriti:2018qty}. In this case, application of the GFT condensates formalism to the cosmological setting has allowed one to recover modified Friedmann equations showing the presence of a bounce in the Planck regime \cite{Gielen:2013kla, Gielen:2014uga, Calcagni:2014tga, Oriti:2016qtz, deCesare:2016rsf}. One could then hope that also for the black hole case singularity resolution could be proven. However, implementation of the GFT dynamics in the black hole context is currently out of reach due to the highly challenging technical difficulties when dealing with generalized condensate states implementing graph connectivity. The main achievement of this manuscript is to show how to successfully implement the LQG dynamics to the case of a spherically symmetric black hole geometry while starting from the full theory and keeping the graph structure.

In this article we adopt the so-called quantum reduced loop gravity (QRLG) \cite{Alesci:2012md, Alesci:2013xd, Alesci:2013xya, Alesci:2014uha, Alesci:2014rra, Alesci:2015nja, Alesci:2016gub, Alesci:2016xqa, Alesci:2017kzc} approach and apply it to spherically symmetric geometries. 
At the heart of this approach lies the old familiar idea that a choice of symmetry compatible coordinate system drastically simplifies the task of solving the Einstein equations. QRLG intends to carry this simplification to the quantum level. To understand this better, it is helpful to briefly review how this framework has been applied to homogeneous anisotropic cosmologies.

In the case of Bianchi I spacetime, the existence of three Killing vector fields allows one to choose a coordinate system in which the metric is diagonal and only dependent on the time variable. The space of Bianchi I metrics is then just a subspace of the full Arnowitt-Deser-Misner (ADM) phase space that consists of homogeneous and diagonal 3-metrics. Note that diagonalizing 3-metrics is always achievable by imposing a partial gauge fixing (i.e. by using the gauge freedom provided by the spatial diffeomorphisms \cite{deturck1984, Grant:2009zz}), which comes at the cost of dealing with second class constraints and selecting a partially reduced phase space. A subspace of the latter is coordinate independent metrics singled out by symmetry. The classical use of a minisuperspace can thus be seen as first reaching the partially reduced phase space and then restricting it to its symmetric sector.
The QRLG approach to cosmology was devoted to access this sector at the quantum level as opposed to LQC, in which the symmetry reduction is performed classically and one is then left with finite-dimensional systems.
The main reason for this extra step is that the fundamental structure of LQG does not permit one to use differential geometry to define the notion of symmetry at the quantum level. In the process of symmetry reduction followed by quantization (as required in LQC) most of the structure of the full Hilbert space is lost and has to be reintroduced via assumptions.
 
The QRLG approach is to revert the process of symmetry reduction and quantization to derive a symmetric sector of LQG in which none of the fundamental structures of the full theory is lost. To achieve this goal, a reduced Hilbert space  is first selected from the full kinematical Hilbert space for which the metric is diagonal and then the symmetry reduction is performed, selecting homogeneous coherent states.
This procedure allows one to work with the complete structure of the full theory, consisting of quantum states of polymeric nature labeled by graphs and $SU(2)$ representations. Moreover, it shows that the minisuperspace effective quantization of LQC can be reproduced at the level of the expectation values of quantum operators acting on the partially gauge fixed Hilbert space. However, the presence of the graphs also leads to some modification in the deep Planckian regime. 

Here we intend to apply the same construction to spherically symmetric geometries. We will do this in four steps: in the first 
step, we implement the gauge fixing condition at the quantum level to define the partially gauge fixed Hilbert space \footnote{Recent studies \cite{Bodendorfer:2014wea, Bodendorfer:2015aca, Bodendorfer:2015qie} attempted to provide a quantization  for the reduced phase space in a radial gauge for the ADM variables different from the one introduced in \cite{Alesci:2018ewg} for connection variables.
 In \cite{Bodendorfer:2015qie} a scheme to implement symmetry reduction at the quantum level was also introduced; however, this analysis relies on a Peldan hybrid connection, yielding a description of the kinematical Hilbert space in terms of point holonomies and a restricted  action of the Hamiltonian constraint to a single point for a given 2-sphere. 
 }. This corresponds 
to the classical reduced phase space for a suitable choice of 
gauge that results in the triad $E^a _i$ having only the five components $E^r _3$, 
$E^{\theta} _{1}$, $E^{\theta} _{2}$, $E^{\phi} _{1}$, $E^{\phi} _{2}$. In terms of the 3-metric, this partial gauge choice amounts to having only the $rr$, $\theta \theta$, $\theta \phi$ and $\phi \phi$ components as nonzero. In the second step, we  project the constraints defined in the full theory to represent the classical gauge unfixed constraints \cite{Alesci:2018ewg}. The third step is to define states belonging to this kinematical Hilbert space where the classical notion of symmetry can 
be defined using spherically symmetric coherent states. Finally, we define the effective constraints by taking the expectation value of the quantum reduced constraints on the symmetry reduced states.

This article is organized as follows. In Sec. \ref{sec.canon} we review the canonical formulation of general relativity when restricted to spherically symmetric geometries. In Sec. \ref{sec:phase-space} we show how this symmetric phase space can be seen as a subspace of a partially gauge fixed phase space. We work out the first class constraint algebra obtained from the gauge unfixing procedure that preserves this subspace. In Sec. \ref{sec:qrlg.Hil} we build the reduced kinematical Hilbert space that implements at the quantum level this partial gauge fixing. We then derive the Hamiltonian constraint operator acting on this Hilbert space in Sec. \ref{sec:rhco}. After this, we build the semiclassical states representing spherically symmetric states  in Sec. \ref{sec:semi.states}. Finally in Sec. \ref{sec:eff.ham} we derive the effective Hamiltonian by taking the expectation value of the reduced Hamiltonian constraint evaluated on the coherent states. Further technical details are presented in  Appendixes \ref{sec:A}, \ref{sec:B}; Appendix  \ref{sec:C} contains an alternative approach to the quantization of the Lorentzian part of the Hamiltonian constraint.

\section{Canonical Formulation in Ashtekar Variables for Geometries in Spherical Symmetry} \la{sec.canon}

The ADM formalism \cite{adm} describes the Hamiltonian evolution of initial data sets in general relativity. In this context, a vacuum initial data set consists of a spacelike Cauchy surface  \footnote{$\Sigma_t$ represents an instant of time in the spacetime $M$. One tacitly assumes that $M = \mathbb{R}\times \Sigma_t$.} $\Sigma_t$ together with its intrinsic metric $q_{ab}$ and a symmetric tensor $\pi^{ab}$  that are required to satisfy
\ba \la{cons-hamil}
&&  \frac{{}^{(3)} R}{4 \kappa^2} - q^{-1} \pi^{ab} \pi_{ab} + \frac{1}{2} q^{-1} \pi^2 =0, \nonumber\\
&& D_a (q^{-1/2} \pi^{ab}) = 0,
\ea 
on $\Sigma_t$. 
Here $D_a$ is the $q_{ab}$ compatible torsion-free derivative operator, $\kappa \equiv 8 \pi G$, $ q \equiv \text{det} (q_{ab})$, and $\pi \equiv q_{ab} \pi^{ab}$. Once embedded in a four-dimensional spacetime, $\pi^{ab}$
is related to the extrinsic curvature $K_{ab}$ of $\Sigma_t$ by 
\be 
\pi^{ab} = \frac{1}{2 \kappa}\sqrt{q} (K^{ab} - K q^{ab}).
\ee 
The canonical phase space variables, namely $q_{ab}$ and $\pi^{ab}$,
are evolved by the relevant Hamiltonian (to be described below) and 
are subject to the following Poisson bracket
\be \la{pb1}
\big \{\pi^{ab} (t, \mathbf{x}) , q_{cd}(t, \mathbf{y}) \big \} = 2\kappa \delta ^a _{(c} \delta ^{b} _{d)} \delta^3 (\mathbf{x}-\mathbf{y})
\ee
at all times. Note that not all components of $q_{ab}$ and $\pi^{ab}$ are independent; of 12 components, 4 can be eliminated by a choice of coordinate gauge and another 4 are eliminated 
by virtue of Eq. \eqref{cons-hamil}. This leaves two propagating degrees of freedom for the gravitational field as expected.  

The intrinsic metrics on all $\Sigma_t$ can be sewed together to 
provide a spacetime metric given by 
\be \la{genmetadm} 
ds^2 = -N^2 dt^2 + q_{ab}(dx^a + N^a dt)(dx^b + N^b dt),
\ee
where $N$ is the lapse function and $N^a$ is the shift vector. 
A choice of $N$ and $N^a$ determines a foliation for the spacetime.
One needs to specify a foliation prior to solving the Hamilton's equations for $q_{ab}$ and $\pi^{ab}$.

Here we are interested in the case where $\Sigma_t$ has the topology $\Sigma_t=\R\times S^2$ and the spacetime geometry is assumed to be spherically symmetric. The most generic spacetime metric is then given by
\be\la{dsspher}
ds^2= -N^2 dt^2 +\Lambda^2 \big(dr+N^r dt\big)^2+R^2 \big(d\theta^2+\sin^2{\theta} \ d\varphi^2 \big)\,,
\ee
where $N, N^r, R, \Lambda$ are functions of $r$ and $t$, with $-\infty<t,r<\infty$. $\Lambda(t,r)$ and $R(t,r)$ are assumed to be positive functions; together with their conjugate momenta they represent the set of phase space canonical variables. Note that it follows from the above equation that the intrinsic metric on the spacelike hypersurfaces  is
\be
d \sigma ^2= \Lambda^2 dt^2 +R^2 \big(d\theta^2+ \sin^2{\theta} \ d\varphi^2 \big).
\ee
Expectedly, once a foliation is chosen, two independent functions are sufficient to describe an arbitrary metric in spherical symmetry.


As it turns out, the canonical quantization program is most conveniently formulated  in terms of the Ashtekar-Barbero connection $A^i _a$ and the densitized triad $E^a _i$ instead of $q_{ab}$ and $\pi^{ab}$. \footnote{See \cite{abhay-canonical} for a thorough exposition on the subject.} This way, deriving the quantum corrected semiclassical Hamiltonian is significantly more straightforward, as will be shown in the subsequent sections. The spatial index $a$ runs over $\{r,\theta,\varphi\}$, while the $SU(2)$ internal index $i \in \{1,2,3\}$.
One then has the following Poisson bracket
\be\la{PB}
\big\{ A_a^i(\mathbf{x}),E^{b}_j(\mathbf{y}) \big\} = \kappa\gamma  \delta^i_j\delta^b_a  \delta^3(\mathbf{x}-\mathbf{y})
\ee
in lieu of Eq. \eqref{pb1}. Note that here $\delta^3(\mathbf{x}-\mathbf{y}) \equiv \delta(r_{\mathbf{x}}-r_{\mathbf{y}})\delta(\theta_{\mathbf{x}}-\theta_{\mathbf{y}})
\delta(\varphi_{\mathbf{x}}-\varphi_{\mathbf{y}})$ and $\gamma$ 
is the Barbero-Immirzi parameter.

To derive the densitized triad, we begin by deriving the tetrad $e^{I} _{\alpha}$ for the metric \eqref{dsspher}. This is done using 
the relation $g_{\alpha \beta}=e^I_{\alpha} e_{I \beta}$. A quick calculation reveals
\baa
g_{tt}&=&-N^2+\Lambda^2 (N^r)^2= -(e^0_t)^2+(e^3_t)^2\,,\\
g_{tr}&=&\Lambda^2 N^r= -e^0_t e^0_r+e^3_t e^3_r\,,\\
g_{rr}&=&\Lambda^2 = -(e^0_r)^2+(e^3_r)^2\,,\\
g_{\theta\theta}&=&R^2=e^1_\theta e_{1\theta}+e^2_\theta e_{2\theta}\,,\\
g_{\varphi\varphi}&=&R^2(\sin{\theta})^2 =e^1_\varphi e_{1\varphi}+e^2_\varphi e_{2\varphi}\,,\\
g_{\theta\varphi}&=&0=e^1_\theta e_{1\varphi}+e^2_\theta e_{2\varphi}\,,
\eaa
from which we read off the complete set of tetrad components 
\begin{subequations}
\ba \la{tet0}
e^0&=& Ndt\,,\\
e^3&=& \Lambda N^r dt + \Lambda dr\,,\\
e^1&=&R\cos{\tilde\alpha} d\theta-R\sin{\theta}\sin{\tilde\alpha}d\varphi\,,\\
e^2&=&R\sin{\tilde\alpha} d\theta+R\sin{\theta}\cos{\tilde\alpha} d\varphi\,.
\ea
\end{subequations}
Here we have left a rotation freedom for the components $e^1$ and $e^2$ described by the angle $\tilde\alpha$, which can have any arbitrary given value. 
The  tetrad components  also satisfy the following equations:
\baa
&&e^0_r e^r_3+e^0_t e^t_3=0\,,\\
&&e^1_\theta e^\theta_{2}+e^1_\varphi e^\varphi_{2}=0\,.
\eaa

If we go to the time gauge where $e^0_a=n_a$, the densitized triad 
\be
E=E^a_i \tau^i \partial_a\,,\quad {\rm where}~~E^a_i=\frac{1}{2}\epsilon^{abc}\epsilon_{ijk}e^j_b e^k_c\,,
\ee
becomes
\be
E=R^2\sin\theta \,\tau^3 \partial_r+\Lambda R\sin{\theta}\big(\cos{\tilde\alpha}\,\tau^1+\sin{\tilde\alpha}\,\tau^2\big)\partial_\theta
+\Lambda R \big(\cos{\tilde\alpha}\,\tau^2-\sin{\tilde\alpha}\,\tau^1 \big)\partial_\varphi\,.\la{SE}
\ee
Here $\tau^i$ are basis vectors in the internal space. \footnote{We use the anti-hermitian basis $\tau_i$, where $[\tau_i,\tau_j]=\epsilon_{ij}\!^k\tau_k$ and $(\tau_i)^2=-\frac{1}{4}\I$ for all $i$'s and $\Tr {\tau_i\tau_j}=-\frac{1}{2}\delta_{ij}$.}

The connection components  $\omega^{IJ}_a=-\omega^{JI}_a$ can be computed from the torsion-free condition $de^I=-\omega^I\!_J\wedge e^J$. The explicit derivation is provided in Appendix \ref{sec:A}. Using the result of Eqs. \eqref{omega1}-\eqref{omega7}, the Ashtekar-Barbero connection $A^i=\Gamma^i+\gamma K^i$, where $\Gamma^i=-\frac{1}{2}\epsilon^i\!_{jk} \omega^{jk}$ and $K^i=\omega^{0i}$, is given by
\ba
A=A^i_a \tau_i dx^a&=&
-\gamma \frac{(\Lambda' N^r +\Lambda {N^r}'- \dot{\Lambda} )}{N} \tau_3 \, dr\n\\
&+& R' \left\{ \left[-\frac{\gamma}{N} \left(N^r-\frac{\dot R}{R'}\right)\sin{\tilde\alpha}+ \frac{1}{\Lambda}\cos{\tilde\alpha}\right]\,\tau_2 +\left[-\frac{\gamma}{N}\left(N^r-\frac{\dot R}{R'}\right)\cos{\tilde\alpha}- \frac{1}{\Lambda}\sin{\tilde\alpha}\right] \,\tau_1\right \}  d\theta\n\\
&+& \sin{\theta}R' \left\{\left[-\frac{\gamma}{N}\left(N^r-\frac{\dot R}{R'}\right)\cos{\tilde\alpha}- \frac{1}{\Lambda}\sin{\tilde\alpha}\right]\,\tau_2-\left[-\frac{\gamma}{N} \left(N^r-\frac{\dot R}{R'}\right)\sin{\tilde\alpha}+ \frac{1}{\Lambda}\cos{\tilde\alpha}\right] \,\tau_1\right\} d\varphi \n\\
&+&\cos{\theta} \tau_3 \,d\varphi \,.\la{Sconn}
\ea

If one is to use the notation of \cite{Bojowald:2005cb}, where the spherically symmetric Ashtekar-Barbero connection and triad are written as
\ba
E&=& E^r(t,r)\sin{\theta} \tau_3\partial_r+\left[E^1(t,r)\tau_1 + E^2(t,r)\tau_2\right]\sin{\theta}\partial_\theta+ \left[E^1(t,r)\tau_2 -E^2(t,r)\tau_1\right]\partial_\varphi \,,\la{SE2}\\
A&=& A_r(t,r)\tau_3 dr+\left[A_1(t,r)\tau_1 + A_2(t,r)\tau_2\right]d\theta + \sin{\theta} \left[A_1(t,r)\tau_2 -A_2(t,r)\tau_1\right]d\varphi
+\cos{\theta}\tau_3 d\varphi\,,\la{Sconn2}
\ea
one has
\ba
&&E^r(t,r)=R^2\,,\quad E^1(t,r)=\Lambda R \cos{\tilde\alpha}\,,\quad E^2(t,r)=\Lambda R \sin{\tilde\alpha}\,,\\
&& A_r(t,r)=-\gamma \frac{(\Lambda' N^r +\Lambda {N^r}'- \dot{\Lambda} )}{N} \,,\\
&&A_1(t,r)=R'  \left[-\frac{\gamma}{N}\left(N^r-\frac{\dot R}{R'}\right)\cos{\tilde\alpha}- \frac{1}{\Lambda}\sin{\tilde\alpha}\right]  \,,\\
&&A_2(t,r)=R'  \left[-\frac{\gamma}{N} \left(N^r-\frac{\dot R}{R'}\right)\sin{\tilde\alpha}+ \frac{1}{\Lambda}\cos{\tilde\alpha}\right]  \,.
\ea


The Poisson bracket \eqref{PB} then takes the form 
\begin{subequations}
\ba
&&\big\{A_r(t,r), E^r(t,r')\big\}=2G\gamma \,\delta(r-r')\,,\\
&&\big\{A_1(t,r), E^1(t,r')\big\}=G\gamma \,\delta(r-r')\,,\\
&&\big\{A_2(t,r), E^2(t,r')\big\}=G\gamma\,\delta(r-r')\,.
\ea
\end{subequations}

\section{Constraints for the Partially Gauge Fixed Phase Space }\la{sec:phase-space}

In this section we derive the classical Hamiltonian, diffeomorphism, and gauge constraints subject to the particular gauge fixing scheme 
of \cite{Alesci:2018ewg}.

The classical phase space is characterized by the following standard seven constraints
\begin{subequations}
\label{eq:constraints:1}
\begin{align}
&\kappa G_i=\partial_a E^a_i+\epsilon\indices{_{ij}^k}A_a^jE_k^a\,, &\text{Gauss constraint}\\
&\kappa H_a=F_{ab}^iE_i^b-A^i_a G_i\,, &\text{Diffeomorphism constraint}\la{DC}\\
&\kappa H=\frac{ E_i^aE_j^b}{\sqrt{\text{det}(E)}}\left[\epsilon\indices{^i^j_k}F_{ab}^k-2(1+\gamma^2)K^i_{[a}K^j_{b]}\right] \,,&\text{Hamiltonian constraint}\la{HC}
\end{align}
\end{subequations}
where
\be
\label{eq:f:1}
F_{ab}^i=\partial_aA_b^i-\partial_bA_a^i+\epsilon\indices{^i_{jk}}A_a^jA_b^k
\ee
are  the curvature components of the Ashtekar-Barbero connection. Notice that when we substitute the triad and the connection given in Eqs. \eqref{SE} and \eqref{Sconn} in the above constraints, we are merely left with the Gauss constraint in the $3=r$ direction, the radial diffeomorphism constraint, and the Hamiltonian constraint.

In \cite{Alesci:2018ewg} we performed the Dirac treatment of this constrained system when the following  radial partial gauge fixing conditions are introduced:
\begin{subequations}
\label{gf}
\begin{align}
&E_I^r=0\,,\quad I=1,2\,,\\
&E_3^A=0\,,\quad A=\theta,\phi\,.
\end{align}
\end{subequations}
These conditions can be seen as additional constraints in the phase space which render the Hamiltonian system to be of second class. Instead of building the Dirac bracket to impose the second class constraints, in \cite{Alesci:2018ewg} we followed an alternative strategy that goes under the name of  ``gauge unfixing'' procedure (GU) \cite{Mitra:1989fg, Mitra:1990mp, Anishetty:1992yk}. 
This procedure, which also requires the inversion of the Dirac matrix, allows us to use the Poisson bracket given in Eq. \eqref{PB} for imposing the remaining three first class constraints. However, while one of these constraints still corresponds to the original $i=3$ component of the Gauss constraint, the other two are given by the reduced phase space part of the radial diffeomorphism constraint given in Eq. \eqref{DC} and the Hamiltonian constraint in Eq. \eqref{HC} plus extra terms. Explicitly, denoting with a tilde the extended\footnote{The connotation ``extended'' refers to the fact that the new expressions for the constraints are obtained by replacing the connection components conjugate to the triad components that we have gauge fixed with their extended versions. This results from solving the second class constraints explicitly. In fact, in \cite{Alesci:2018ewg} it was shown that this corresponds  to adding a linear combination of second class constraints to the original $H_r$ and $ H$ so that the new expressions preserve the gauge conditions. Effectively, this gauge unfixing procedure amounts to having a new set of first class constraints, equivalent to the initial one, but now written as functionals of only the reduced phase space coordinates.} representation of these remaining constraints, we have
\ba
 \tilde{G}_3[\alpha_3] &=& {}^{R} G_3 [\alpha_3]=\frac{1}{\kappa}\int d^3 x \  \alpha_3 \ \Big[\partial_r E^r _3 + \epsilon_{3I} \ ^{J} A^I _B E^ B _J\Big],   \label{eq:G:tilde}\\
\tilde{H}_r[N^r]&=&\frac{1}{\kappa}{}^{\va R} H_r[N^r]+\frac{1}{\kappa}\int d^3 x\,\left(\partial_AN^r\right)\left[\frac{\epsilon^{IJ}E_I^A\partial_BE_J^B}{E_3^r} 
+\frac{\delta^{IJ}E_I^AE_J^B\mathcal{I}_B}{(E_3^r)^2}\right]\,,\label{eq:hr:tilde}\\
\tilde{H}_{\va E}[N]
&=&\frac{2}{\kappa}\int d^3 x\, \frac{N}{\sqrt{\text{det}(E)}}\Bigg[{}^{\va R} H_{\va E}\n\\
&-&\gamma\left\{\frac{\mathcal{I}_AE_I^A \mathcal{I}_BE^{IB}}{(E_3^r)^2}
+\epsilon^{IJ}\frac{E_I^A(\partial_BE^B_J)\mathcal{I}_A}{E_3^r}
+\epsilon^{IJ} E_I^A\mathcal{I}_B \partial_A\left(\frac{E_J^B}{E_3^r}\right)
-E_3^rE_I^A \partial_A\left(\frac{\partial_BE^{IB}}{E_3^r}\right)
\right\}\Bigg]
\,,\label{eq:hh:tilde}\\
\tilde{H}_{\va L}[N]&=&-2\frac{(1+\gamma^2)}{\kappa}\int d^3 x\, \frac{N {}^{\va R} H_{\va L}}{\sqrt{\text{det}(E)}}, \n\\
\ea
where $\alpha_3$ is the $i=3$ component of the $\alpha_i$ smearing function associated with the Gauss constraint ,${}^{\va R} H_r$ is the {\it reduced radial diffeomorphism} 
\be
\label{eq:barh}
{}^{\va R} H_r=\left(\partial_rA_A^I\right)E_I^A-A_r^3\partial_rE_3^r\,,
\ee
 ${}^{\va R} H_{\va E}$ is the {\it reduced Euclidean Hamiltonian}
\ba
\label{eq:barh:2}
{}^{\va R} H_{\va E}=E_3^rE_I^A\epsilon\indices{^I_J}\partial_rA_A^J+E_I^AE_J^BA^I_{[A}A^J_{B]}+E_3^rE_I^AA_r^3A_A^I\,,
\ea
and ${}^{\va R} H_{\va L}$ is the {\it reduced Lorentzian Hamiltonian}
\ba
\label{eq:barh:3}
{}^{\va R} H_{\va L}=E_I^AE_J^B K^I_{[A}K^J_{B]}+E_3^rE_I^A K_r^3 K_A^I\,.
\ea

To shorten the notation, in the expressions above  we have defined
\ba
D_A&\equiv& E^B_I\partial_AA_B^I-\partial_B\left(A_A^IE^B_I\right)\,,\\
\mathcal{I}_A&\equiv& \int_0^r dr'\left[D_A+E_3^r\partial_A A_r^3\right]_{r'}\,.
\ea

Notice that the Lorentzian part of the Hamiltonian constraint does not pick up any extra contribution in addition to the reduced term. This is because, once projected on the gauge surface described by Eq. \eqref{gf}, the Lorentzian term given in Eq. \eqref{HC} does not contain any extrinsic curvature component conjugate to the flux (i.e. triad) components that we have gauge fixed; therefore, in the gauge unfixing procedure of \cite{Alesci:2018ewg}, we do not have to perform any substitution of extended momenta inside $H_{\va L}$.

The Hamiltonian constraint that appears in Eq. \eqref{HC} can also be written as
\be\la{LHR}
\kappa H=\frac{s}{\gamma^2}\left[\frac{\epsilon\indices{^i^j_k} E_i^aE_j^b F_{ab}^k}{\sqrt{\text{det}(E)}}+\left(1-\frac{\gamma^2}{s}\right)\sqrt{\text{det}(E)} R\right] \,,
\ee
where $s=\pm$ is the spacetime metric signature, $R$ is the Ricci scalar given by 
\be
R=R_{ab}\!^{ij}e^a_i e^b_j=-\epsilon_{ijk}R^k_{ab}e^a_i e^b_j
\,,
\ee
with 
\be
\quad R^k_{ab}=2\partial_{[a} \Gamma^k_{b]}+\epsilon^k\!_{lm} \Gamma^l_a  \Gamma^m_b\,.
\ee

As we will see in Sec. \ref{sec:qrlg.Hil}, our aim is to build the QRLG quantum theory where we will access the reduced
phase space using projectors from the full theory.
Therefore we are not going to perform the reduced phase space
quantization as attempted in the existing literature. This means
that if we had the physical Hilbert space of the full theory, namely the kernel of the quantum operators corresponding to the constraints given in Eq. \eqref{eq:constraints:1}, we would simply
transform it to the Hilbert space of QRLG. Unfortunately, while
the structure of the solutions to the Gauss and the vector constraints are very well understood in the full theory \cite{Thiemann:2007zz, Ashtekar:2004eh, rovelli_2004}, very little is known about the structure of the Hamiltonian constraint \cite{Thiemann:1996aw, Alesci:2015wla} and its kernel \cite{Alesci:2011ia,Alesci:2013kpa}.
Our strategy is then to use the projector on the kernel of the Gauss and the vector constraints and instead quantize the reduced Hamiltonian constraint. In particular our final aim in this paper 
is not to find the full set of solutions but just to derive the 
effective reduced Hamiltonian constraint when the symmetry reduction is imposed at the level of coherent states.   
To this end, we will compute in the next section the reduced Hamiltonian constraint  for spherically symmetric geometry.

\subsection{Symmetric subspace of the reduced phase space: Hamiltonian}\la{sec:Schw}

Using the curvature components  of the Ashtekar-Barbero connection that we worked out in Eq. \eqref{FA},
 the spherically symmetric Euclidean part of the Hamiltonian constraint reduces to
\ba
 H_{\va sph}^{\va E}[N]&=&\frac{1}{\kappa}\int_\Sigma d^3x N(x) \frac{E^a_i E^b_j}{\sqrt{{\rm det}(E)}} \epsilon^{ij}\!_{k}F^k_{ab}(A)\n\\
&=&\frac{2}{\kappa}\int_\Sigma d^3x \frac{ N(x) }{\sqrt{(E^\theta_1 E^\varphi_2-E^\theta_2 E^\varphi_1 )E^r_3}}
\Bigg[E^r_3
\Big(E^\theta_1F^2_{r\theta}(A)-E^\theta_2F^1_{r\theta}(A)
+E^\varphi_1F^2_{r\varphi}(A)-E^\varphi_2F^1_{r\varphi}(A)
\Big)\n\\
&+&\Big(E^\theta_1 E^\varphi_2-E^\theta_2 E^\varphi_1\Big)F^3_{\theta\varphi}(A)
\Bigg]
\n\\
&=&\frac{2}{\kappa}\int_\Sigma d^3x\frac{ N(x) \sin\theta}{\sqrt{((E^1)^2 +(E^2)^2  )E^r}}
\Bigg[2E^r A_r\left(E^1 A_1+E^2A_2
\right)
+2E^r (E^1A'_2-E^2A'_1)\n\\
&+&\Big((E^1)^2 +(E^2)^2 \Big)\Big(
(A_1^2+A^2_2)-1\Big)
\Bigg]\,.\la{HH-spher}
\ea

On the other hand, the reduced Euclidean Hamiltonian given in Eq. \eqref{eq:barh:2} yields
\ba\la{EHHR}
{}^{\va R}H^{\va E}[N]&=&\frac{2}{\kappa}\int_\Sigma d^3{x}\, \frac{N}{\sqrt{\text{det}(E)}} \left[E_3^rE_I^AA_r^3A_A^I+ E_3^rE_I^A\epsilon\indices{^I_J}\partial_rA_A^J+E_I^AE_J^BA^I_{[A}A^J_{B]}\right]\n\\
&=&\frac{2}{\kappa}\int_\Sigma d^3x\frac{ N(x) \sin\theta}{\sqrt{((E^1)^2 +(E^2)^2  )E^r}}
\Bigg[2E^r A_r\left(E^1 A_1+E^2A_2
\right)
+2E^r (E^1A'_2-E^2A'_1)\n\\
&+&\Big((E^1)^2 +(E^2)^2 \Big)\left(
A_1^2+A^2_2\right)
\Bigg]\,,
\ea
while the correction terms in the extended version of the Euclidean Hamiltonian constraint given in Eq. \eqref{eq:hh:tilde}, resulting from the gauge unfixing procedure, reduce to
\ba
&-&\frac{2}{\kappa}\int_\Sigma d^3{x}\, \frac{N}{\sqrt{\text{det}(E)}}
\left\{\frac{\mathcal{I}_AE_I^A \mathcal{I}_BE^{IB}}{(E_3^r)^2}
+\epsilon^{IJ}\frac{E_I^A(\partial_BE^B_J)\mathcal{I}_A}{E_3^r}
+\epsilon^{IJ} E_I^A\mathcal{I}_B \partial_A\left(\frac{E_J^B}{E_3^r}\right)
-E_3^rE_I^A \partial_A\left(\frac{\partial_BE^{IB}}{E_3^r}\right)
\right\}\n\\
&=&\frac{2}{\kappa}\int_\Sigma d^3{x}\, \frac{N}{\sqrt{\text{det}(E)}} E_I^A \partial_A\partial_BE^{IB}\n\\
&=&-\frac{2}{\kappa}\int_\Sigma d^3x\frac{ N(x) \sin\theta}{\sqrt{((E^1)^2 +(E^2)^2  )E^r}}
\Big((E^1)^2 +(E^2)^2\Big)\,.\la{EHext}
\ea
We thus see that
\be
 H_{\va sph}^{\va E}[N]= \tilde H^{\va E}[N]\,,
\ee
as expected; namely, the role of the extra terms in the extended Euclidean Hamiltonian is to provide the contribution given by $\partial_\theta A^3_\varphi$ inside $F^3_{\theta\varphi}$ in $H_{\va sph}^{\va E}[N]$, since this is the only term where a connection component not belonging to the reduced phase space appears.

For the spherically symmetric Lorentzian part  of the Hamiltonian constraint given in Eq. \eqref{eq:barh:3}, we have
\ba
{H}_{\va L}[N]&=&-2\frac{(1+\gamma^2)}{\kappa}\int_\Sigma d^3x\, \frac{N(x)}{\sqrt{\text{det}(E)}}\left( E_I^AE_J^B K^I_{[A}K^J_{B]}+E_3^rE_I^A K_r^3 K_A^I\right).
\ea
Let us also write the Lorentzian Hamiltonian constraint in terms of the Ricci scalar. By means of Eq. \eqref{RE}, we have 
\ba\la{H_L}
{H}_{\va L}[N]&=&-\frac{1}{\kappa}\left(1+\frac{1}{\gamma^2}\right)\int_\Sigma d^3x\,N(x) \sqrt{\text{det}(E)}R\n\\
&=&\frac{1}{\kappa}\left(1+\frac{1}{\gamma^2}\right)\int_\Sigma d^3x\, \frac{N(x)}{2 (E^r _3)^{5/2} \sin^{1/2}{\theta} \ [(E^{\varphi} _1 )^2+(E^{\varphi} _2 )^2]^{3/2}} \n\\
&\times & \bigg[-2 \sin^2 {\theta} (E^{\varphi} _1)^4 \Big[(\partial_{\theta} E^r _3)^2 - E^r _3 \partial^2 _{\theta} E^r _3 \Big] - 2 \sin{\theta} E^r _3 (E^{\varphi} _1)^3 [\partial_{\theta} E^r _3 \partial_{\theta} E^{\theta} _2 + E^r _3 \partial^2 _{\theta} E^{\theta} _2] \n\\
&-& 2 E^r _3 E^{\varphi} _1 \Big(E^{\varphi} _2  \big[(\sin{\theta} \ E^{\varphi} _2 \partial_{\theta} E^r _3 - 4 E^r _3 \partial_{\theta} E^{\theta} _1) \partial_{\theta} E^{\theta} _2 + \sin{\theta} \ E^r _3 E^{\varphi} _2 \partial^2_{\theta} E^{\theta} _2\big] \n\\
&+& 2 (E^r _3)^2 \partial_r E^{\varphi} _1 \partial_r E^r _3 \Big) + (E^{\varphi} _1)^2 \Big( -4 \sin^2{\theta} (E^{\varphi} _2)^2 \big[(\partial_{\theta} E^r _3)^2 - E^r _3 \partial^2 _{\theta} E^r _3\big]\n\\
&+& 2\sin{\theta} E^r _3 E^{\varphi} _2 \big[\partial_{\theta} E^r _3 \partial_{\theta} E^{\theta} _1 + E^r _3 \partial^2 _{\theta} E^{\theta} _1\big] + (E^r _3)^2 \big[4 (\partial_{\theta} E^{\theta} _1)^2 + (\partial_{r}E^r _3)^2 \n\\
&+&  4 E^r _3 \partial^2 _r E^r _3 \big] \Big) + E^{\varphi} _2 \Big( -2 \sin^2 {\theta} \ (E^{\varphi} _2)^3 \big[(\partial_{\theta} E^r _3)^2 - E^r _3 \partial^2 _{\theta} E^r _3\big] + 2 \sin{\theta} \n\\
&\times & E^r _3 (E^{\varphi} _2)^2 \big[\partial_{\theta} E^r _3 \partial_{\theta} E^{\theta} _1 + E^r _3 \partial^2_{\theta} E^{\theta} _1\big] - 4 (E^r _3)^3 \partial_r E^{\varphi} _2 \partial_r E^r _3 + (E^r _3)^2 E^{\varphi} _2 \n\\
& \times &  \big[4 (\partial_{\theta} E^{\theta} _2)^2 + (\partial_r E^r _3)^2 + 4 E^r_3  \partial^2 _r E^r _3\big]\Big) \bigg] \n\\
&=&-\frac{2}{\kappa}\left(1+\frac{1}{\gamma^2}\right)\int_\Sigma d^3x\, N(x) \sin\theta
\left[\frac{2R}{\Lambda}\left(\frac{R'\Lambda'}{\Lambda}-R''\right)
+\Lambda\left(1-\frac{(R')^2}{\Lambda^2}\right)\right]\,.
\ea

The expression in terms of the fluxes that appears in the second equality above is the one that we are going to use to quantize the Lorentzian Hamiltonian constraint. In fact, given the diagonal action of the reduced flux operators, we can compute its expectation value in a lengthy but straightforward manner, without having to rely on any recoupling theory. Despite the fact that we used spherical symmetry to arrive at this expression, such simplifications would in any case be enforced by the coherent states that we build in the next section. This way we have simplified the calculation of the expectation value of the Lorentzian Hamiltonian without the need to sacrifice any relevant quantum corrections (see  Appendix \ref{sec:A} for more details on the symmetry assumptions used to derive the expression above).

\section{Quantum Reduced Loop Gravity Kinematical Hilbert Space} \la{sec:qrlg.Hil}

In this section we construct step by step a reduced kinematical Hilbert space $\mathcal{H}^{R}$ that implements at the quantum level the radial partial gauge fixing in Eq. \eqref{gf}. We  start with the standard LQG kinematical Hilbert space $\mathcal{H}^{K}$ representing quantum holonomy-flux algebra. We then perform a weak imposition for the quantum version of Eq. \eqref{gf} that  restricts the non-gauge-invariant spin network basis states arriving at $\mathcal{H}^{R}$.
Symbolically, $P \mathcal{H}^K = \mathcal{H}^{R}$, where $P$ is the projection operator defined below. 

The first ingredient of the quantum reduction process consists of a choice of spatial manifold triangulation adapted to the topology of interest and selecting a subclass of graphs labeling the spin network basis of $\mathcal{H}^{R}$. A natural choice for a spherically symmetric geometry is to restrict to cuboidal triangulations, where at each vertex we have three directions, one corresponding to the radial direction and the other two to the angular directions on the 2-spheres foliating the spatial manifold. 

In what follows we present the technical details of our construction.

\subsection{Reduced spin network states}

\begin{figure}[h]
\centerline{ \(
\begin{array}{c}
\includegraphics[height=5cm]{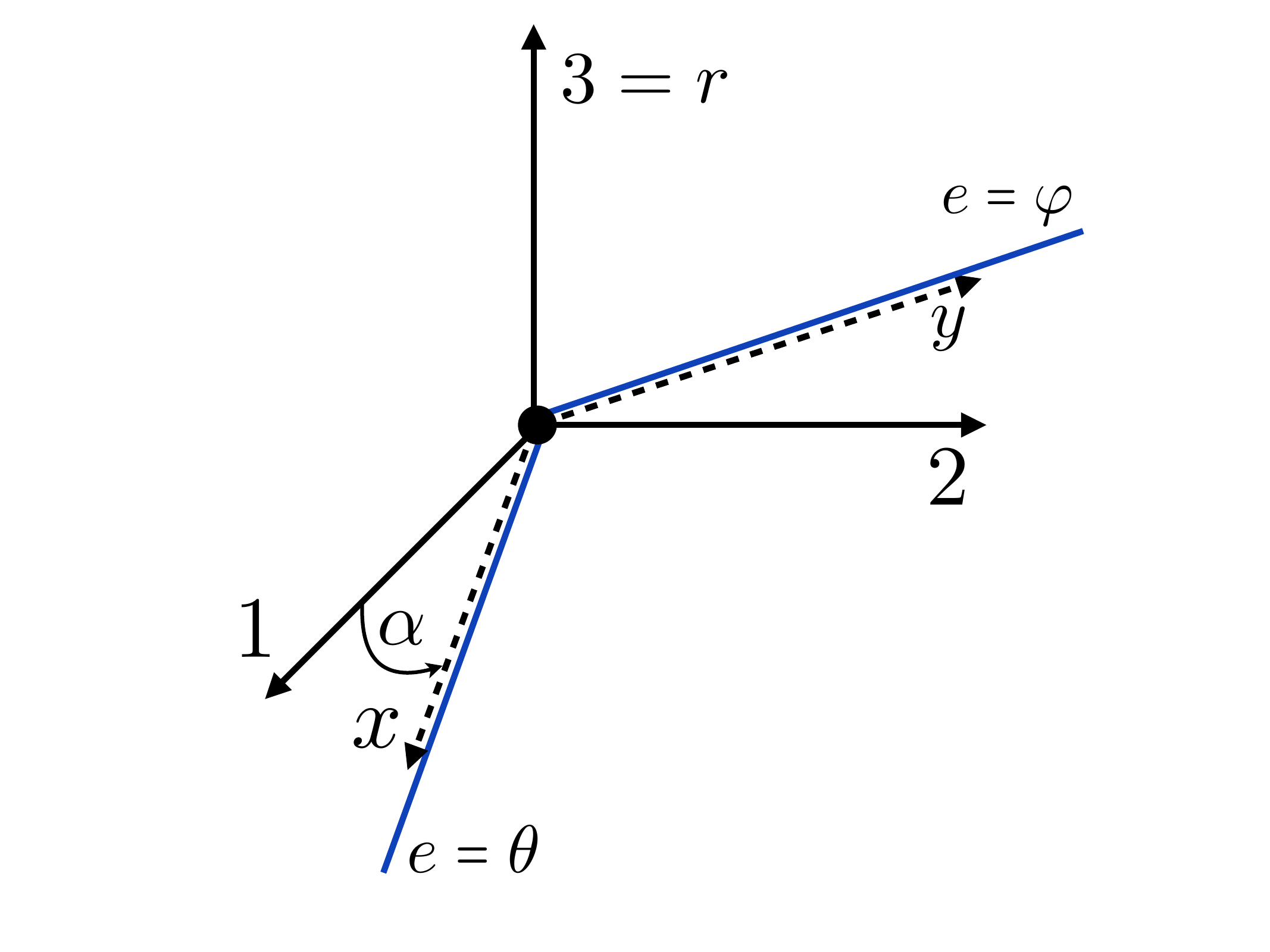}
\end{array}\) } \caption{Tangent and internal directions.}
\label{directions1}
\end{figure}

Given a two-dimensional surface $S^a$ with  normal  vector
\be
n_a=\frac{\partial x^b}{\partial \sigma_1}\frac{\partial x^c}{\partial \sigma_2}\epsilon_{abc}\,,
\ee 
where $\sigma_1$ and $ \sigma_2$ are local coordinates on $S^a$, we restrict our choice of fluxes to the surfaces whose normal vectors  are aligned with the three tangent directions $r,\theta,\varphi$.
We choose two $su(2)$ orthonormal bases labeled by $\{x,y,z\}$
and $\{1,2,3\}$ for which the basis elements $3$ and $z$ coincide [$\{1,2\}$ differ from $\{x,y\}$ by an $SO(2)$ rotation $\alpha$].   
Consistently with our classical gauge fixing, we take the direction $r$ to be aligned with the internal direction $3$ while the directions $\theta$ and $\varphi$ are aligned with $x$ and $y$ (see Fig. \ref{directions1}).
Therefore, our gauge fixing reads
\ba
&&E_3(S^\theta)=0=E_3(S^\varphi)\,,\la{GF1}\\
&&E_1(S^r)=0=E_2(S^r)\,,\la{GF2}
\ea
where
\be\la{E}
E_i(S^a)=\int E^a_i n_a d\sigma_1 d\sigma_2\,.
\ee

In order to implement the above partial gauge fixing at the quantum level, we select cuboidal graphs $\Gamma$'s 
adapted to this set of coordinates. This means that the edges $\ell_e$ are aligned with the three directions $\{r, \theta, \varphi\}$ such that $ \dot{\ell}_e^a\propto \delta^a_e$. Along these directions the $SU(2)$ holonomies are
\be
g_e=\mathcal {P} e^{\int_{\ell_e} \tau_i A^i_a \dot{\ell}_e^a (s) ds}.
\ee 
The set of discretized 2-spheres at a given value of $r$ is equipped with a grid of plaquettes with edges labeled by the tangential coordinates $ \theta$ and $\varphi$.  In a graphical representation, this is given by
\be \la{6valvert}
 \begin{array}{c}
\includegraphics[width=6cm]{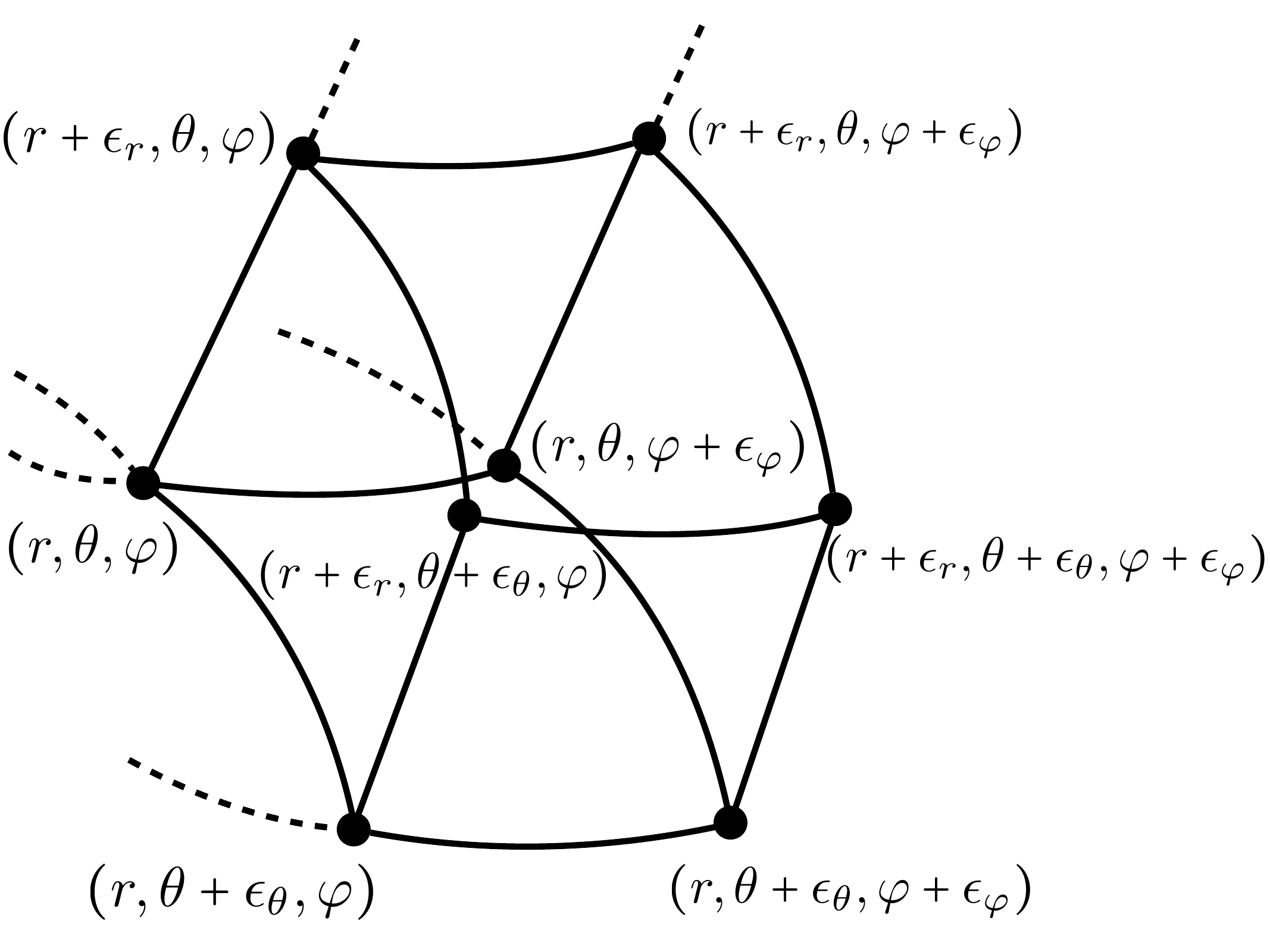} \end{array}
\ee
with $\epsilon_r$, $\epsilon_\theta$, and $ \epsilon_\varphi$ being coordinate lengths in the tangential directions. We designate this set as the set of reduced graphs. Our projection operator $P$ acts on $\mathcal{H}^K = \oplus_\gamma \mathcal{H}^K _{\gamma} $, where $\gamma$ are arbitrary graphs labeling $\mathcal{H}^{K}$, in two steps: first  by restricting $\gamma$ to $\Gamma$, where $\Gamma$ is given by cubulation of at most 6-valent vertices as shown in Eq. \eqref{6valvert}, and then projecting $\mathcal{H}^K _{\Gamma}$ to its reduced subspace $\mathcal{H}^R _{\Gamma}$.
The kinematical Hilbert space $\mathcal{H}^{R}$ is then obtained from the direct sum over all reduced graphs of this form that we construct from the union of $\Gamma$, namely
\be
\mathcal{H}^{\va R}=\oplus_\Gamma \mathcal{H}^{\va R}_{\va \Gamma}\,.
\ee
$\mathcal{H}^R _{\Gamma}$ is defined by assigning to 
each link in a given tangent direction the 
following basis elements 
 \begin{subequations}\la{D}
 \ba
&&{}^x\!D^{j_x}_{\bar m_x \bar n_x}(g_\theta)=\langle \bar m_x, \vec{u}_x| D^{j_x}(g_\theta)|\bar n_x, \vec{u}_x \rangle = 
\langle g_{\theta}|x,j_x, \tilde{m}_x, \tilde{n}_x \rangle \,,\la{Dx}\\
&&{}^y\!D^{j_y}_{\bar m_y \bar n_y}(g_\varphi)=\langle \bar m_y, \vec{u}_y| D^{j_y}(g_\varphi)|\bar n_y, \vec{u}_y \rangle = \langle g_{\varphi}| y,j_y, \tilde{m}_y, \tilde{n}_y\rangle \,,\la{Dy}\\
&&D^{j_z}_{\bar m_z \bar n_z}(g_r)=\langle \bar m_z, j_z| D^{j_z}(g_r)|j_z, \bar n_z \rangle = \langle g_{r}| r, j_r, \tilde{m}_z, \tilde{n}_z\rangle\,,
\ea
\end{subequations}
where $|j, m \rangle$ is an element of the spin basis that diagonalizes both  $\tau^i\tau_i$ and $\tau_3$, and $\bar m_x, \bar n_x=\pm j_x$ , $\bar m_y, \bar n_y=\pm j_y$, and $\bar m_z, \bar n_z=\pm j_z$. We denote two orthogonal unit vectors in the arbitrary internal directions $I \in \{x,y\}$ on the (1,2)-plane by
$\vec{u}_I$. Then
\be
|\bar n_I, \vec{u}_I \rangle= D^{j_I}(\vec{u}_I)|j_I, \bar n_I \rangle=\sum_{m} |j_I, m \rangle D^{j_I}_{m \bar n_I}(u_I)
\ee
is an $SU(2)$ coherent state having maximum or minimum magnetic number along $\vec{u}_I$ ($D^j_{mn}(g)$ are the standard  Wigner matrices in the spin basis $|j,m\rangle$).

The basis elements given in Eqs. \eqref{Dx} and \eqref{Dy} can also be written as
\be\la{states}
{}^I\!D^{j_I}_{\bar m_I \bar n_I}(g)=D^{j_I -1}_{\bar m_I m}(u_I) D^{j_I}_{mn}(g)D^{j_I }_{n \bar n_I}(u_I)\,,
\ee
where $u_I$ is an $SU(2)$ group element that rotates the $3$-axis into $\vec{u}_I$ and repeated indices are summed over. Given the convention shown in Fig. \ref{directions1}, we parametrize the rotation group elements as
\ba\la{ux}
&&u_x=R\left(\alpha, \frac{\pi}{2},0\right)
=e^{\alpha \tau_3 }e^{\frac{\pi}{2}\tau_2}\,,\\
&&u_y=R\left(\alpha+\frac{\pi}{2}, \frac{\pi}{2},-\frac{\pi}{2}\right)
=e^{(\alpha+\frac{\pi}{2}) \tau_3 }e^{\frac{\pi}{2}\tau_2}e^{-\frac{\pi}{2}\tau_3}\la{uy} \,.
\ea
The angle $\alpha$ above that enters the construction of the reduced states and the operators that we construct below is {\it a priori} independent of $\tilde{\alpha}$ that appears in the classical solutions for the connection and the densitized triad of Sec. \ref{sec.canon}
[the relation between the couples of  internal directions (1,2) and $(x,y)$ can be chosen independently for the classical solution and the quantum construction]. In order to implement the residual $U(1)$ gauge invariance, we are going to integrate over the angle $\alpha$ in the reduced states, while the $\tilde \alpha$ appearing in the classical solutions  given in Eqs. \eqref{SE2} and \eqref{Sconn2} for triad and connection, around which our semiclassical states are peaked, is held fixed.


Notice that unlike the reduced states built for cosmological applications in \cite{Alesci:2014uha}, in Eq. \eqref{D} we also include the off-diagonal terms, i.e. states peaked on maximum-minimum magnetic numbers and not just maximum-maximum or minimum-minimum. As we  will show below, these states are in fact allowed by the radial gauge fixing given in Eq. \eqref{gf}. They are important in the black hole context since the symmetry reduced Ashtekar-Barbero connection contains general off-diagonal terms (see Sec. \ref{sec:Schw}).

We can now show that on $\mathcal{H}^{R}$ the gauge fixing conditions  given in Eqs. \eqref{GF1} and \eqref{GF2} are weakly satisfied. 
Let us concentrate on 3-cells with surfaces $S^a$ that intersect the respective dual edges $\dot{\ell}_e^a$ only once. Hence $S^a$ are three  orthogonal faces of the cube dual to a 6-valent node of the reduced graph.
This provides a regularization of the reduced fluxes. 
By means of the Baker-Hausdorff formula, let us first derive the following relations:
\begin{subequations}\la{uzxy}
\ba
&&u_x \tau_3 u^{-1}_x
=\tau_1\cos{\alpha}+\tau_2\sin{\alpha}=\tau_x\,,\la{3x}\\
&&u_y \tau_3 u^{-1}_y=-\tau_1\sin{\alpha}+\tau_2\cos{\alpha}=\tau_y\,,\la{3y}\\
&&u^{-1}_x \tau_3 u_x=\tau_1\,,\la{x3}\\
&&u^{-1}_y \tau_3 u_y=-\tau_2,,\la{y3}\\
&&u^{-1}_x \tau_1 u_x=\tau_3 \cos{\alpha}-\tau_2 \sin{\alpha}\,,\la{1x}\\
&&u^{-1}_y \tau_1 u_y=-\tau_3 \sin{\alpha}+\tau_1 \cos{\alpha}\,,\la{1y}\\
&&u^{-1}_x \tau_2 u_x=\tau_3 \sin{\alpha}+\tau_2 \cos{\alpha}\,,\la{2x}\\
&&u^{-1}_y \tau_2 u_y=\tau_3 \cos{\alpha}+\tau_1 \sin{\alpha}\,.\la{2y}\\ \nonumber
\ea
\end{subequations}

It is then immediate  that
\ba\la{GFq1}
\langle  j' _x, \bar{m}'_x, \bar{n}'_x, x|\hat E_3(S^\theta)|x, j_x, \bar{m}_x, \bar{n}_x \rangle &=&-i \kappa\gamma o(\ell_3, S^\theta) \int dg \,\overline{{}^{x}\! D^{j'_x}_{\bar m'_x \bar n'_x}(g)}\, {}^x\!D^{j_x}_{\bar m_x m}(\tau_3) \,{}^x\! D^{j_x}_{m \bar n_x}(g) \n\\
&=& -i\kappa\gamma  o(\ell_3, S^\theta) \,{}^x\! D^{j_x}_{\bar m_x \bar m_x}(\tau_3)=0\,,
\ea
and similarly for the other gauge fixing conditions.
Notice that this relation is only valid for $j \neq 1/2$ since 
in that case the off-diagonal matrix elements will not be necessarily vanishing. However, this case is not relevant for our construction because we are only interested in the semiclassical limit of the effective Hamiltonian through coherent states which provide 
good approximation to classical geometry only in the limit $j \gg 1$ (see Sec. \ref{sec:semi.states}). Since these states peak the spin quantum numbers on large values, the spin 1/2 contributions to the semiclassical expectation values we compute below are largely suppressed and we do not have to worry about them for the purposes of our analysis. 

For the other components of the fluxes which are classically nonvanishing we get the following expectation values:
\begin{subequations}\la{GFq2}
\ba
\langle \hat E_3(S^r) \rangle &=&-i\kappa\gamma  o(\ell_3, S^r) D^{j_z}_{\bar m_z \bar m_z}(\tau_3)= -\kappa\gamma o(\ell_3, S^r)\bar m_z\la{E3r}\,,\\
\langle \hat E_1(S^\theta) \rangle &=&-i\kappa\gamma o(\ell_x, S^\theta)\, {}^x\!D^{j_x}_{\bar m_x \bar m_x}(\tau_1)= -\kappa\gamma o(\ell_x, S^\theta)\bar m_x \cos{\alpha}\,,\la{E1t}\\
\langle \hat E_1(S^\varphi) \rangle &=&-i
\kappa\gamma o(\ell_y, S^\varphi) \, {}^y\!D^{j_y}_{\bar m_y \bar m_y}(\tau_1)=
\kappa\gamma o(\ell_y, S^\varphi)  \bar m_y \sin{\alpha}\,,\la{E1v}\\
\langle \hat E_2(S^\theta) \rangle &=&-i\kappa\gamma o(\ell_x, S^\theta)\, {}^x\!D^{j_x}_{\bar m_x \bar m_x}(\tau_2)
=-\kappa\gamma o(\ell_x, S^\theta)\bar m_x \sin{\alpha}\,,\la{E2t}\\
\langle \hat E_2(S^\varphi) \rangle &=&-i
\kappa\gamma o(\ell_y, S^\varphi)\, {}^y\!D^{j_y}_{\bar m_y \bar m_y}(\tau_2)=-
\kappa\gamma o(\ell_y, S^\varphi) \bar m_y \cos{\alpha}\,,\la{E2v}
\ea
\end{subequations}
where $o(\ell, S)$ is a sign denoting the orientation between a link and its dual face. All off diagonal matrix elements vanish for these operators as well as long as  $j \neq 1/2$, which, as just explained, is not relevant for the semiclassical limit.

\subsection{Quantum Gauss constraint}

We proceed by projecting the kernel of the full theory Gauss constraint. As expected, this will be performed by the operation 
\be 
P^{\dagger} \hat{G}_i P,
\ee
where $P$ is the projection operator defined previously. 
The kernel of $\hat{G}_i$ is given by the well-known gauge-invariant 
spin network states obtained by contraction with the $SU(2)$ intertwiners at the nodes of the graphs. Operator $P$ is then 
restricting the intertwiners in the way explained in Sec. \ref{sec:3valver}. The operation $P^{\dagger} \hat{G}_i P$ 
maps the kernel of $\hat{G}_i$ to the kernel of  ${}^{R}\hat{G}_3$, 
representing the classical phase space reduction given in Eq. \eqref{eq:G:tilde}. The states annihilated by ${}^{R}\hat{G}_3$ are
now $\alpha$ invariant, where $\alpha$ is parameter of rotation around the internal $3$-axis that has been aligned to the $r$-axis by the projector $P$. 

\subsection{3-valent vertex state} \la{sec:3valver}

Let us write the resolution of the identity in terms of coherent states:
\ba
\I_j&=&\sum_m|j,m\rangle\langle m, j| = 
 d_j\int_{S^2} d\vec{u}  |\bar m,\vec{u}\rangle\langle \vec{u}, \bar m|\n\\
 &=&d_j\sum_{m,n}\int_{S^2} d\vec{u} D^j_{m \bar m}(u) \overline{D^j_{n\bar m}(u)} |j,m\rangle\langle n, j|\,.
\ea

The gauge-invariant version of our states given in Eq. \eqref{states} can be obtained from the standard gauge-invariant spin network states as
\ba
&&\iota^{*m}_{j_I} D^{j_I}_{mn}(g)\iota^n_{j_I}=d_{j_I}^2\int_{S^2} du^1_I du^2_I\iota^{*m}_{j_I} \langle m, j_I| {\bar m_I}, \vec{u}^1_I\rangle\langle \vec{u}^1_I, \bar m_I | D^{j_I}(g)|{\bar n_I},\vec{u}^2_I\rangle\langle\vec{u}^2_I,  \bar n_I|{j_I},n\rangle 
\iota^n_{j_I}\n\\
&&=d_{j_I}^2\sum_{a,b,c,d}\int_{S^2} du^1_I du^2_I\iota^{*m}_{j_I} D^{j_I}_{a{\bar m_I}}( u^1_I) \overline{D^{j_I}_{b{\bar m_I}}( u^1_I)}\langle m, {j_I}| {j_I},a\rangle\langle b, {j_I} | D^{j_I}(g)|{j_I},c\rangle\langle d, {j_I}|{j_I},n\rangle 
D^{j_I}_{c{\bar n_I}}( u^2_I) \overline{D^{j_I}_{d{\bar n_I}}( u^2_I)}
\iota^n_{j_I}\n\\
&&=d_{j_I}^2\int_{S^2} du^1_I du^2_I\iota^{*m}_{j_I} D^{j_I}_{m{\bar m_I}}( u^1_I) D^{j_I}_{{\bar m_I}{\bar n_I}}\left( (u^1_I)^{-1} g\,
 u^2_I\right) \overline{D^{j_I}_{n{\bar n_I}}( u^2_I)}
\iota^n_{j_I}
\ea
where $\iota^{*m}_{j_I}$ is an intertwiner in the $3$ basis. Setting $u^1_I=u^2_I=u_I$ amounts to projection on $\mathcal{H}^{R}$. The reduced holonomies are then obtained by restricting the Haar measure $du$ only to $U(1)$ rotation around the $z$-axis, namely
\be
 {}^{\va R}\!D^{j_I}_{mn}(g)=d_{j_I}^2\int_0^{2\pi} \!d\alpha\, D^{j_I}_{m{\bar m_I}}( u_I(\alpha)) D^{j_I}_{{\bar m_I}{\bar n_I}}\left( (u_I(\alpha))^{-1} g\, u_I(\alpha)\right) \overline{D^{j_I}_{n{\bar n_I}}( u_I(\alpha))}\,.
\ee

In a graphical notation, this is
\be\la{DR}
 {}^{\va R}\!D^{j_I}_{mn}(g)=\int_0^{2\pi} \!d\alpha\,
 \begin{array}{c}
\includegraphics[width=6.5cm]{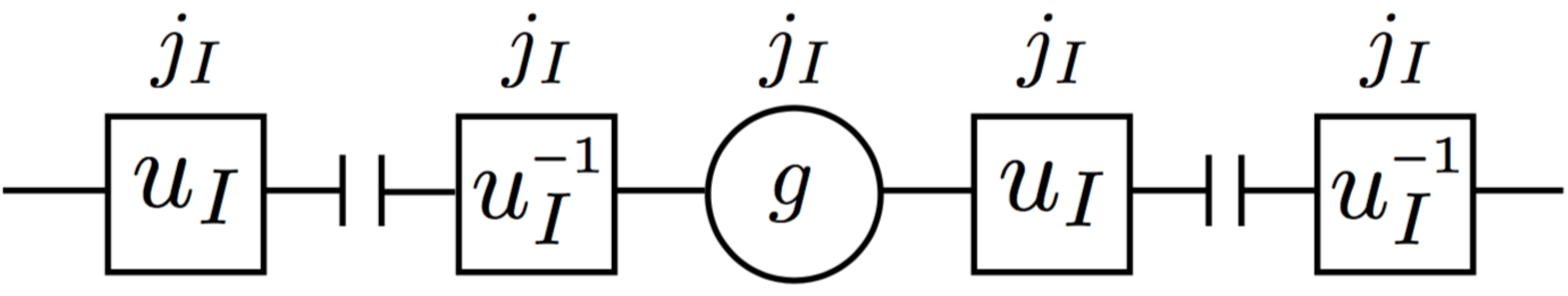}. \end{array}
\ee
Here we denoted the projection on the highest or lowest magnetic number as
\be
 \langle \bar m_I, j_I|=
 \begin{array}{c}
\includegraphics[width=1cm]{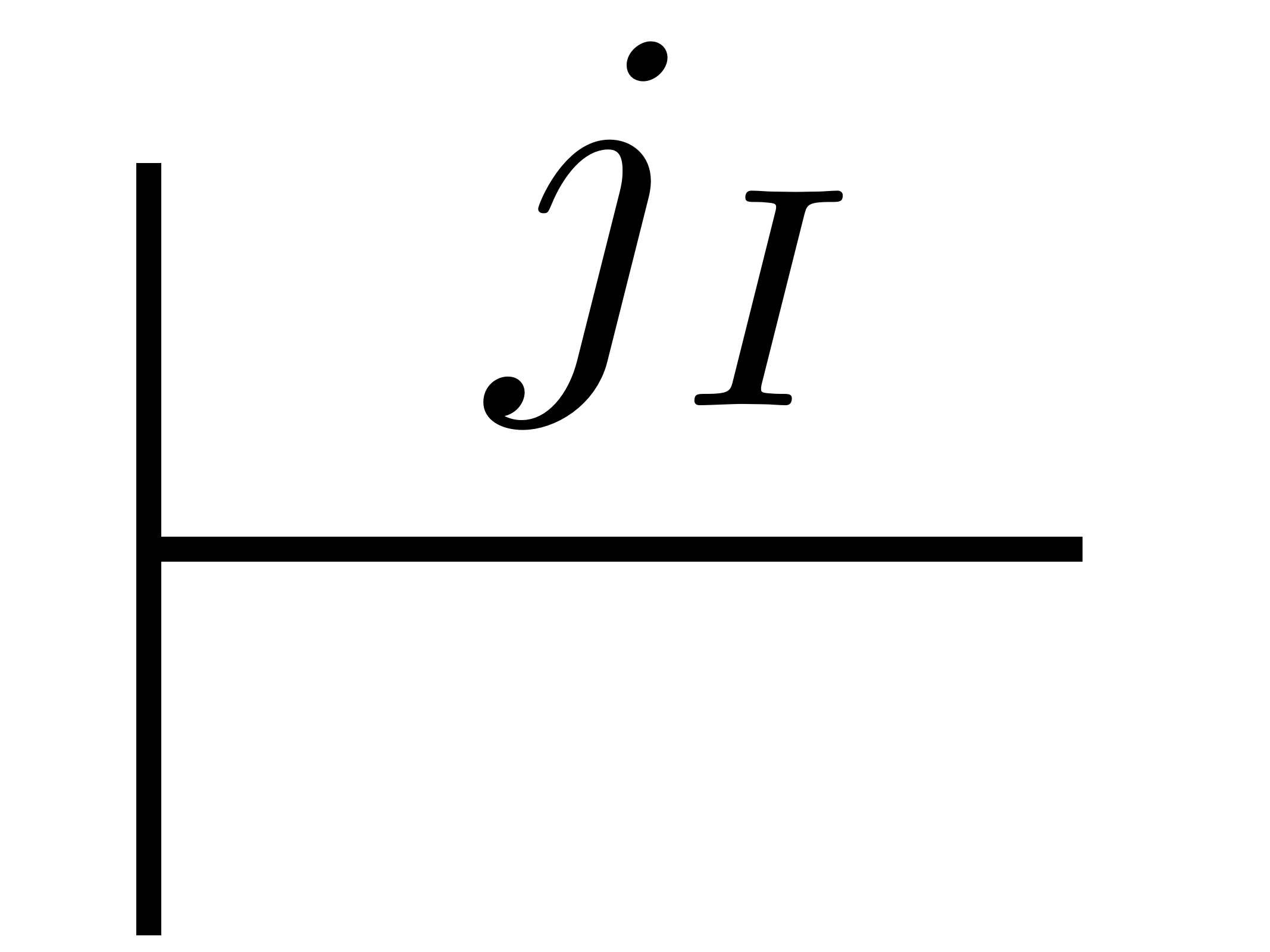}\,,\end{array}
\quad
| j_I, \bar m_I\rangle=
 \begin{array}{c}
\includegraphics[width=1cm]{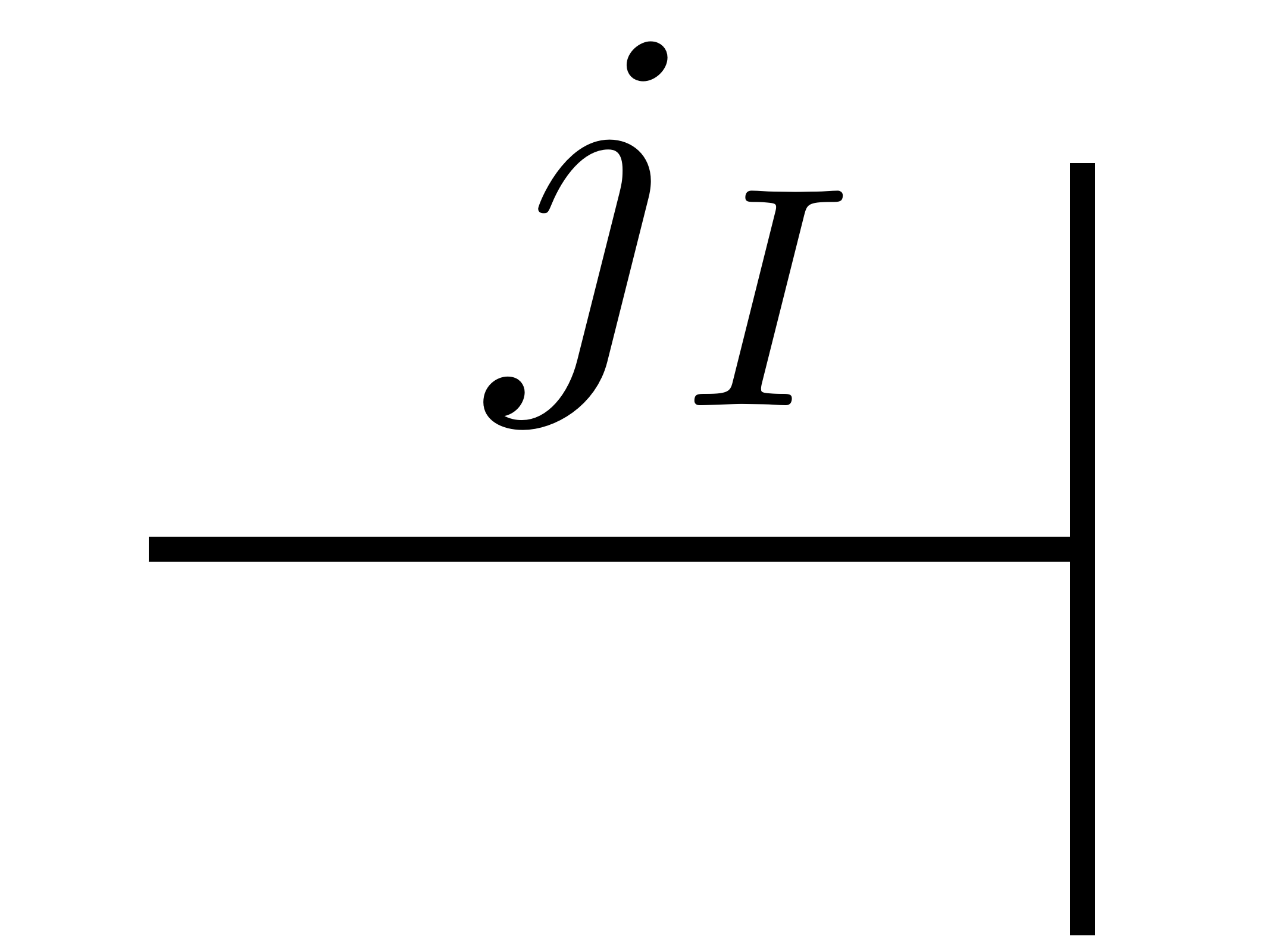}\,, \end{array}
\ee
and the Wigner matrices in the $|j,m\rangle$ basis for the $u_I$ rotations and a generic $SU(2)$ group element $g$ respectively as
\baa
&&D^{j_I}_{mn}(u_I)=
 \begin{array}{c}
\includegraphics[width=1.5cm]{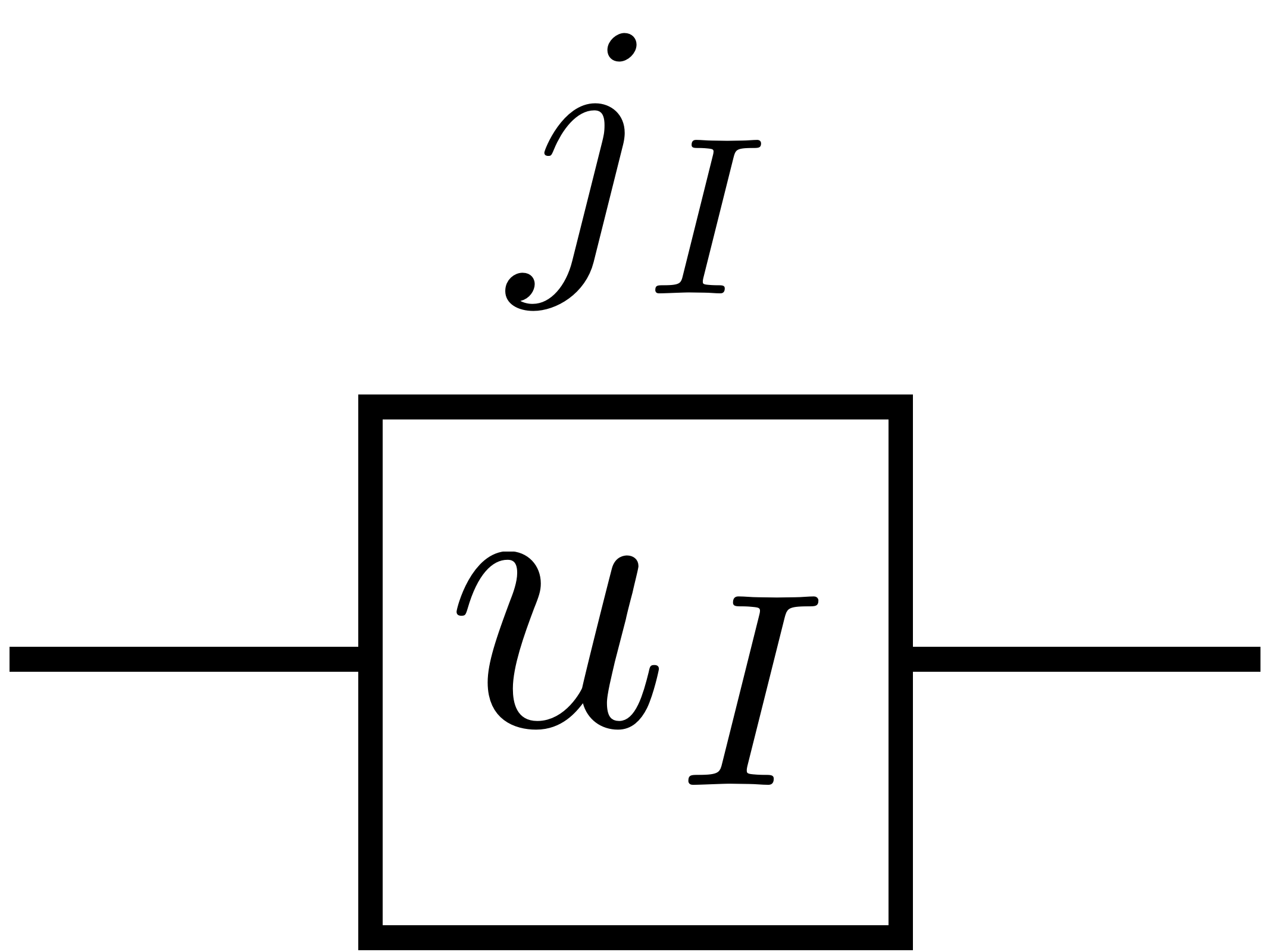}\,,\end{array}\\
&&D^{j}_{mn}(g)=
 \begin{array}{c}
\includegraphics[width=1.5cm]{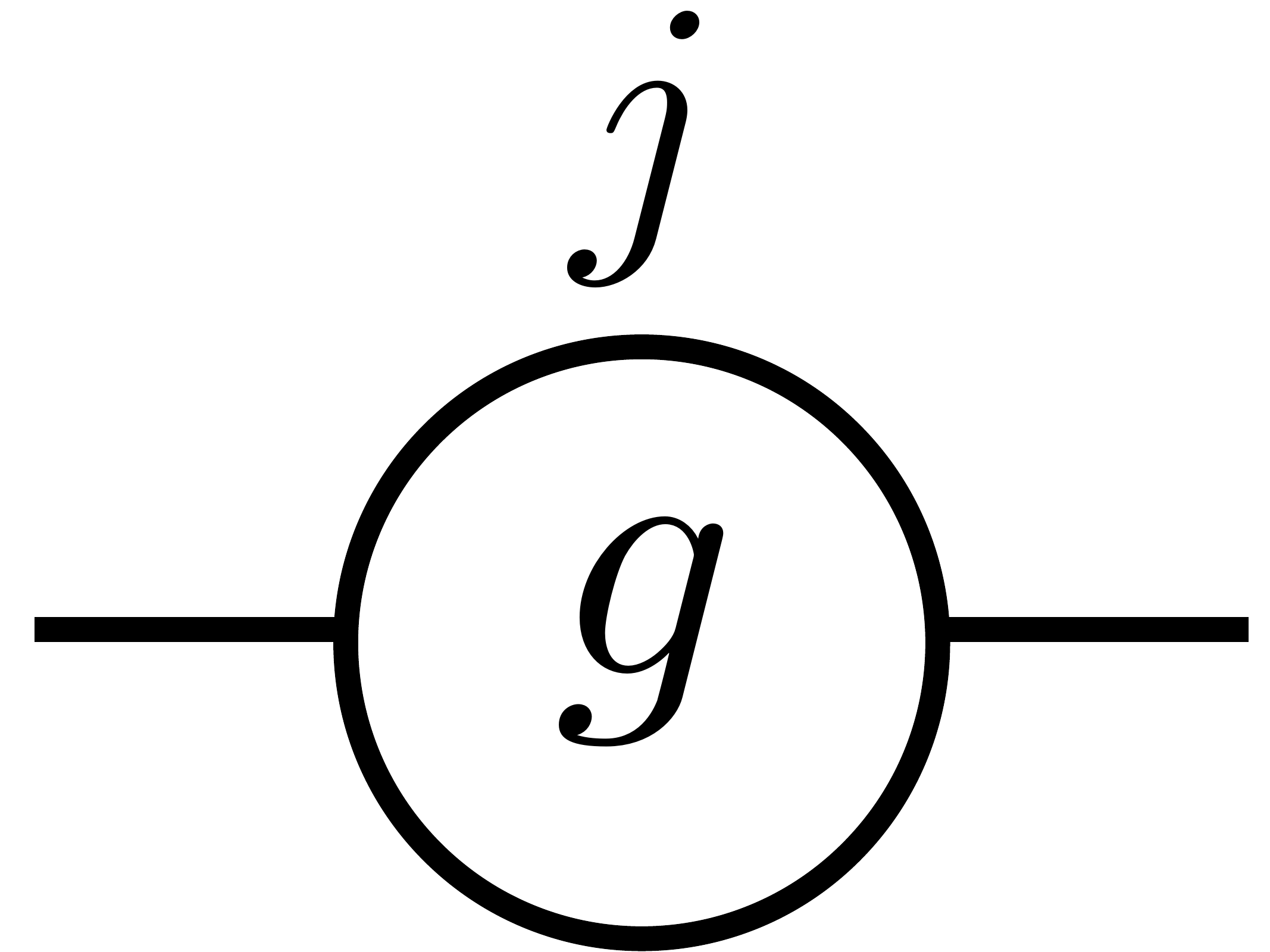}\,.\end{array}
\eaa
Using this notation, we can represent a reduced 3-valent vertex state as
\be\la{3v}
|v_3^{\va R}(j)\rangle= \int_0^{2\pi} \!d\alpha\,
 \begin{array}{c}
\includegraphics[width=8cm]{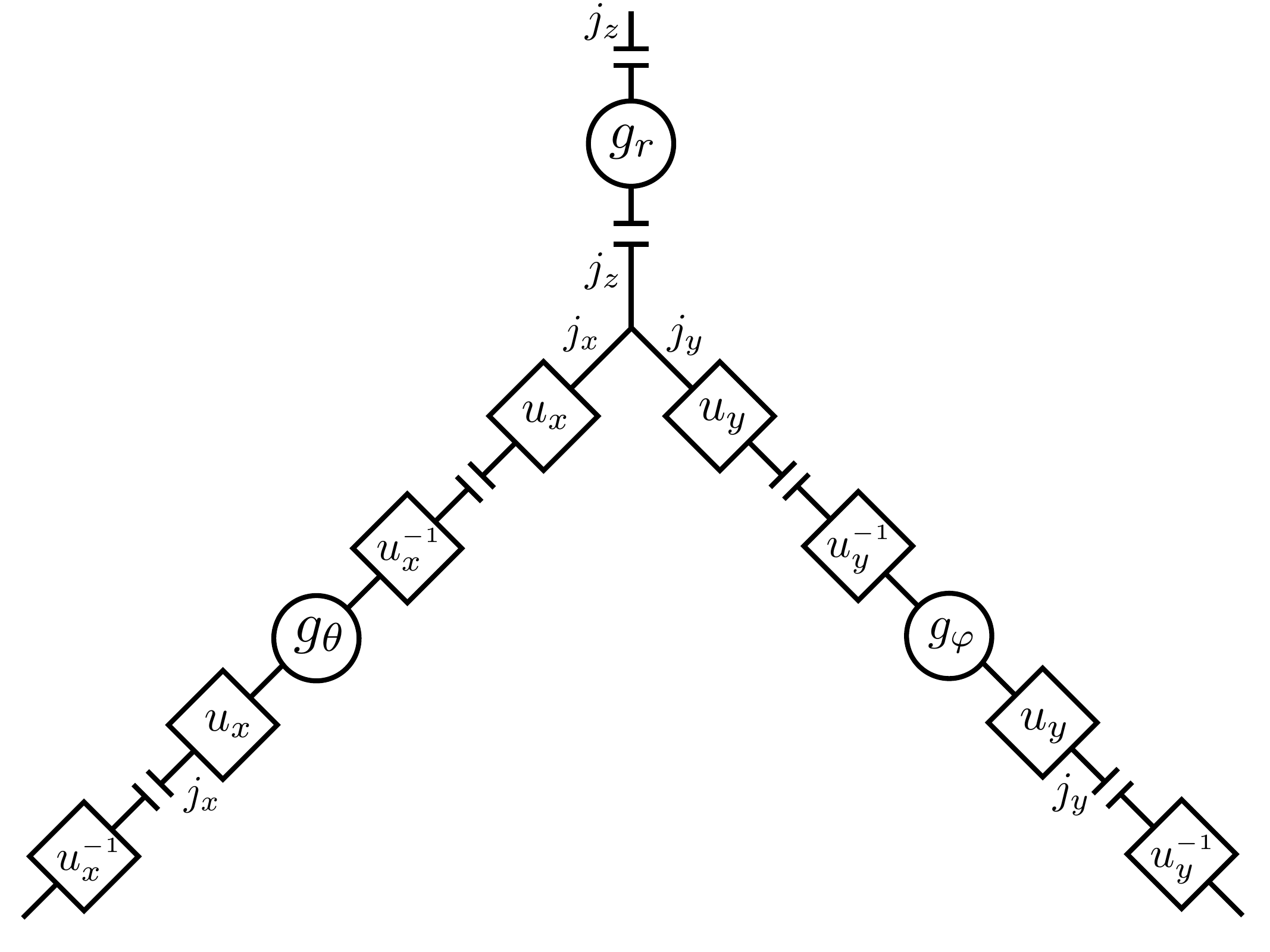}\,,\end{array}
\ee
where the 3-valent node represents the standard $3-j$ symbol and 
its contraction with $SU(2)$ coherent states defines the reduced intertwiners.

\subsection{Geometric operators}\la{sec:geo-op}

On $\mathcal{H}^{R}$ that we constructed above, we can now define the action of geometric operators such as the area and the volume operators by importing the regularization techniques of the full theory \cite{Rovelli:1994ge, Ashtekar:1996eg, Ashtekar:1997fb}.
These geometric operators are constructed out of the reduced flux operators defined as follows. Let us introduce the projectors
\ba
\hat P^z&=&\sum_{\bar m_z=\pm j_z}|j_z, \bar m_z\rangle\langle \bar m_z, j_z|\,,\\
\hat P^I&=&\sum_{\bar m_I=\pm j_I}| \vec{u}_I, \bar m_I\rangle\langle  \bar m_I, \vec{u}_I|\,,
\ea
where we recall $I \in \{x,y\}$. The reduced flux operators are then
\ba
{}^{\va R}\hat E_i(S^r)=\hat P^z \hat E_i(S^r) \hat P^z\,,\la{Er}\\
{}^{\va R}\hat E_i(S^\theta)=\hat P^x \hat E_i(S^\theta) \hat P^x\,,\la{Et}\\
{}^{\va R}\hat E_i(S^\varphi)=\hat P^y \hat E_i(S^\varphi) \hat P^y\la{Ev}\,.
\ea
In a similar fashion as one can prove the validity of the partial gauge fixing condition given in Eq. \eqref{GFq1} (as well as all the others), it is immediate to see that the reduced flux operators defined above are diagonal on the reduced quantum states constructed in the previous part of this section. Nonvanishing eigenvalues correspond only to those components associated to the remaining reduced phase space densitized triad variables as written in Eq. \eqref{SE2}, and they are given by the relations that appear in Eq. \eqref{GFq2}. This is a key result of  our construction, which will allow us to compute the expectation value of the Hamiltonian constraint in a relatively straightforward manner. But let us first exploit this property of the reduced fluxes to compute the spectrum of the main geometrical operators in LQG.

We can concentrate on a 3-valent node with three links departing along the three directions $\{r, \theta, \varphi\}$ that appears in Eq. \eqref{3v}. 
If we consider three surfaces $S^a$ that intersect the dual links $\ell_a$ once, we can construct the associated area operators in terms of the reduced fluxes given in Eqs. \eqref{Er}, \eqref{Et}, and \eqref{Ev}. The action of these reduced area operators on the reduced states associated with the dual links that they intersect is

\ba
&& {}^{\va R}\hat A(S^\theta)  {}^{\va R}\!D^{j_x}_{mn}(g_\theta)
=\sqrt{{}^{\va R} \hat E_i(S^\theta)  {}^{\va R}\hat E^{i}(S^\theta)} \, {}^{\va R}\!D^{j_x}_{mn}(g_\theta)
 =\kappa\gamma j_x {}^{\va R}\!D^{j_x}_{mn}(g_\theta)\,,\\
 && {}^{\va R}\hat A(S^\varphi) {}^{\va R}\!D^{j_y}_{mn}(g_\varphi)
 =\sqrt{ {}^{\va R}\hat E_i(S^\varphi) {}^{\va R}\hat E^{i}(S^\varphi) }\, {}^{\va R}\!D^{j_y}_{mn}(g_\varphi)
 =\kappa\gamma j_y {}^{\va R}\!D^{j_y}_{mn}(g_\varphi)\,,\\
 && {}^{\va R}\hat A(S^r)  {}^{\va R}\!D^{j_z}_{mn}(g_r)
 =\sqrt{ {}^{\va R}\hat E_i(S^r)  {}^{\va R}\hat E^{i}(S^r) }\,  {}^{\va R}\!D^{j_z}_{mn}(g_r)
=\kappa\gamma j_z {}^{\va R}\!D^{j_z}_{mn}(g_r)\,.
\ea
 
Next, we can consider the volume of the region containing only the 3-valent node $v$. The associated reduced volume operator regularized on the cube dual to $v$ acts diagonally on the reduced 3-valent vertex state given in Eq. \eqref{3v} as
 \ba
  {}^{\va R}\hat V(v) |v_3^{\va R}(j)\rangle
&=& \sqrt{\left|\frac{1}{3!} \epsilon_{abc} \epsilon^{ijk } {}^{\va R}\hat E_i(S^a)  {}^{\va R}\hat E_j(S^b) {}^{\va R} \hat E_j(S^c)\right|}  |v_3^{\va R}(j)\rangle\n\\
&=& \sqrt{ \left| \left( {}^{\va R}\hat E_1(S^\theta)  {}^{\va R}\hat E_2(S^\varphi)- {}^{\va R}\hat E_2(S^\theta) {}^{\va R} \hat E_1(S^\varphi)\right)  {}^{\va R}\hat E_3(S^r)\right|}  |v_3^{\va R}(j)\rangle\n\\
 &=&(\kappa\gamma)^{\frac{3}{2}}\sqrt{\left|j_xj_y j_z\right|}|v_3^{\va R}(j)\rangle\,.
 \ea
This diagonal action of the volume operator on the reduced 3-valent vertex state represents a key characteristic of this construction which enables us to considerably simplify the calculation of the Hamiltonian constraint action.

\subsection{Recoupling theory}\la{sec:rec-rul}


In order to study the action of holonomy operators in $\mathcal{H}^R$, we need to define the recoupling theory for our states.
To this end, for a product of states along the same direction we introduce the following ``reduced'' recoupling rule:
\ba\la{rec}
 \begin{array}{c}
\includegraphics[width=7.cm]{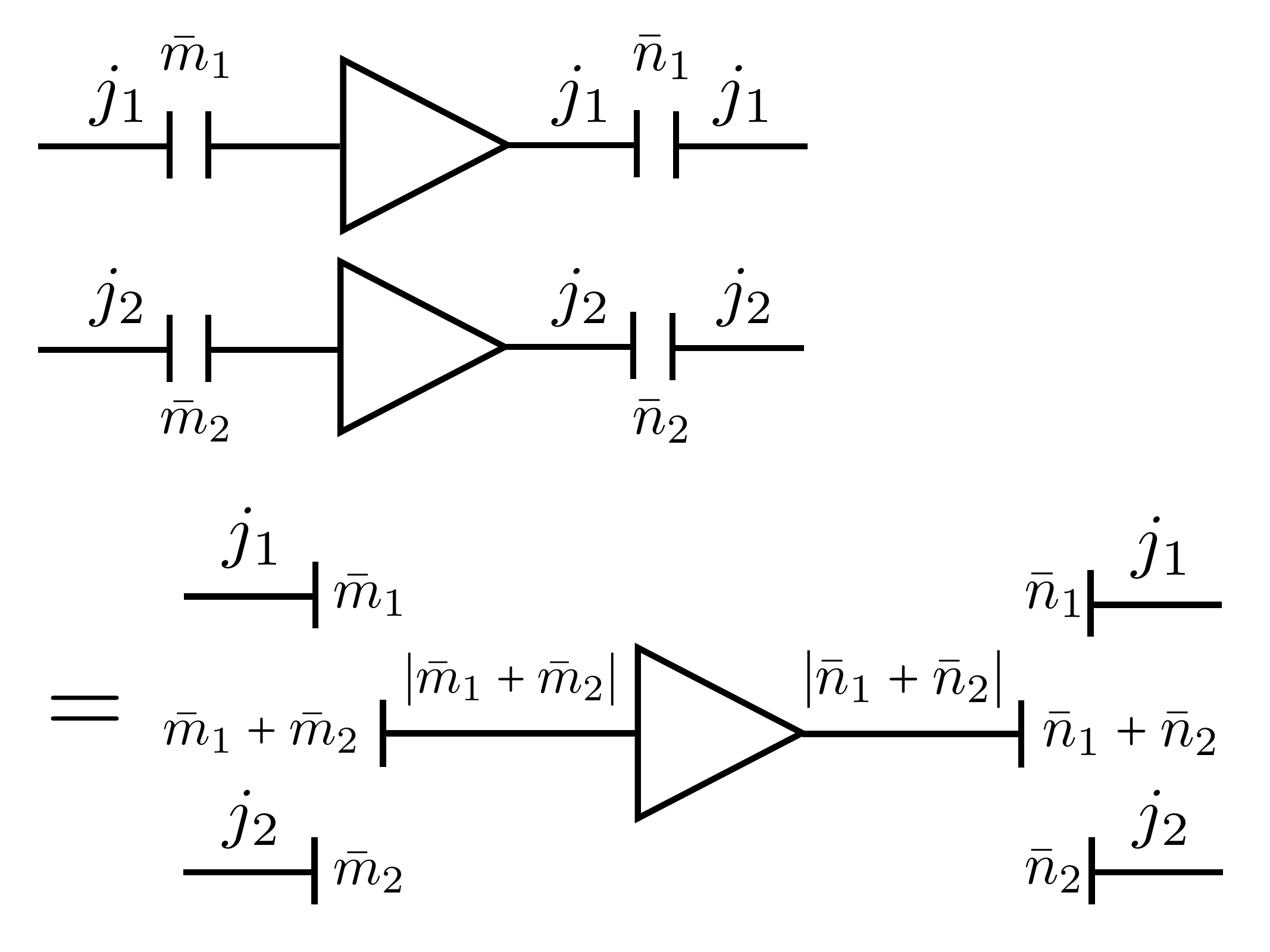}\end{array}
\ea
where a triangle denotes the Wigner matrix of either a given rotation or a generic $SU(2)$ element. Let us point out that in the case ${\rm sign}(\bar m_1\bar m_2)={\rm sign}( \bar n_1 \bar n_2)=+$, the explicit expression of the  two Clebsch-Gordan coefficients $C^{Kk}_{j_1 \bar m_1,j_2 \bar m_2}$ and $C^{Kk}_{j_1 \bar n_1,j_2 \bar n_2}$ are nonvanishing only for $K=|\bar m_1+\bar m_2|, k= \bar m_1+\bar m_2$ and $K=|\bar n_1+\bar n_2|, k= \bar n_1+\bar n_2$ respectively. Hence, in this case the reduction is naturally implemented in this product rule. However, in the case where the two magnetic numbers have opposite signs, there is a tower of spins $K$ that are allowed.  In this case we need to restrict to the lowest spin in the tower in order for the reduction to be implemented in our recoupling rule.
Moreover, in the case where the reduced states in the tensor product contain a $U(1)$ integral, as in Eq. \eqref{DR}, we first need  to align the two states by fixing the same value for the $U(1)$ angles and then project by integrating the tensor product of the two states over this angle, in order to rewrite the product in terms of original reduced states on the links.
We can now use these recoupling rules to compute the product of two 3-valent reduced states. Doing this we find
\ba
|v_3^{\va R}(j)\rangle\otimes |v_3^{ \va R}(j')\rangle&=& \int_0^{2\pi} \!d\alpha \,
 \begin{array}{c}
\includegraphics[width=8.7cm]{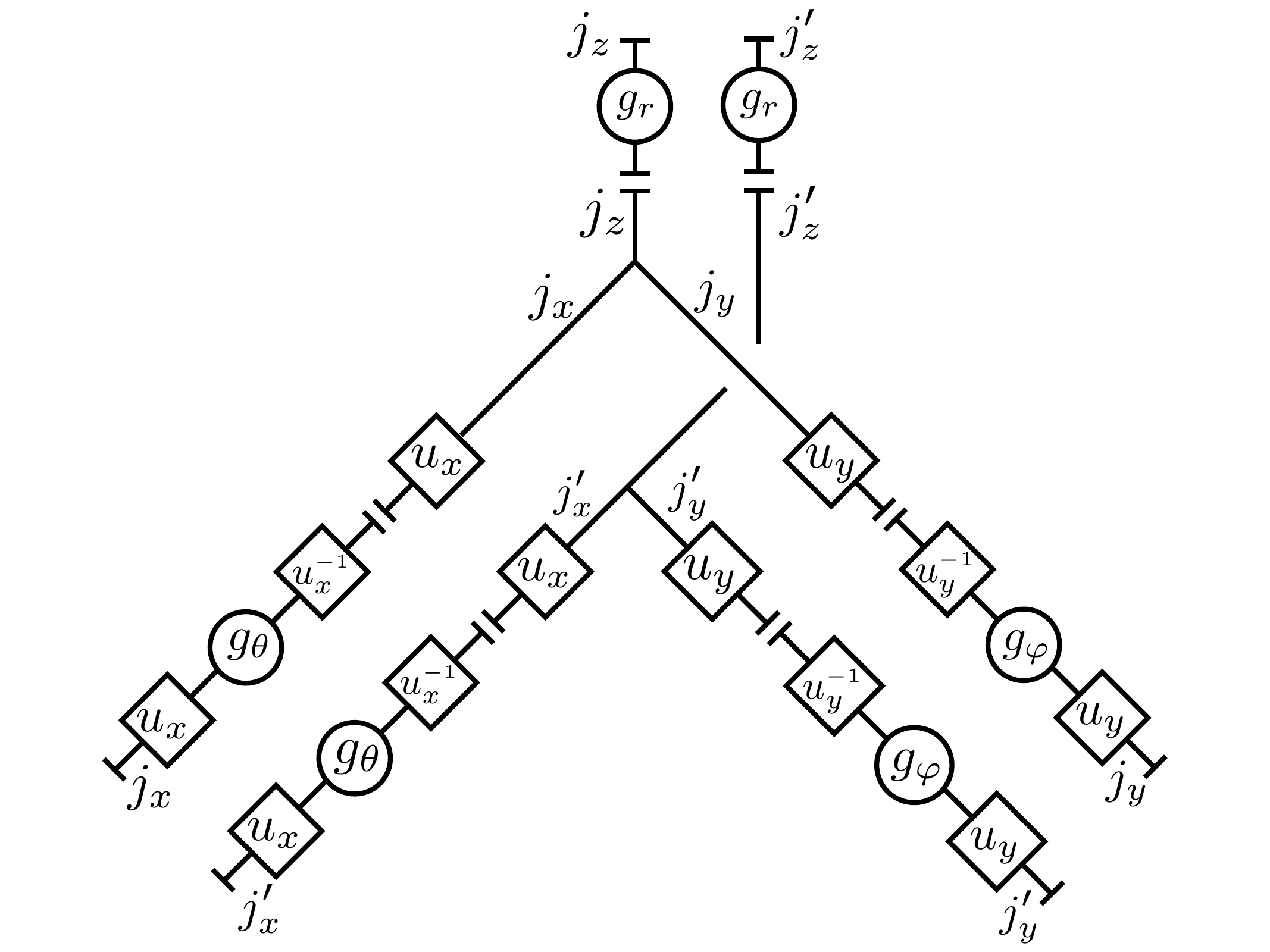}\end{array}\n\\
&=& \int_0^{2\pi} \!d\alpha \, 
\begin{array}{c}
\includegraphics[width=8.7cm]{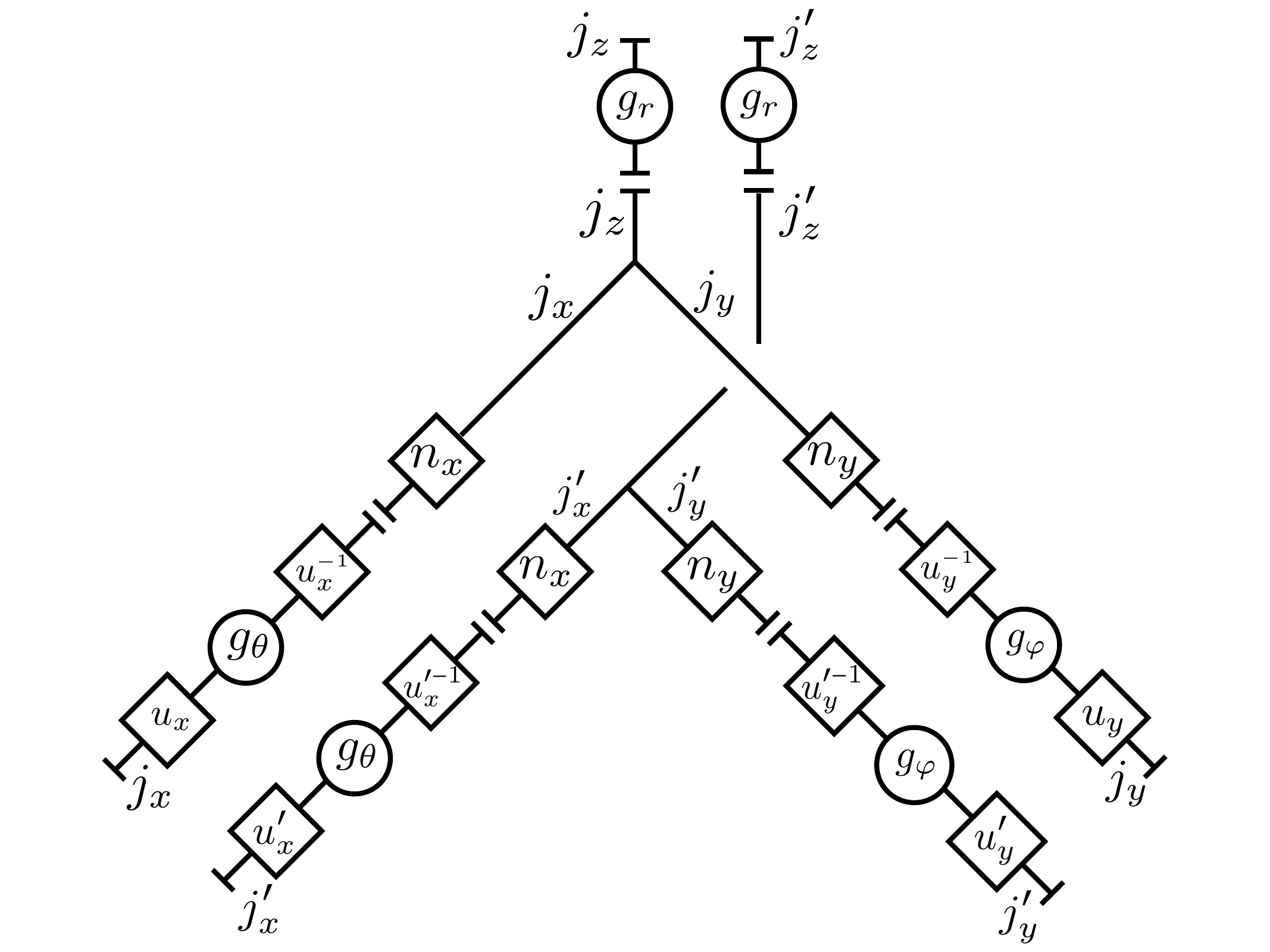}\end{array}\n\\
&=& \int_0^{2\pi} \!d\alpha 
\begin{array}{c}
\includegraphics[width=8.8cm]{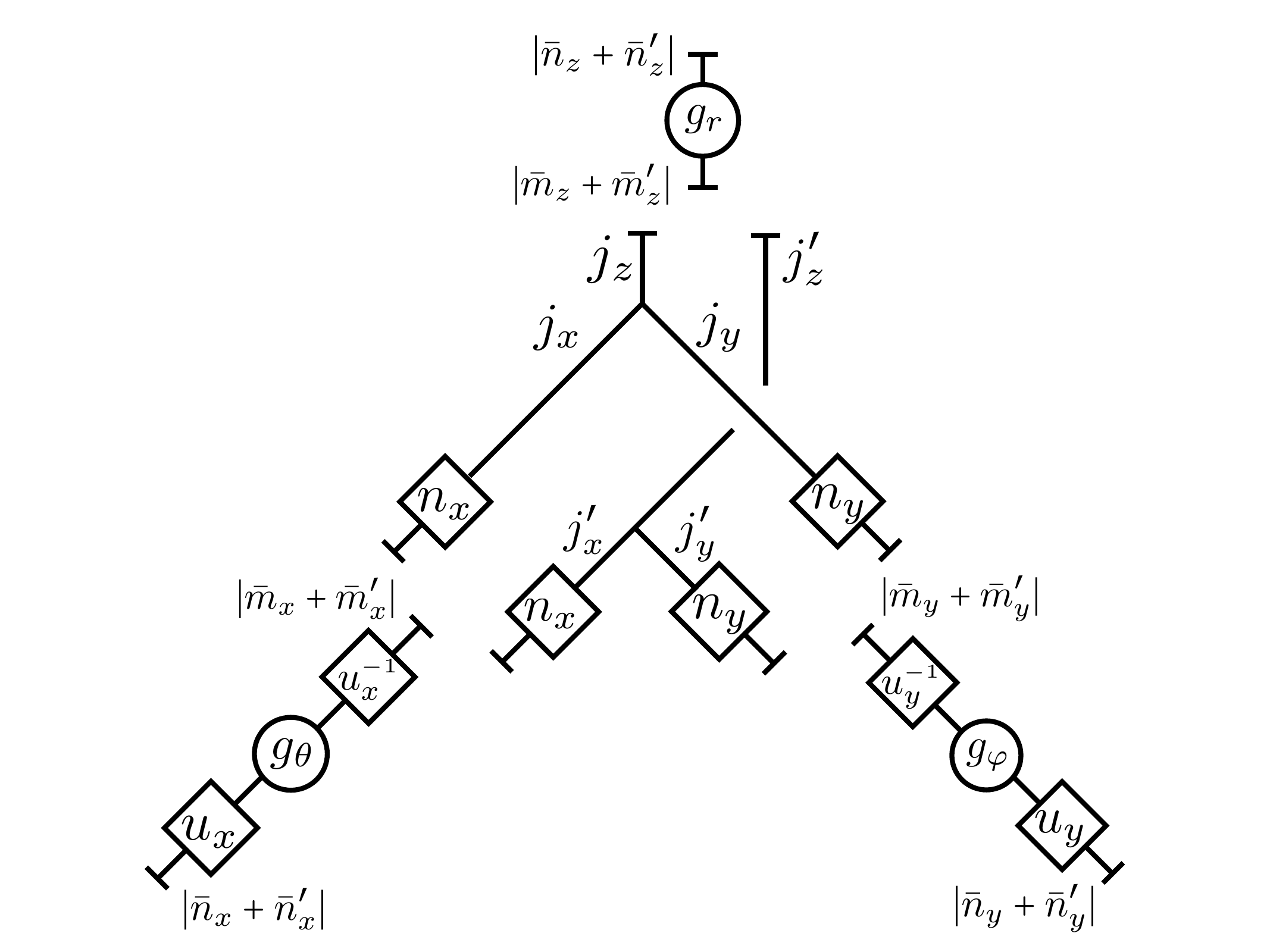}\end{array},\la{prod1}
\ea
where  we have introduced the graphical notation
\be
\begin{array}{c}
\includegraphics[width=1.8cm]{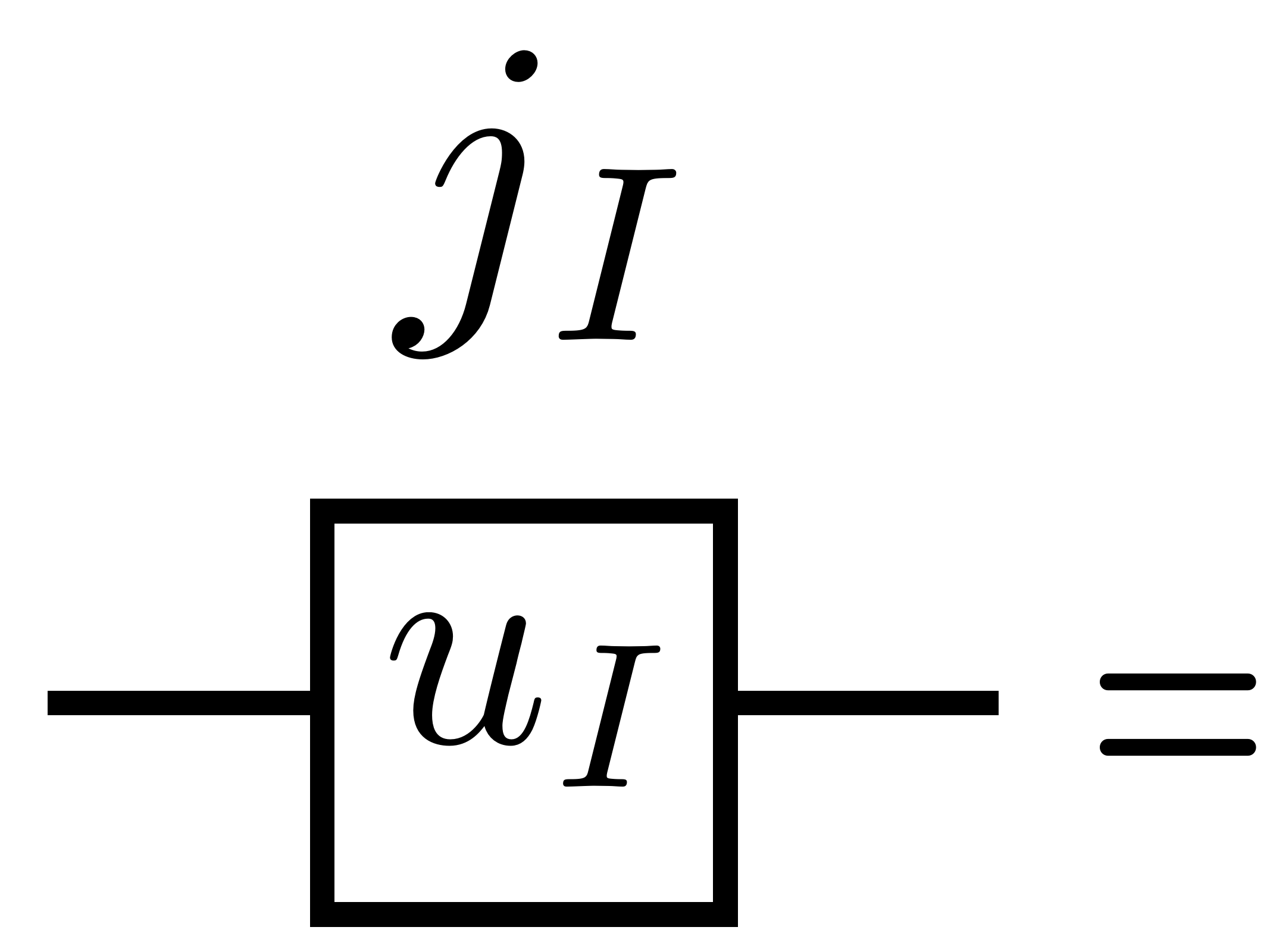}\,\,
\includegraphics[width=2.cm]{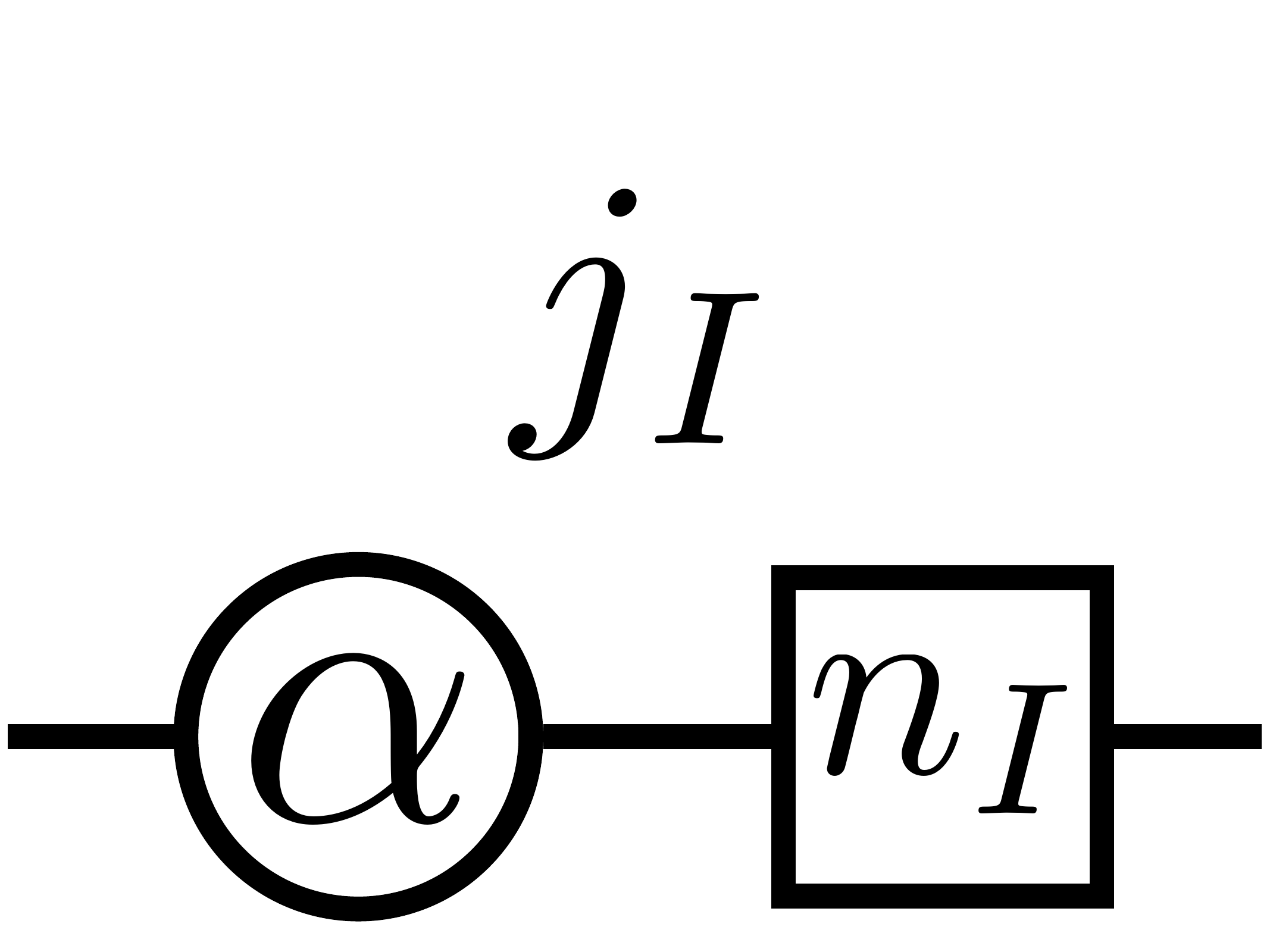}\end{array}\\
\ee
to use the fact that, due to the gauge invariance at the 3-valent node of $|v_3^{ \va R}(j')\rangle$, the $\alpha$ dependence of the rotation group elements $u_x, u_y$ [see Eqs. \eqref{ux} and \eqref{uy}] acting near the node can be reabsorbed and then the integral  sees only these group elements on the links of the reduced state.



Finally, we can rewrite the product of  two 3-valent reduced states as
\ba
|v_3^{\va R}(j)\rangle\otimes |v_3^{ \va R}(j')\rangle
&=&
\begin{Bmatrix}
    j_x  & j'_x & |\bar m_x+\bar m'_x| \\[0.3em]
    j_y  & j'_y & |\bar m_y+\bar m'_y| \\[0.3em]
    j_z  & j'_z &|\bar m_z+\bar m'_z|
\end{Bmatrix}
 \int_0^{2\pi} \!d\alpha \!\!
\begin{array}{c}
\includegraphics[width=8.7cm]{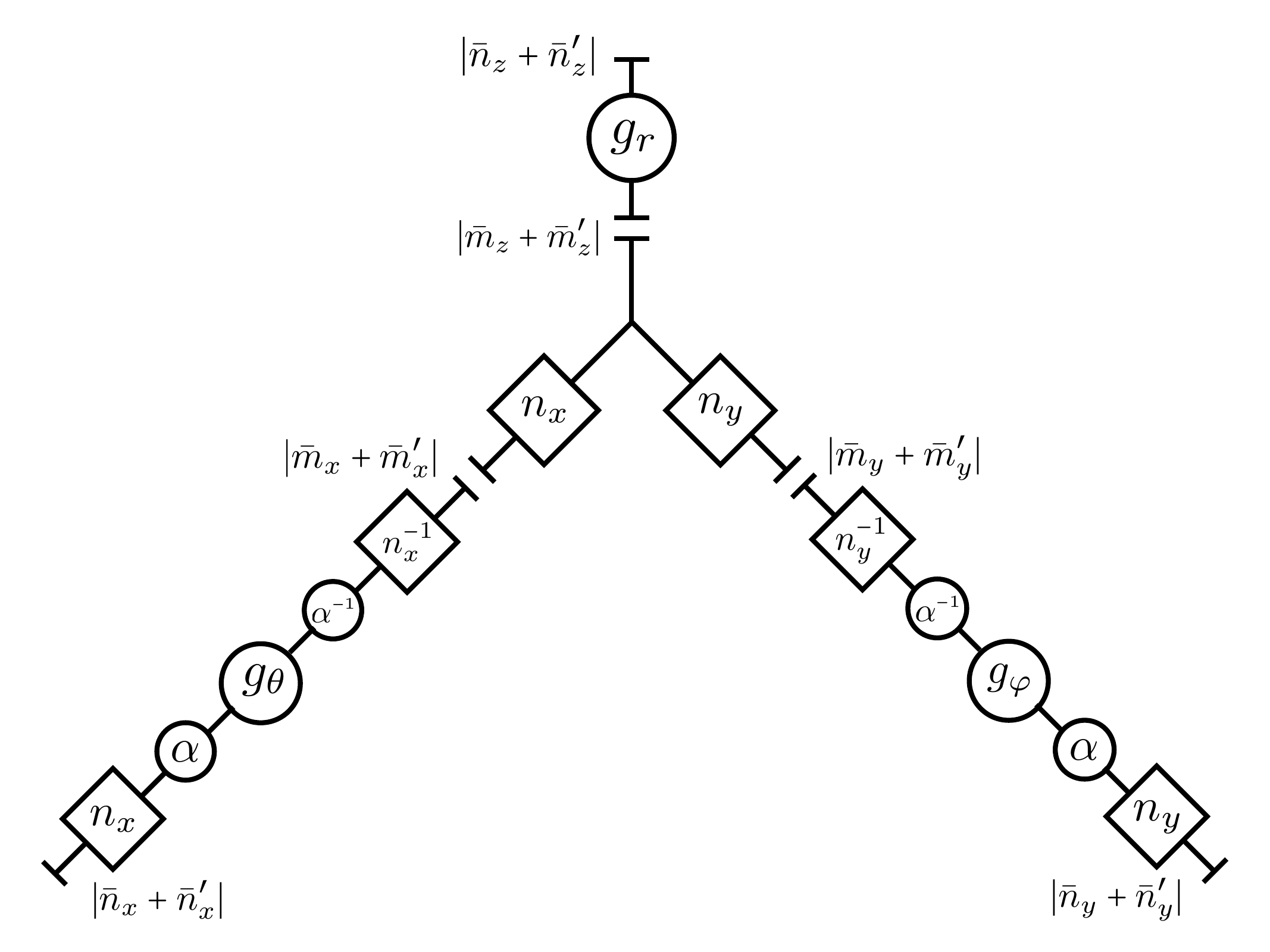}\end{array}\!\!\!\!\!\!,\la{prod2}
\ea
where we have used once more the recoupling rule of Eq. \eqref{rec},   as well as the standard $SU(2)$ recoupling theory, to rewrite the product in terms of an original reduced 3-valent vertex state times a $9j$ symbol.

If we now compute the norm of the 3-valent vertex state \eqref{3v} through a scalar product consistent with the  reduced recoupling rule introduced above, namely  by again first aligning the bra and ket states, then performing the integration over the $SU(2)$ group elements through standard recoupling theory and finally performing the integral over $\alpha$, we get
\be\la{norm}
|v_3^{\va R}(j)|\equiv \sqrt{|\langle v_3^{\va R}(j)|v_3^{\va R}(j) \rangle |}=(2\pi)^{1/8}\sqrt{
\left(\begin{array}{c}
\includegraphics[width=1.5cm]{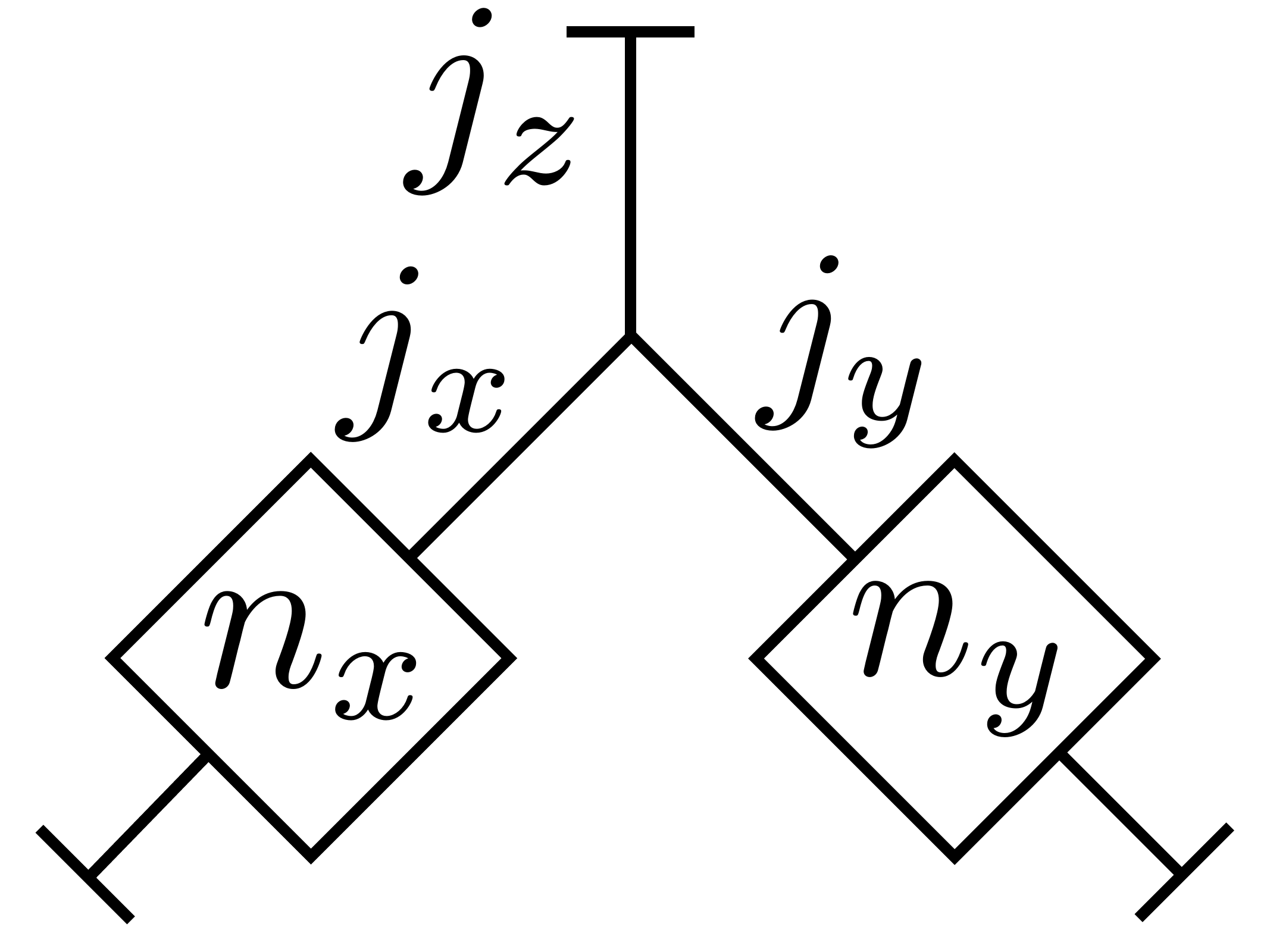}\end{array}\right)^*
\begin{array}{c}
\includegraphics[width=1.5cm]{norm1.pdf}\end{array}\,,
}
\ee
where the factor of 1/8 for the $2\pi$ coming from the integral over $\alpha$ is due to the fact that the same angle $\alpha$ is eventually shared by four 3-valent vertices around a cell in the $(x,y)$-plane;
 similarly for $v_3^{ \va R}(j')$. Therefore, comparing \eqref{prod1} with \eqref{prod2}, we see that
\be
|v_3^{\va R}(j)| |v_3^{ \va R}(j')| = (2\pi)^{1/8} \{9j\} |v_3^{ \va R}(|\bar m+\bar m'|)|\,,
\ee
where the $9j$ symbol is the one in \eqref{prod2}. 
It follows that, in terms of normalized 3-valent vertex states
\be
|\widetilde{v_3^{\va R}}(j)\rangle= \frac{|v_3^{\va R}(j)\rangle}{|v_3^{\va R}(j)|}\,,
\ee
we get the product rule
\be\la{prodrule}
|\widetilde{v_3^{\va R}}(j)\rangle\otimes |\widetilde{v_3^{\va R}}(j')\rangle=\frac{1}{(2\pi)^{1/8}}|\widetilde{v_3^{\va R}}(| \bar{m}+ \bar{m}'|) \rangle\,.
\ee
Namely, by working with normalized intertwiners, the $9j$ symbol gets reabsorbed in the product rule and this provides a great simplification in the expectation value of the Hamiltonian constraint that we compute below.

%

Within this construction, the kinematical Hilbert space encodes the information of a radial metric tensor [as defined by the partial gauge fixing in Eqs. \eqref{GF1} and \eqref{GF2}]. The residual gauge freedom left is encoded by the $U(1)$ internal rotations around the $3$-direction and radial diffeomorphisms preserving the reduced graphs structure, in accordance with the classical analysis of \cite{Alesci:2018ewg}. The latter can be implemented by standard group averaging techniques, defining the dual Hilbert space in terms of a sum  over all the reduced graphs related by a (reduced) radial diffeomorphism with a shift smearing function depending only on the $r$ coordinate.

\subsection{Quantum vector constraint}
The quantum vector constraint is imposed in the full theory
by group averaging the spin network states over spatial diffeomorphisms as described in the seminal paper \cite{Ashtekar:1995zh}. One introduces a Gelfand triple
$\text{Cyl} \subset \mathcal{H}^{K} \subset \text{Cyl}^*$
where $\text{Cyl}$ is the space of cylindrical functions. 
The vector constraint has a well-defined action on $\text{Cyl}^*$. 
Let $\mathcal{U}[\phi]$ be an operator acting on $\text{Cyl}$ with $\phi$ being a self-diffeomorphism 
of $\Sigma_t$. We then have 
\be 
\mathcal{U}[\phi] \psi_{\gamma}(A) = \psi_{\phi^{-1} \gamma}(A),
\ee 
where $\psi_{\gamma} \in \text{Cyl}$ and $A$ is the Ashtekar-Barbero connection. However, since $\mathcal{U}[\phi]$ is not weakly continuous, it cannot be produced by a self-adjoint infinitesimal generator. Therefore one has to restrict attention to finite spatial diffeomorphisms when searching for diffeomorphism invariant states. 
Solving the ``finite version'' of the vector constraint equation boils down to searching for all $\psi$ that satisfy
\be \la{udiffeq}
\mathcal{U}[\phi] \psi = \psi.
\ee
This equation, however, has no nontrivial solutions in $\mathcal{H}^K$. Nonetheless, it can be solved for $\psi \in \text{Cyl}^*$. Formally, the solution is given as the averaging of the dual states over the group of spatial diffeomorphisms,
\be 
(\psi_{[\gamma]}| = \sum_{\phi \in \text{Diff}(\Sigma_t)} \langle \psi _{\phi \gamma}|,
\ee
where $[\gamma]$ is the equivalence class of graphs. 

In principle one should try to find the kernel of $\hat{\tilde{H}}_r$. But this goal is too ambitious at the moment. 
However, we know that classically and in the case where the shift vector $N^r$ does not depend on the angular coordinates [see Eq. \eqref{eq:hr:tilde}], $\tilde{H}_r = {}^{R} \tilde{H}_r$. Therefore it may not be implausible to assume that the kernels of the corresponding  quantized operators coincide. We expect that averaging the kinematical states constructed here over the group of radial diffeomorphisms will provide the required solutions to Eq. \eqref{udiffeq}.

\section{Reduced Hamiltonian Constraint Operator} \la{sec:rhco}

A  regularized expression of the  Hamiltonian constraint operator in LQG was introduced in \cite{Thiemann:1996aw} with an action defined on a  graph-dependent triangulation of the spacelike hypersurfaces. This construction can be easily adapted to the cubulation used here to define $\mathcal{H}^{R}$. 

\subsection{Euclidean term}\la{sec:rhco-A}

Importing techniques developed for the cosmological case \cite{Alesci:2013xd, Alesci:2015nja}, we can define the Euclidean part of the reduced Hamiltonian constraint regularized on the faces of a cubic cell dual to a 6-valent node $v$ (modulo an overall constant) as
\be\la{HE}
{}^{\va R}\hat H^{\va E}_{\va \square}[N]=-\frac{4}{\kappa^2\gamma}N(v)\epsilon^{ijk}\tr\left[\left( {}^{\va R}\hat g_{\alpha_{ij}}-  {}^{\va R}\hat g^{-1}_{\alpha_{ij}}\right) {}^{\va R}\hat g^{-1}_{s_k} [{}^{\va R}\hat g_{s_k}, {}^{\va R}\hat V(v)]\right]\,,
\ee
where $N(v)$ is the lapse function at the node $v$; the reduced holonomies $ {}^{\va R}\hat g$ are taken in the fundamental representation; and the internal indices $i,j,k$ take values over $3,x,y$. In the regularization above, the link $s_k$ corresponds to one of the six edges $\ell_3, \ell_x, \ell_y$ (two per direction, both denoted with the same $\ell_i$) departing from the node $v$, while $\alpha_{ij}$ corresponds to a loop in the plane $(ij)$. Thus $\alpha_{ij}=\ell_i \circ \ell_j\circ \ell_i^{-1}\circ \ell_j^{-1}$ and we consider the non-graph-changing version of Thiemann's regularization.
We take the Hamiltonian operator in the fundamental representation. 

To elucidate the action of the operator given in Eq. \eqref{HE}, it is enough to concentrate on a reduced 3-valent vertex state that is defined in Eq. \eqref{3v} for the case $k=z, i=x, j=y$. The other components of Eq. \eqref{HE} act in a similar fashion. The extra structure of the 6-valent vertex state, namely the 6-valent reduced  intertwiner, is not directly affected by the action of ${}^{\va R}\hat H^{\va E}_{\va \square}[N]$. In fact, in the process of computing  the expectation value of ${}^{\va R}\hat H^{\va E}_{\va \square}[N]$ those coefficients cancel due to normalization.
By means of the reduced recoupling rules introduced in Sec. \ref{sec:rec-rul}, the reduced Hamiltonian constraint action can be computed  analogously to the cosmological case \cite{Alesci:2014uha} and it yields
\ba \la{HE12}
&&-\frac{4}{\kappa^2\gamma}N(v)\tr\left[\left( {}^{\va R}\hat g_{\alpha_{xy}}-  {}^{\va R}\hat g^{-1}_{\alpha_{xy}}\right) {}^{\va R}\hat g^{-1}_{s_z} [{}^{\va R}\hat g_{s_z}, {}^{\va R}\hat V(v)]\right]|v_3^{\va R}(j)\rangle\n\\
&&=-8\pi \sqrt{\frac{\gamma}{\kappa}}N(v)
\sum_{\mu=\pm 1/2}s(\mu)\sqrt{j_xj_y (j_z+\mu)}\int_0^{2\pi} \!d\alpha\,
\begin{array}{c}
\includegraphics[width=8.cm]{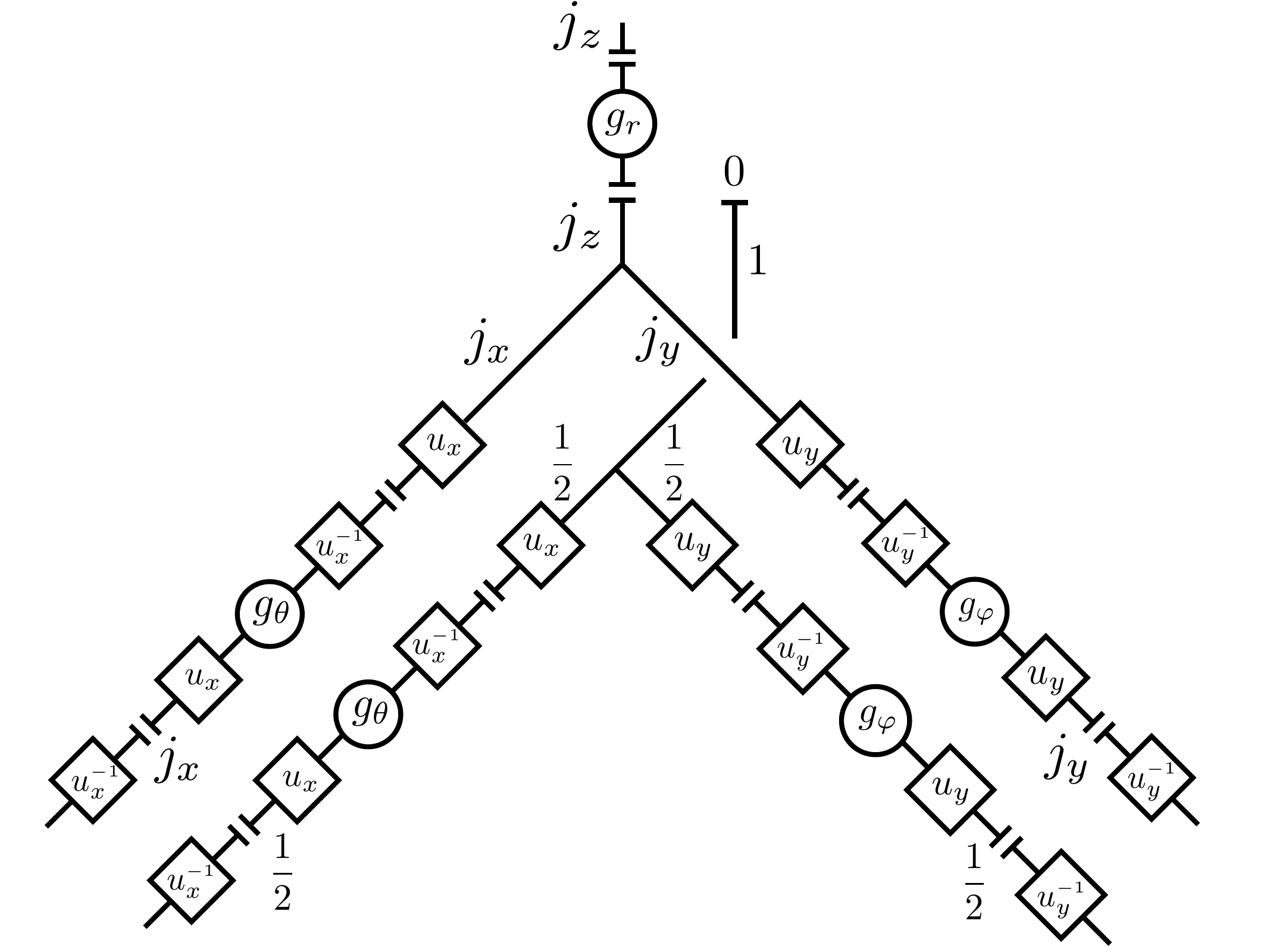}\end{array}\n\\
&&= -8\pi \sqrt{\frac{\gamma}{\kappa}}N(v) 
\sum_{\mu,\mu_x^{\va m},\mu_x^{\va n},\mu_y^{\va m},\mu_y^{\va n}=\pm 1/2}s(\mu)\sqrt{j_xj_y (j_z+\mu)}\int_0^{2\pi} \!d\alpha\,
\begin{array}{c}
\includegraphics[width=8.cm]{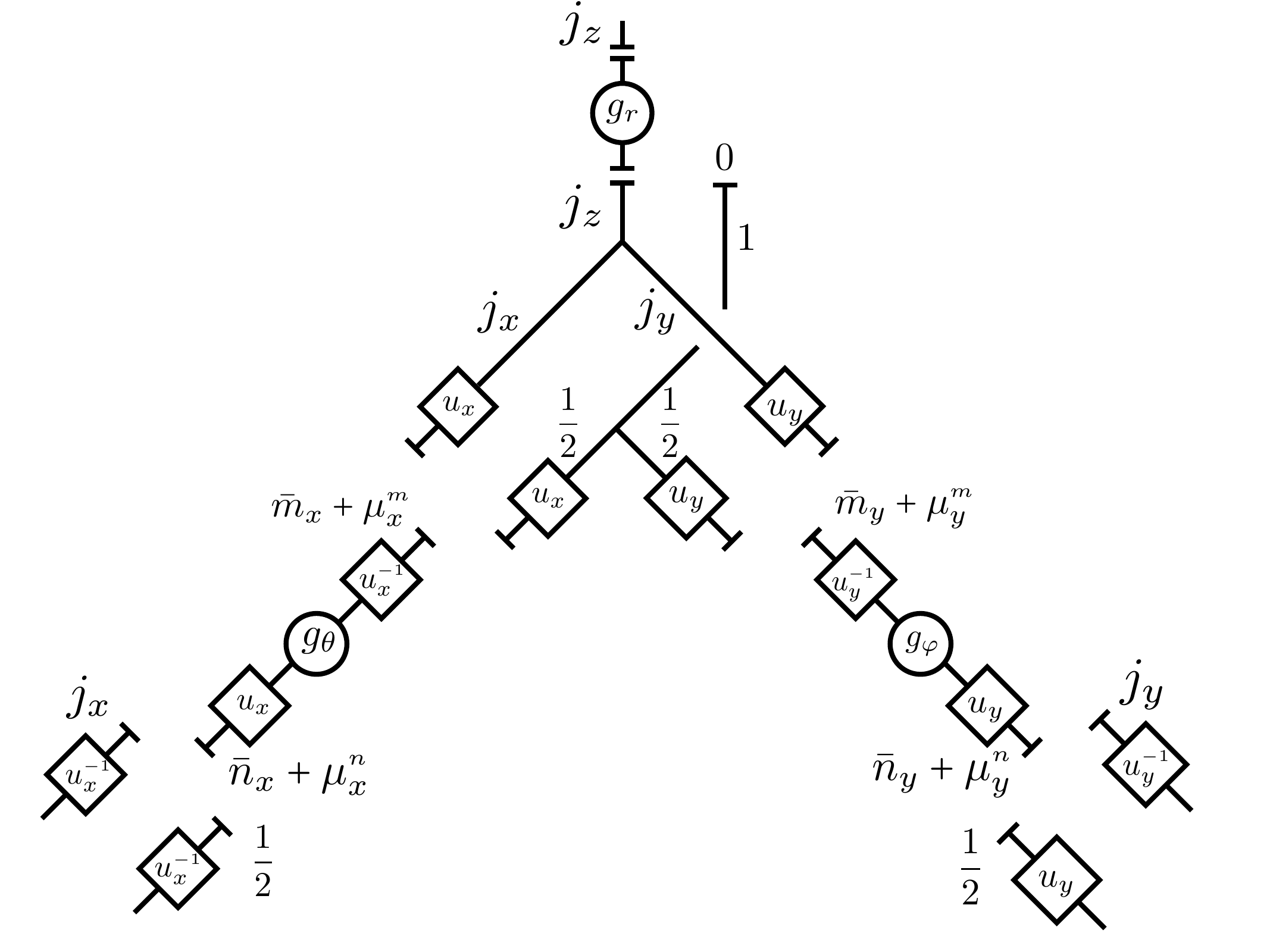}\end{array}\!\!\!\!\!\!\!.\la{Hv3}
\ea

If we now include also the other three 3-valent nodes of the graph that close the loop attached by the Hamiltonian constraint and we denote the associated state as $|v_{\va \square}^{\va R}\rangle$, we obtain
\ba
&&-\frac{4}{\kappa^2\gamma}N(v)\tr\left[\left( {}^{\va R}\hat g_{\alpha_{xy}}-  {}^{\va R}\hat g^{-1}_{\alpha_{xy}}\right) {}^{\va R}\hat g^{-1}_{s_z} [{}^{\va R}\hat g_{s_z}, {}^{\va R}\hat V(v)]\right]|v_{\va \square}^{\va R}\rangle\n\\
&&
= -8\pi \sqrt{\frac{\gamma}{\kappa}}N(v)
\sum_{\mu,\mu_x^{\va m},\mu_x^{\va n},\mu_y^{\va m},\mu_y^{\va n},\mu_x^{\va m'},\mu_x^{\va n'},\mu_y^{\va m'},\mu_y^{\va n'}=\pm 1/2}s(\mu)\sqrt{j_xj_y (j_z+\mu)}\int_0^{2\pi} \!d\alpha\,\n\\
&&\begin{array}{c}
\includegraphics[width=10.cm]{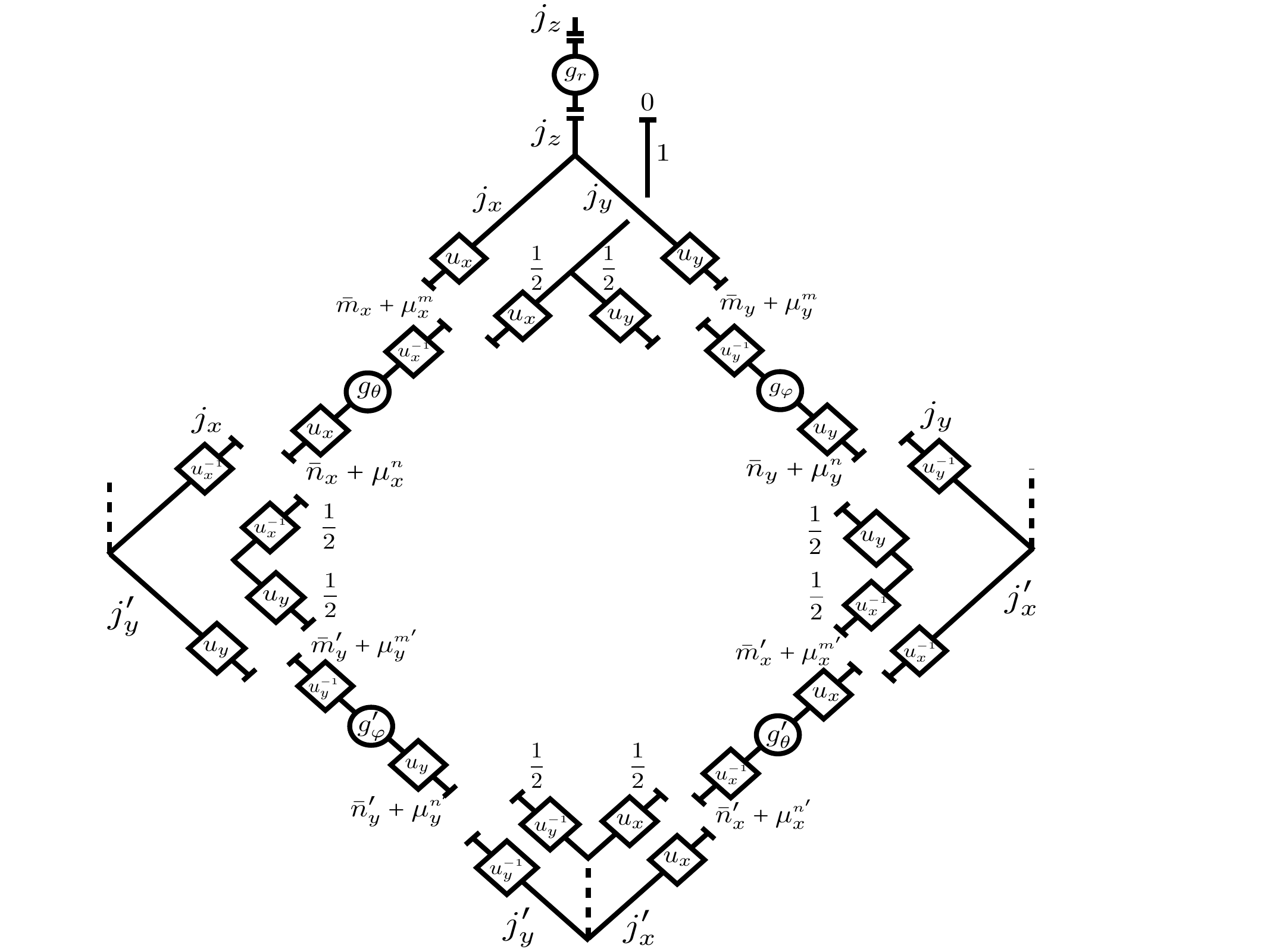}\end{array},\la{Hv3-loop}
\ea
where $s(\mu)$ denotes the sign of $\mu$ and   the dashed lines indicate  the radial links of the other nodes where the constraint has not acted. Notice that we have not used the recoupling theory at the nodes since, in the expectation value of the Hamiltonian constraint, the intertwiners of the states will get reabsorbed in the normalization of the vertices [see Eq. \eqref{norm}].

The complete action of the operator given in Eq. \eqref{HE}
can now be deduced from Eq. \eqref{HE12} with the following consideration. 
The loop operator in the $ij$ plane, being made of ${}^{i}D$ and 
${}^{j}D$ reduced states, will introduce the 2-valent 
intertwiners projected in the $ij$ direction and a coupling of the spins. The expression $ {}^{\va R}\hat g^{-1}_{s_k} [{}^{\va R}\hat g_{s_k}, {}^{\va R}\hat V(v)]$  will now just 
introduce a 3-valent intertwiner oriented along 
$ijk$ and  a coefficient  $\sqrt{j_i j_j (j_k +1/2)} - \sqrt{j_i j_j (j_k - 1/2)}$. The resulting expression is given by the sum over cyclic permutations of the second line of Eq. \eqref{Hv3-loop}.

\subsection{Lorentzian term} \la{slterm}

As anticipated above, we are going to quantize the Lorentzian term of the Hamiltonian constraint starting from its classical expression in terms of the Ricci scalar as it appears in Eq. \eqref{LHR}. In fact, by replacing the spin connection components in terms of the fluxes and their first and second derivatives, as follows from solving the torsion-free condition, we obtain the  Lorentzian term in the form given in Eq. \eqref{H_L}. In the final operator only the reduced fluxes and their derivatives appear. In the $\mathcal{H}^R$ construction that was outlined in  Sec. \ref{sec:geo-op}, we saw how the reduced fluxes have a representation which is diagonal on the reduced spin network states. The spatial derivatives of the reduced fluxes $ {}^{\va R}E_i(S^b(x))$ can be quantized in terms of discrete differences of reduced flux operators acting on neighboring links. More precisely, for first and second order derivatives we have, respectively,
\ba
&&\partial_a {}^{\va R}\hat{E}_i(S^b(v))\equiv {}^{\va R}\hat{E}_i(S^b(v+\epsilon_a))-{}^{\va R}\hat{E}_i(S^b(v))\,,\la{partialE}\\
&&\partial^2_a {}^{\va R}\hat{E}_i(S^b(v))\equiv {}^{\va R}\hat{E}_i(S^b(v+2\epsilon_a))-2{}^{\va R}\hat{E}_i(S^b(v+\epsilon_a))+{}^{\va R}\hat{E}_i(S^b(v))\,,\la{partial2E}
\ea
where $v$ denotes the node whose departing links are dual to the surfaces $S^b$ where the fluxes are smeared and $v+\epsilon_a\, (v+2\epsilon_a)$ are the neighboring  (next to the neighboring) nodes in the direction $a$, taking into account the spatial manifold orientation.

In this way, the lengthy expression in Eq. \eqref{H_L} can be quantized in a straightforward manner, without having to rely on Thiemann's regularization techniques  \cite{Thiemann:1996aw} for the Lorentzian term expressed in terms of the extrinsic curvature. We will use the quantization scheme given in Eqs. \eqref{partialE} and \eqref{partial2E} to compute the expectation value of the Lorentzian Hamiltonian operator in Sec. \ref{sec:Lor-term} below (as well as for the extension term of the Euclidean part in Sec. \ref{sec:Euc-term}).  See Appendix \ref{sec:C} for an alternative quantization scheme for the 3D Ricci scalar.

\section{Semiclassical states} \la{sec:semi.states}

The construction of semiclassical states in $\mathcal{H}^R$ follows the prescription outlined in \cite{Alesci:2014uha}, which is in turn based on the definition introduced in \cite{Thiemann:2000ca}. \footnote{See also \cite{Dasgupta:2005yu} for a previous attempt to investigate singularity resolution through the use of LQG coherent states for a Schwarzschild spacetime.} The key ingredient is the heat kernel of the Laplace
operator for each edge $\ell$ of the graph acting on the $\delta$-function of the $SU(2)$ group element $g_\ell$ associated to the link. Explicitly 
\ba
K_\lambda(g_\ell, g)&=& e^{-\frac{\lambda}{2}\Delta_{g_\ell}}\delta(g_\ell, g)\n\\
&=& \sum_{j_\ell}(2j_\ell+1)e^{-\frac{\lambda}{2}j_\ell(j_\ell+1)}\chi_{j_\ell}(g_\ell^{-1} g)\,,
\ea
with $\lambda$ being a positive real number controlling the fluctuations of the state and $\chi_{j_\ell}$ the $SU(2)$ character in the irreducible representation $j_\ell$.
Here we are introducing the convention that the index $\ell$ indicates both the tangent coordinate and the associated internal direction, namely $\ell \in \{(r,z), (\theta,x), (\varphi,y)\}$.
Coherent semiclassical states are then obtained by analytic continuation from $g\in SU(2)$ to $G\in SL(2,\C)$, namely
\be
\psi^\lambda_{G}(g_\ell)=K_\lambda(g_\ell, G)\,,
\ee
where the complexifier $G$ reads
\be\la{G}
G=g\exp{\left(i\frac{ \lambda}{\kappa\gamma }E_i(S^\ell)\tau^i\right)}\,,
\ee
and $E_i(S^\ell)$ is the flux across the surface dual to the link $\ell$ of area $\delta^2_\ell$ in the fiducial metric.

Therefore, using the classical expressions for the fluxes and connection components found in the previous sections, the semiclassical states for the three directions of the cellular decomposition read [see Appendix \ref{sec:B} for more details on the derivation]
\ba
\psi^\lambda_{G}(g_r)&=& \sum_{j_z=0}^\infty\sum_{\bar m_z, \bar n_z}(2j_z+1)e^{-\frac{\lambda}{2}j_z(j_z+1)} D^{j_z}_{\bar m_z \bar n_z}(g_r^{-1})D^{j_z}_{\bar n_z \bar m_z }(e^{\epsilon_r A_r\tau_3} e^{i\frac{\lambda \delta^2_r}{\kappa\gamma }E^{r}\sin{\theta}\tau_3})\n\\
&=&\sum_{j_z=0}^\infty \sum_{\bar m_z}(2j_z+1)e^{-\frac{\lambda}{2}j_z(j_z+1)} e^{\lambda\bar m_z \frac{ \delta^2_r  E^{r}\sin{\theta}}{\kappa\gamma }}
D^{j_z}_{\bar n_z \bar m_z }(e^{\epsilon_r A_r\tau_3}) \,D^{j_z}_{\bar m_z \bar n_z}(g_r^{-1})\,,\la{psir}\\
\psi^\lambda_{G}(g_\theta)&=& \sum_{j_x=0}^\infty\sum_{\bar m_x, \bar n_x}(2j_x+1)e^{-\frac{\lambda}{2}j_x(j_x+1)} {}^x\!D^{j_x}_{\bar m_x \bar n_x}(g_\theta^{-1}){}^x\!D^{j_x}_{\bar n_x \bar m_x}\left(e^{\epsilon_\theta(A_1\tau_1 + A_2\tau_2)} e^{i\frac{\lambda \delta^2_\theta}{\kappa\gamma }(E^1\tau_1+E^2\tau_2)\sin{\theta}}\right)\n\\
&=& \sum_{j_x=0}^\infty\sum_{\bar m_x, \bar n_x}(2j_x+1)e^{-\frac{\lambda}{2}j_x(j_x+1)} 
 e^{\lambda\bar m_x \frac{ \delta^2_\theta E^x}{\kappa\gamma }}
{}^x\!D^{j_x}_{\bar n_x \bar m_x}\left(e^{\epsilon_\theta(A_1\tau_1 + A_2\tau_2)} \right)
 {}^x\!D^{j_x}_{\bar m_x \bar n_x}(g_\theta^{-1})
 \,,\la{psit}\\
\psi^\lambda_{G}(g_\varphi)&=& \sum_{j_y=0}^\infty\sum_{\bar m_y, \bar n_y}(2j_y+1)e^{-\frac{\lambda}{2}j_y(j_y+1)} {}^y\!D^{j_y}_{\bar m_y \bar n_y}(g_\varphi^{-1}){}^y\!D^{j_y}_{\bar n_y \bar m_y}\left(e^{\epsilon_\varphi\left((A_1\tau_2 -A_2\tau_1)\sin{\theta}
\right)} e^{i\frac{\lambda \delta^2_\varphi}{\kappa\gamma }(E^1\tau_2-E^2\tau_1)}\right)\n\\
&=& \sum_{j_y=0}^\infty\sum_{\bar m_y, \bar n_y}(2j_y+1)e^{-\frac{\lambda}{2}j_y(j_y+1)} 
e^{\lambda\bar m_y\frac{ \delta^2_\varphi E^y }{\kappa\gamma }}
{}^y\!D^{j_y}_{\bar n_y \bar m_y}\left(e^{\epsilon_\varphi\left((A_1\tau_2 -A_2\tau_1)\sin{\theta}
\right)} \right)
{}^y\!D^{j_y}_{\bar m_y \bar n_y}(g_\varphi^{-1}) 
\,,\la{psiv}
\ea
where we have used the property
\be
\chi_{j_\ell}(g_\ell^{-1} G)={}^\ell\!D^{j_\ell}_{\bar m_\ell \bar m_\ell}(g_\ell^{-1} G)=
\sum_{n=-j_\ell}^{j_\ell}{}^\ell\!D^{j_\ell}_{\bar m_\ell n}(g_\ell^{-1} ) {}^\ell\!D^{j_\ell}_{n \bar m_\ell}(G)
={}^\ell\!D^{j_\ell}_{\bar m_\ell \bar n_\ell }(g_\ell^{-1} ) {}^\ell\!D^{j_\ell}_{\bar n_\ell  \bar m_\ell}(G)\,,
\ee
and the following  relations that we derived in Appendix \ref{sec:B},
\ba
E^x&=&(E^1\cos{\tilde\alpha}+E^2\sin{\tilde\alpha})\sin{\theta}=\Lambda R\sin{\theta}\,,\\
E^y&=&E^1\cos{\tilde\alpha}+E^2\sin{\tilde\alpha}=\Lambda R\,.
\ea
Notice that in the coherent state associated with the $\varphi$-direction we have not included the $A^3_\varphi=\cos\theta$ component of the connection since this does not enter the reduced phase space (it is conjugate to a flux component that we have gauge fixed)  and thus it is not part of the reduced Hilbert space $\mathcal{H}^R$ either.
Its contribution to the spherical Euclidean Hamiltonian constraint given in Eq. \eqref{HH-spher} is encoded in the extra terms that appear in Eq. \eqref{eq:hh:tilde}, as already pointed out above.

We can see from Eq. \eqref{co-st} that for $j_x, j_y,  j_z\gg 1$ the coefficients $\psi^\lambda_{G}(j_\ell) $ in the coherent states  become Gaussian weights for the fluxes peaked around the semiclassical values
$\widetilde j_\ell=\delta^2_\ell j^0_\ell $, with $j^0_\ell$ given by
\begin{subequations}\la{j0}
\ba
j^{\va 0}_x&=&\frac{\Lambda R\sin{\theta}}{\kappa\gamma}=\frac{\sqrt{g_{RR} g_{\varphi\varphi}}}{\kappa\gamma}
=\frac{E_1^\theta}{\kappa\gamma\cos\tilde\alpha}=\frac{E_2^\theta}{\kappa\gamma\sin\tilde\alpha}
=\frac{1}{\kappa\gamma}\left(E_1^\theta\cos\tilde\alpha+E_2^\theta\sin\tilde\alpha\right)
\,,\\
j^{\va 0}_y&=&\frac{\Lambda R}{\kappa\gamma}=\frac{\sqrt{g_{RR} g_{\theta\theta}}}{\kappa\gamma}
=-\frac{E_1^\varphi}{\kappa\gamma\sin\tilde\alpha}=\frac{E_2^\varphi}{\kappa\gamma\cos\tilde\alpha}
=\frac{1}{\kappa\gamma}\left(-E_1^\varphi\sin\tilde\alpha+E_2^\varphi\cos\tilde\alpha\right)
\,,\\
j^{\va 0}_z&=&\frac{ R^2\sin{\theta}}{\kappa\gamma}=\frac{g_{\varphi\varphi}}{\kappa\gamma\sin{\theta}}=\frac{E^r_3}{\kappa\gamma}\,,
\ea
\end{subequations}
 and $\delta^2_x=\epsilon_r \epsilon_\varphi, \delta^2_y=\epsilon_r \epsilon_\theta, \delta^2_z=\epsilon_\theta \epsilon_\varphi$.

%

Let us  write the quantum reduced coherent states in the compact notation
\be\la{co-st}
\psi^\lambda_{G}(g_\ell)= \sum_{j_\ell=0}^\infty\sum_{\bar m_\ell, \bar n_\ell=\pm j_\ell}(2j_\ell+1)(\psi^\lambda_{G})^{j_\ell}_{ \bar n_\ell \bar m_\ell} \,{}^\ell\!D^{j_\ell}_{\bar m_\ell \bar n_\ell}(g_\ell^{-1})\,,
\ee
with the matrix coefficients $(\psi^\lambda_{G})^{j_\ell}_{ \bar n_\ell \bar m_\ell} $ explicitly given by Eqs. \eqref{psir}, \eqref{psit}, and \eqref{psiv}.

Finally, we can define the normalized quantum reduced coherent states as
\be\la{co-st-norm}
\widetilde{ \psi^\lambda_{G}}(g_\ell)=\frac{\psi^\lambda_{G}(g_\ell)}{|\psi^\lambda_{G}(g_\ell)|}  \,,
\ee
where
\be
|\psi^\lambda_{G}(g_\ell)|=\sqrt{\sum_{j_\ell=0}^\infty\sum_{\bar m_\ell, \bar n_\ell=\pm j_\ell}|(\psi^\lambda_{G})^{j_\ell}_{ \bar n_\ell \bar m_\ell}|^2}\,.
\ee

Let us conclude this section by defining the coherent state associated to the reduced spin network state used in Eq. \eqref{Hv3-loop} to compute the action of the reduced Euclidean Hamiltonian constraint on it. By including also the faces in the $(r,\theta)$-plane and $(r,\varphi)$-plane, the reduced spin network state reads
\be
|v_{\va \square}^{\va R}\rangle=\begin{array}{c}
\includegraphics[width=10.cm]{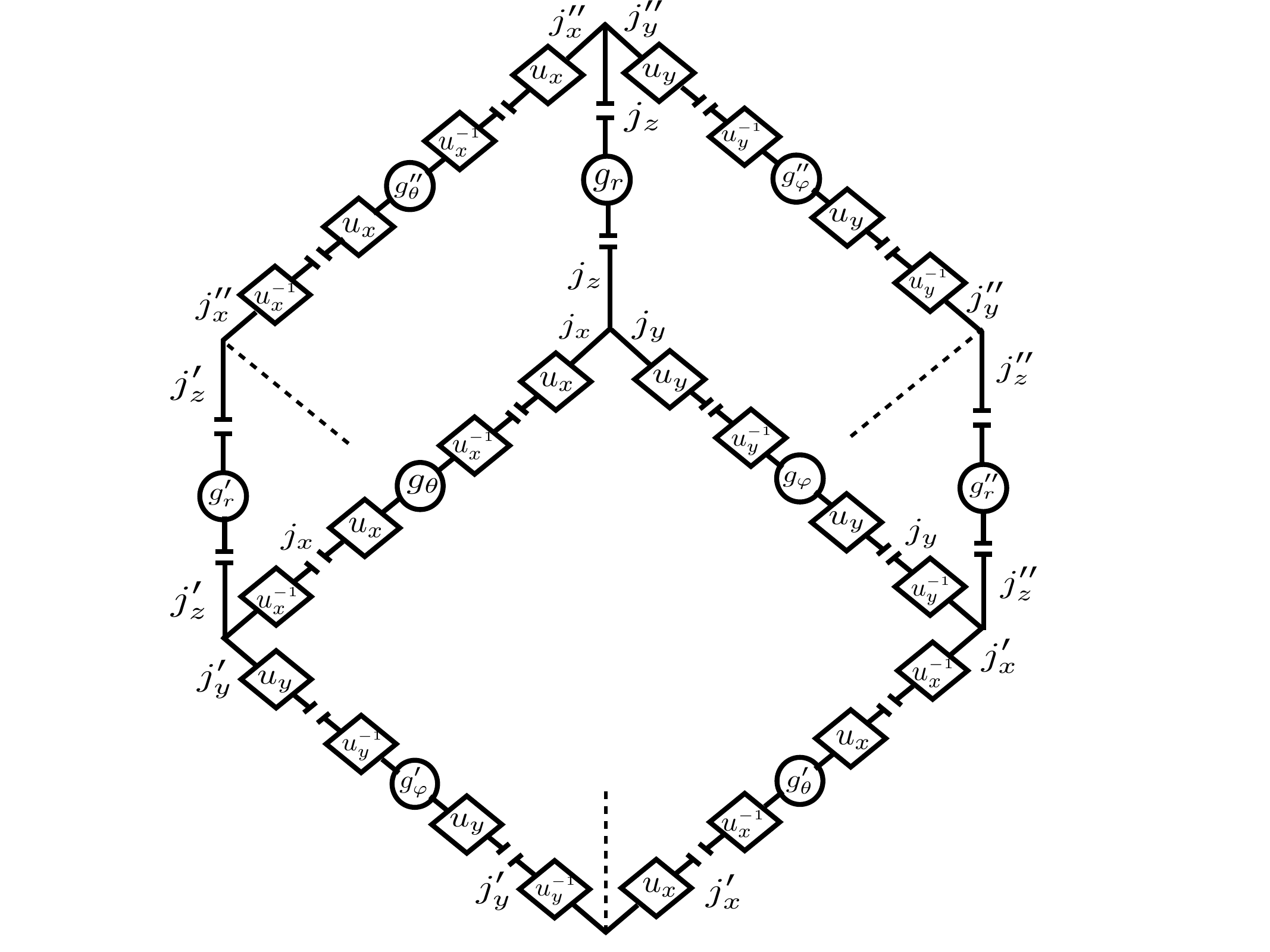}\end{array}.\la{cube}
\ee
The associated normalized quantum reduced coherent state is then given by
\be
|\widetilde{\psi^\lambda_{\va \square}}\rangle=\prod_{\ell=x,y,z}\sum_{j_\ell, j'_\ell, j''_\ell=0}^\infty\sum_{\bar m_\ell, \bar n_\ell =\pm j_\ell}\sum_{\bar m'_\ell, \bar n'_\ell =\pm j'_\ell}
\sum_{\bar m''_\ell, \bar n''_\ell =\pm j''_\ell}
\!\!\!\begin{array}{c}
\includegraphics[width=9.cm]{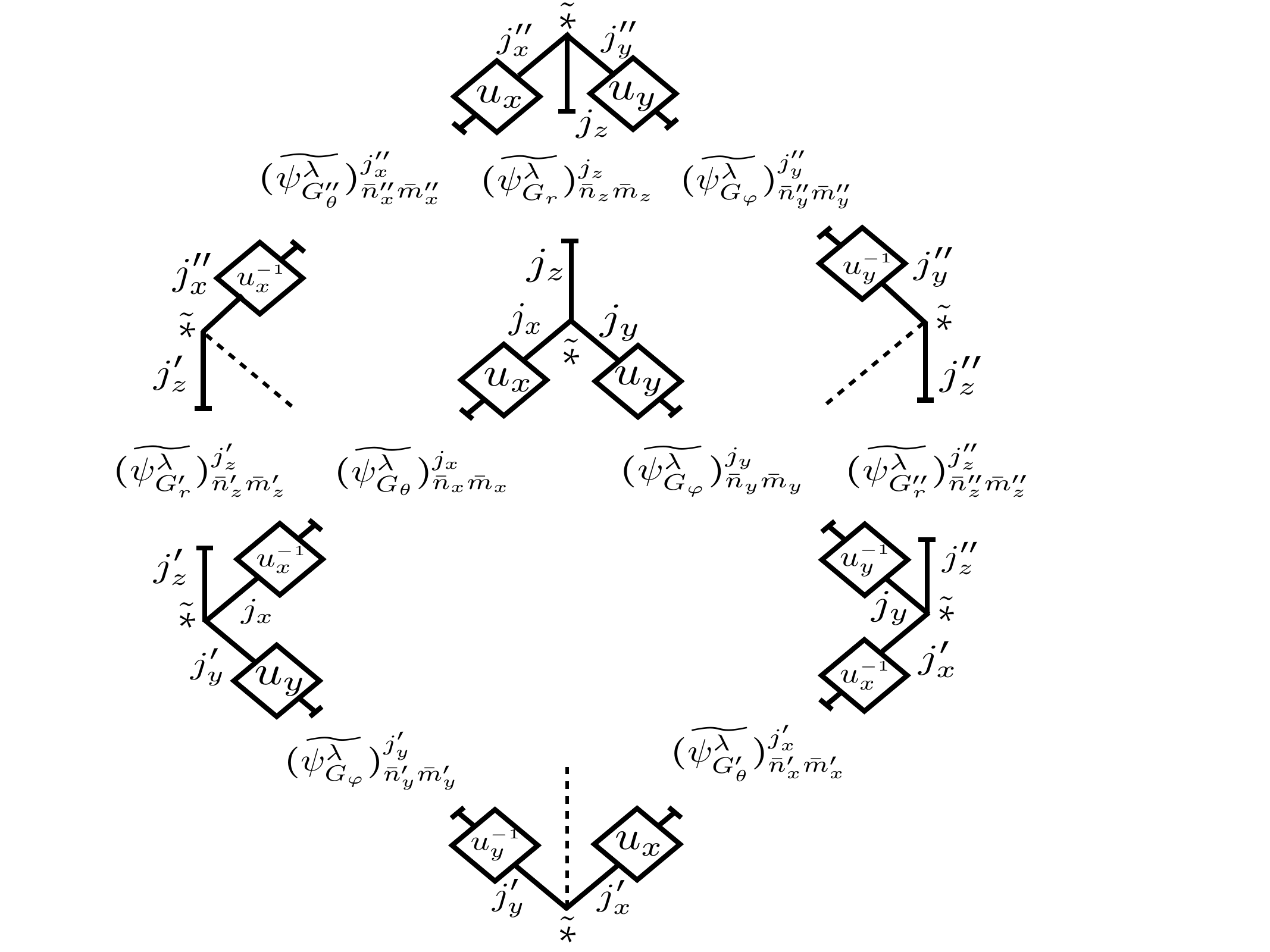}\end{array}
\!\!\!\!\!\!\!\!\!\!\!\!|\widetilde{v_{\va \square}^{\va R}}\rangle\,,\la{coherent-cube}
\ee
where for each 3-valent vertex in the state defined in Eq. \eqref{cube} we have used the norm given in Eq. \eqref{norm} and we have introduced the notation
\be
\begin{array}{c}
\includegraphics[width=1.5cm]{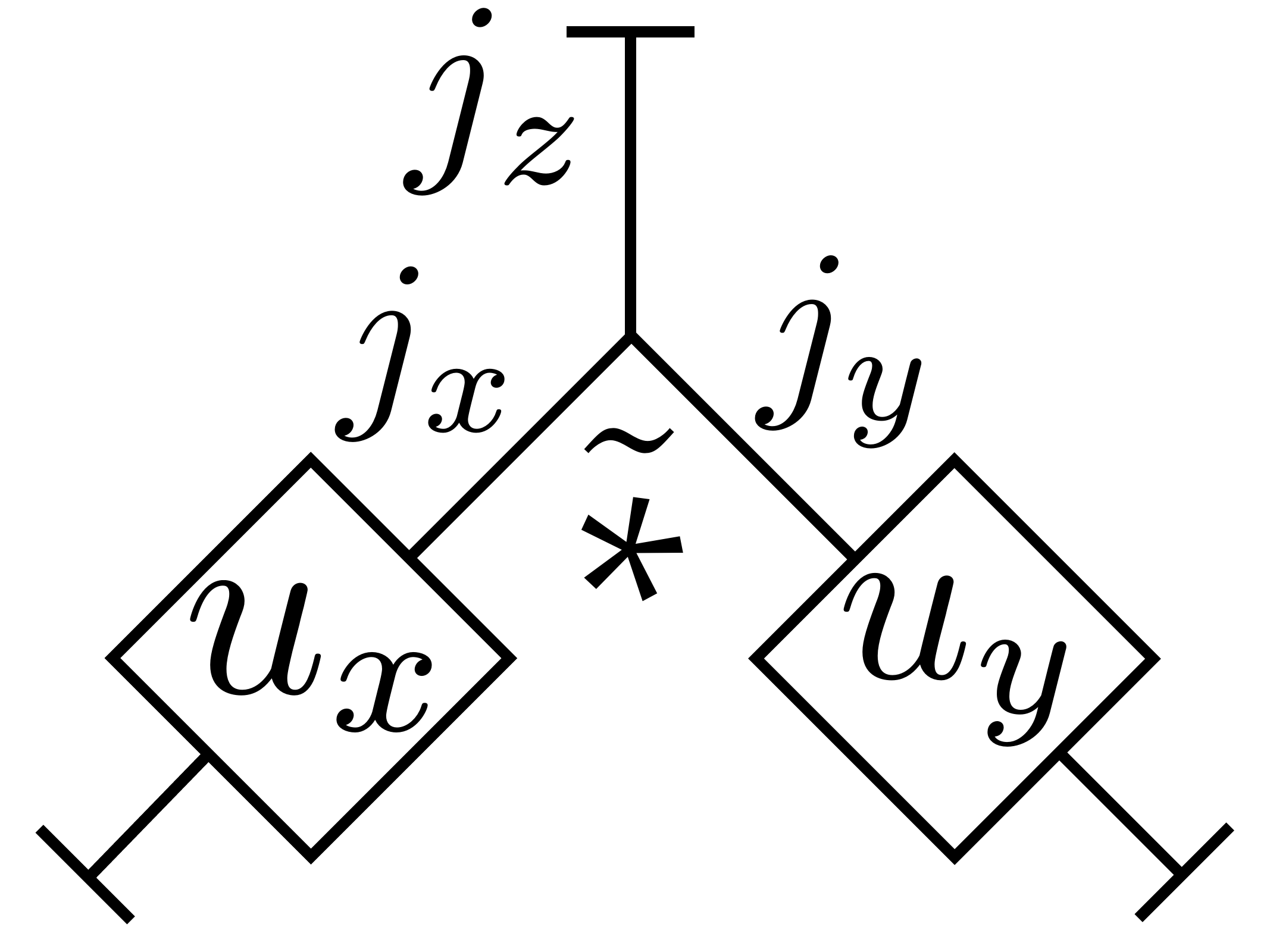}\end{array}
=\frac{\left(\begin{array}{c}
\includegraphics[width=1.5cm]{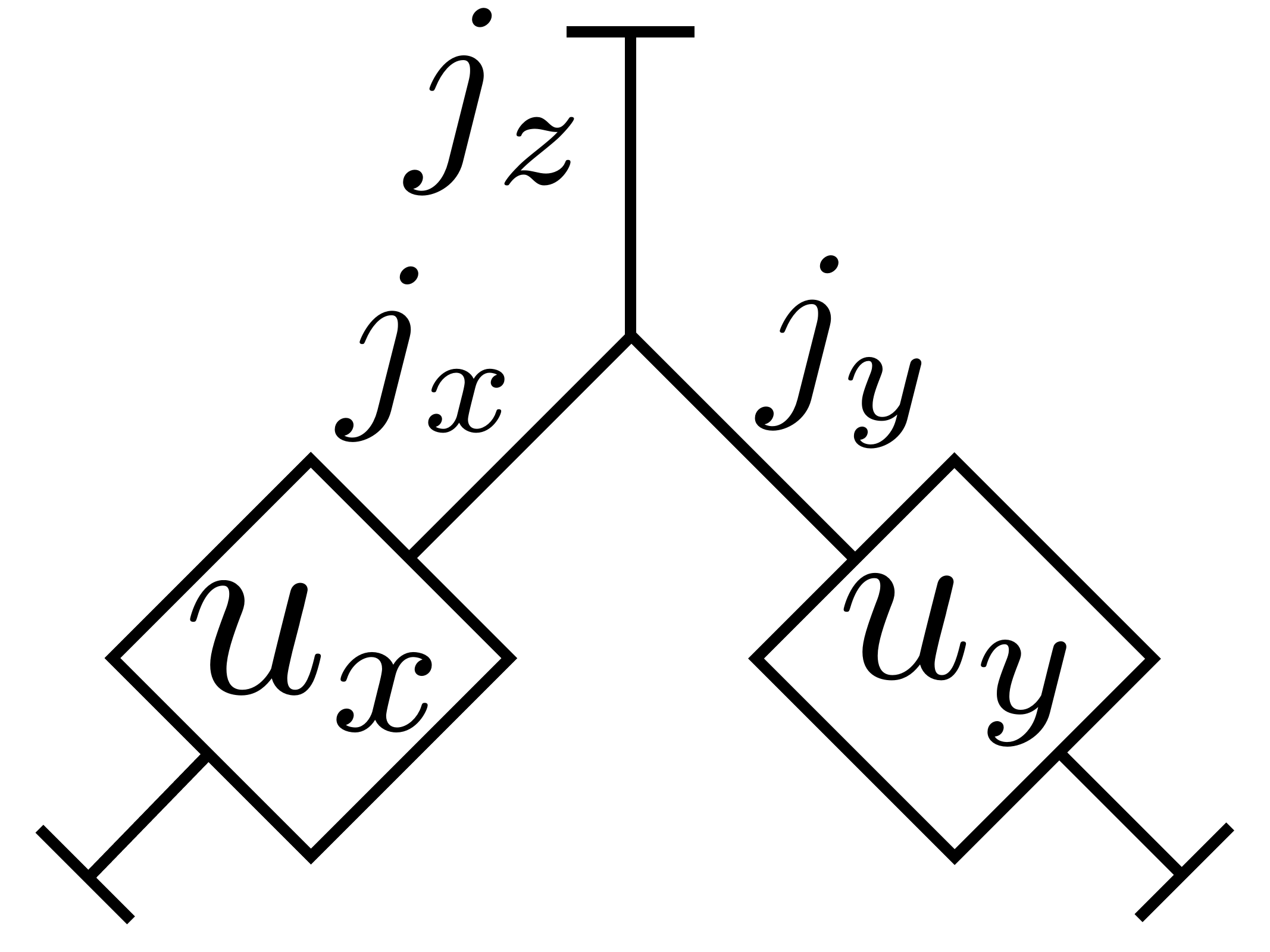}\end{array}\right)^*
}{\sqrt{
\left(\begin{array}{c}
\includegraphics[width=1.5cm]{norm1.pdf}\end{array}\right)^*
\begin{array}{c}
\includegraphics[width=1.5cm]{norm1.pdf}\end{array}
}
}\,,
\ee
in order to include the full normalization of the coherent state wave function.

\section{Effective Hamiltonian}\la{sec:eff.ham}

We are now ready to compute the expectation value of the reduced  Hamiltonian constraint on the coherent state that we define in Eq. \eqref{coherent-cube} based on a single cuboidal cell of the spatial manifold triangulation by means of the action given in Eq. \eqref{Hv3-loop} and its analogue on the other two orthogonal planes. 

\subsection{Euclidean term}\la{sec:Euc-term}

Let us start with the Euclidean Hamiltonian constraint operator whose action on the basis of $\mathcal{H}^R$ was computed in Eq. \eqref{Hv3-loop} for one choice of tangent loop. For the other two possibilities it is straightforward to see that a similar result holds, as already explained at the end of Sec. \ref{sec:rhco-A}. If we now use this action to compute the expectation value on the normalized coherent state of the form given in Eq. \eqref{coherent-cube}, by means of the normalized coherent state wave function properties we obtain the following result:
\ba
&&\langle \widetilde{\psi^\lambda_{\va \square}}| {}^{\va R}\hat H^{\va E}_{\va \square}[N] |\widetilde{\psi^\lambda_{\va \square}}\rangle\approx
-4 \sqrt{\frac{\gamma}{\kappa}}N(v)
\sum_{\mu=\pm1/2}{s(\mu)}\n\\
&&\times \Bigg(
\sqrt{\widetilde j_z\widetilde  j_y (\widetilde j_x+\mu)}\tr\left[\tau_x e^{\epsilon_\varphi\left((A_1(r)\tau_2 -A_2(r)\tau_1)\sin{\theta}
\right)} 
 e^{\epsilon_r A_r(r)\tau_3} 
 e^{-\epsilon_\varphi\left((A_1(r+\epsilon_r)\tau_2 -A_2(r+\epsilon_r)\tau_1)\sin{\theta}
 \right)} 
e^{-\epsilon_r A_r(r+\epsilon_r)\tau_3} \right]\n\\
&&-\sqrt{\widetilde j_z\widetilde  j_x (\widetilde j_y+\mu)}\tr\left[\tau_y e^{\epsilon_\theta(A_1(r)\tau_1 + A_2(r)\tau_2)} e^{\epsilon_r A_r(r)\tau_3} 
e^{-\epsilon_\theta(A_1(r+\epsilon_r)\tau_1 + A_2(r+\epsilon_r)\tau_2)} 
e^{-\epsilon_r A_r(r+\epsilon_r)\tau_3} \right]\n\\
&&+\sqrt{\widetilde j_x\widetilde j_y (\widetilde j_z+\mu)}\tr\Big[ \tau_3 e^{\epsilon_\theta(A_1(r)\tau_1 + A_2(r)\tau_2)} 
e^{\epsilon_\varphi \left((A_1(r)\tau_2 -A_2(r)\tau_1)\sin{(\theta+\epsilon_\theta)}
\right)}\n\\
&&\times \,\,
e^{-\epsilon_\theta(A_1(r)\tau_1 + A_2(r)\tau_2)} 
e^{-\epsilon_\varphi\left((A_1(r)\tau_2 -A_2(r)\tau_1)\sin{\theta}
\right)} 
\Big]
\Bigg)\n\\
&&\approx
-4\sqrt{\frac{\gamma}{\kappa}}\frac{N(v)}{ \sqrt{ j^0_x  j^0_yj^0_z }}
\n\\
&&\times \Bigg(
 \epsilon_\theta j^0_z   j^0_y\,\tr\left[\tau_x e^{\epsilon_\varphi\left((A_1(r)\tau_2 -A_2(r)\tau_1)\sin{\theta}
 \right)} 
 e^{\epsilon_r A_r(r)\tau_3} 
 e^{-\epsilon_\varphi\left((A_1(r+\epsilon_r)\tau_2 -A_2(r+\epsilon_r)\tau_1)\sin{\theta}
 \right)} 
e^{-\epsilon_r A_r(r+\epsilon_r)\tau_3} \right]\n\\
&&-\epsilon_\varphi j^0_z  j^0_x  \,\tr\left[\tau_y e^{\epsilon_\theta(A_1(r)\tau_1 + A_2(r)\tau_2)} e^{\epsilon_r A_r(r)\tau_3} 
e^{-\epsilon_\theta(A_1(r+\epsilon_r)\tau_1 + A_2(r+\epsilon_r)\tau_2)} 
e^{-\epsilon_r A_r(r+\epsilon_r)\tau_3} \right]\n\\
&&+\epsilon_r j^0_x j^0_y \,\tr\Big[ \tau_3 e^{\epsilon_\theta(A_1(r)\tau_1 + A_2(r)\tau_2)} 
e^{\epsilon_\varphi \left((A_1(r)\tau_2 -A_2(r)\tau_1)\sin{(\theta+\epsilon_\theta)}
\right)}\n\\
&&\times \,\,
e^{-\epsilon_\theta(A_1(r)\tau_1 + A_2(r)\tau_2)} 
e^{-\epsilon_\varphi\left((A_1(r)\tau_2 -A_2(r)\tau_1)\sin{\theta}
\right)} 
\Big]
\Bigg)\n\\
&&=
-\frac{4}{\kappa} \frac{N(v)}{ \sqrt{{\rm det}(E)}}
\n\\
&&\times \Bigg[
- \epsilon_\theta E^r_3 \left(-E_1^\varphi\sin\tilde\alpha+E_2^\varphi\cos\tilde\alpha\right)\,
\bigg[ie^{-i\tilde\alpha-\frac{i}{2}(A_r(r)+A_r(r+\epsilon_r))\epsilon_r -\frac{1}{2}\left( \sqrt{-A_1^2(r)-A_2^2(r)} +\sqrt{-A_1^2(r+\epsilon_r)-A_2^2(r+\epsilon_r)}   \right) \sin{\theta}\epsilon_\varphi}
\bigg]\n\\
&&\times\Bigg(
\left( 1+e^{\sqrt{-A_1^2(r)-A_2^2(r)}\sin{\theta}\epsilon_\varphi}   \right)
\left( -1+e^{\sqrt{-A_1^2(r+\epsilon_r)-A_2^2(r+\epsilon_r)}\sin{\theta}\epsilon_\varphi}   \right)
\sqrt{-A_1^2(r)-A_2^2(r)}\n\\
&&\times
\bigg(
\left( e^{2i\tilde\alpha} -e^{i(A_r(r)+A_r(r+\epsilon_r))\epsilon_r} \right)
A_1(r+\epsilon_r)
-i\left( e^{2i\tilde\alpha} +e^{i(A_r(r)+A_r(r+\epsilon_r))\epsilon_r} \right)
A_2(r+\epsilon_r)
\bigg)\n\\
&&+
\left( -1+e^{\sqrt{-A_1^2(r)-A_2^2(r)}\sin{\theta}\epsilon_\varphi}   \right)
\left( 1+e^{\sqrt{-A_1^2(r+\epsilon_r)-A_2^2(r+\epsilon_r)}\sin{\theta}\epsilon_\varphi}   \right)\n\\
&&\times
\bigg(
-e^{i(2\tilde\alpha+A_r(r)\epsilon_r)} (A_1(r)-iA_2(r)) 
+e^{iA_r(r+\epsilon_r)\epsilon_r} (A_1(r)+iA_2(r)) 
\bigg)\sqrt{-A_1^2(r+\epsilon_r)-A_2^2(r+\epsilon_r)}
\Bigg)\n\\
&&\Bigg/
\Bigg(8\sqrt{-A_1^2(r)-A_2^2(r)} \sqrt{-A_1^2(r+\epsilon_r)-A_2^2(r+\epsilon_r)}\Bigg)\n\\
&&+ \epsilon_\varphi E^r_3 \left(E_1^\theta\cos\tilde\alpha +E_2^\theta\sin\tilde\alpha\right)\,
\bigg[ie^{-i\tilde\alpha-\frac{i}{2}(A_r(r)+A_r(r+\epsilon_r))\epsilon_r -\frac{1}{2}\left( \sqrt{-A_1^2(r)-A_2^2(r)} +\sqrt{-A_1^2(r+\epsilon_r)-A_2^2(r+\epsilon_r)}   \right)\epsilon_\theta}
\bigg]\n\\
&&\times \Bigg(
\left( 1+e^{\sqrt{-A_1^2(r)-A_2^2(r)}\epsilon_\theta}   \right)
\left( -1+e^{\sqrt{-A_1^2(r+\epsilon_r)-A_2^2(r+\epsilon_r)}\epsilon_\theta}   \right)
\sqrt{-A_1^2(r)-A_2^2(r)}\n\\
&&\times
\bigg(
-\left( e^{2i\tilde\alpha} -e^{i(A_r(r)+A_r(r+\epsilon_r))\epsilon_r} \right)
A_1(r+\epsilon_r)
+i\left( e^{2i\tilde\alpha} +e^{i(A_r(r)+A_r(r+\epsilon_r))\epsilon_r} \right)
A_2(r+\epsilon_r)
\bigg)\n\\
&&-
\left( -1+e^{\sqrt{-A_1^2(r)-A_2^2(r)}\epsilon_\theta}   \right)
\left( 1+e^{\sqrt{-A_1^2(r+\epsilon_r)-A_2^2(r+\epsilon_r)}\epsilon_\theta}   \right)\n\\
&&\times
\bigg(
-e^{i(2\tilde\alpha+A_r(r)\epsilon_r)} (A_1(r)-iA_2(r)) 
+e^{iA_r(r+\epsilon_r)\epsilon_r} (A_1(r)+iA_2(r)) 
\bigg)
\sqrt{-A_1^2(r+\epsilon_r)-A_2^2(r+\epsilon_r)}\
\Bigg)\n\\
&&\Bigg/
\Bigg(8\sqrt{-A_1^2(r)-A_2^2(r)} \sqrt{-A_1^2(r+\epsilon_r)-A_2^2(r+\epsilon_r)}\Bigg)\n\\
&&+ \epsilon_r \left(E_1^\theta\cos\tilde\alpha +E_2^\theta\sin\tilde\alpha\right) \left(-E_1^\varphi\sin\tilde\alpha+E_2^\varphi\cos\tilde\alpha\right)
\frac{1}{8} e^{-\frac{1}{2}\sqrt{-A_1^2(r)-A_2^2(r)} \left(2\epsilon_\theta + (\sin{\theta} +\sin{(\theta+\epsilon_\theta)})\epsilon_\varphi\right)  }\n\\
&&\times
\left( -1+e^{2\sqrt{-A_1^2(r)-A_2^2(r)}\epsilon_\theta}   \right)
\left( -1+ e^{\sqrt{-A_1^2(r)-A_2^2(r)} \left(\sin{\theta} +\sin{(\theta+\epsilon_\theta)}\right)\epsilon_\varphi  }
 \right)
\Bigg]
\,,\la{HEeff}
\ea
where we have used Eq. \eqref{j0} of the coherent states to peak the fluxes around their semiclassical values.

If we now expand the lengthy expression above to third order in the $\epsilon$'s and use Eq. \eqref{j0}, 
 we get
\ba
\langle \widetilde{\psi^\lambda_{\va \square}}| {}^{\va R}\hat H^{\va E}_{\va \square}[N] |\widetilde{\psi^\lambda_{\va \square}}\rangle&\approx&\frac{\epsilon_r\epsilon_\theta\epsilon_\varphi }{\kappa}\frac{2 N(v) }{\sqrt{{\rm det}(E)}}
\Bigg[E^r_3 A_r\left(E^\theta_1 A_1+E^\theta_2A_2
\right)
+E^r_3 (E^\theta_1A'_2-E^\theta_2A'_1)\n\\
&+&\sin\theta \left(E^r_3 A_r\left(E^\varphi_2 A_1-E^\varphi_1A_2
\right)+E^r_3 (E^\varphi_2 A'_2+E^\varphi_1A'_1)
\right)\n\\
&+&\sin\theta(E^\theta_1 E^\varphi_2-E^\theta_2 E^\varphi_1)\left(
A_1^2+A^2_2\right)
\Bigg] + o(\epsilon^4)\,,
\ea
which matches exactly the classical expression given in Eq. \eqref{EHHR} in the limit $\epsilon_r, \epsilon_\theta, \epsilon_\varphi \rightarrow 0$, once summed over all the vertices.

In order to obtain the full expression of the spherically symmetric Euclidean Hamiltonian constraint given in Eq. \eqref{HH-spher}, we also need to quantize the extra terms in Eq. \eqref{eq:hh:tilde} coming from the phase space extension of the gauge unfixing procedure. The quantization of all these extra terms would result in a rather complicated operator. However, we know from the classical analysis that only the last term in Eq. \eqref{EHext} will remain. Therefore, in order to simplify the construction of the full Euclidean Hamiltonian constraint, let us just quantize the term
\be
{}^{\va R}H^{\va ext}[N]=\frac{2}{\kappa}\int_\Sigma d^3{x}\, \frac{N}{\sqrt{\text{det}(E)}} E_I^A \partial_A\partial_BE^{IB}\,.
\ee
Following the quantization prescription given in Eq. \eqref{partial2E} for the second derivatives of the fluxes,  the quantum version of the Euclidean Hamiltonian constraint regularized on a  cubic cell dual to a 6-valent node $v$  with  coordinates $\{r, \theta, \varphi\}$ and its neighbouring ones is given by
\ba
&&{}^{\va R}\hat H^{\va ext}_{\va \square}[N]=\frac{2}{\kappa} \frac{N(v)}{{}^{\va R}\hat V(v)} 
\Bigg[
{}^{\va R}\hat E_I(S^\theta(r,  \theta, \varphi))
\left(  {}^{\va R}\hat E_I(S^\theta(r, \theta+2\epsilon^\theta, \varphi))
-2{}^{\va R}\hat E_I(S^\theta(r, \theta+\epsilon^\theta, \varphi))
+{}^{\va R}\hat E_I(S^\theta(r, \theta, \varphi))
\right)\n\\
&&+
{}^{\va R}\hat E_I(S^\varphi(r,  \theta, \varphi))
\left(  {}^{\va R}\hat E_I(S^\theta(r, \theta+\epsilon^\theta, \varphi+\epsilon^\varphi))
-{}^{\va R}\hat E_I(S^\theta(r, \theta+\epsilon^\theta, \varphi))
-{}^{\va R}\hat E_I(S^\theta(r, \theta, \varphi+\epsilon^\varphi))
+{}^{\va R}\hat E_I(S^\theta(r, \theta, \varphi))
\right)\n\\
&&+
{}^{\va R}\hat E_I(S^\theta(r,  \theta, \varphi))
\left(  {}^{\va R}\hat E_I(S^\varphi(r, \theta+\epsilon^\theta, \varphi+\epsilon^\varphi))
-{}^{\va R}\hat E_I(S^\varphi(r, \theta, \varphi+\epsilon^\varphi))
-{}^{\va R}\hat E_I(S^\varphi(r, \theta+\epsilon^\theta, \varphi))
+{}^{\va R}\hat E_I(S^\varphi(r, \theta, \varphi))
\right)\n\\
&&+
{}^{\va R}\hat E_I(S^\varphi(r,  \theta, \varphi))
\left(  {}^{\va R}\hat E_I(S^\varphi(r, \theta, \varphi+2\epsilon^\varphi))
-2{}^{\va R}\hat E_I(S^\varphi(r, \theta, \varphi+\epsilon^\varphi))
+{}^{\va R}\hat E_I(S^\varphi(r, \theta, \varphi))
\right)
\Bigg]\,.\la{HRext}
\ea
When computing its expectation value on the coherent state defined in Eq. \eqref{coherent-cube}, it is immediate to see that only the first line on the right-hand side of the expression above contributes. Therefore we get
\ba
\langle \widetilde{\psi^\lambda_{\va \square}}| {}^{\va R}\hat H^{\va ext}_{\va \square}[N] |\widetilde{\psi^\lambda_{\va \square}}\rangle&\approx&\frac{2}{\kappa}\frac{\epsilon_r\epsilon_\varphi }{ \epsilon_\theta}\frac{N(v) }{\sqrt{{\rm det}(E)}}
(E^\theta_1 E^\varphi_2-E^\theta_2 E^\varphi_1)\Big[\sin{(\theta+2\epsilon_\theta)}
-2\sin{(\theta+\epsilon_\theta)}+\sin{\theta}\Big]\n\\
&\approx&-2\frac{\epsilon_r \epsilon_\theta \epsilon_\varphi}{\kappa }\frac{N(v) }{\sqrt{{\rm det}(E)}}
(E^\theta_1 E^\varphi_2-E^\theta_2 E^\varphi_1)\sin{\theta}+ o(\epsilon^4)\,.\la{EHext-exp}
\ea

Therefore, at the leading order we recover
\be
\sum_{\square}\langle \widetilde{\psi^\lambda_{\va \square}}|   \left( {}^{\va R}\hat H^{\va E}_{\va \square}[N]+ {}^{\va R}\hat H^{\va ext}_{\va \square}[N]\right) |\widetilde{\psi^\lambda_{\va \square}}\rangle\approx H_{\va sph}^{\va E}[N]  + o(\epsilon^4)\,,
\ee
showing how our construction exhibits the correct semiclassical limit.

We can now write the first order correction to the classical expression of the reduced Euclidean Hamiltonian constraint above. This is obtained by looking at the terms of order four in  $\epsilon$ in Eq. \eqref{HEeff} as well as in  Eq. \eqref{HRext}. We find
\ba
\langle \widetilde{\psi^\lambda_{\va \square}}| {}^{\va R}\hat H^{\va E}_{\va \square}[N]+ {}^{\va R}\hat H^{\va ext}_{\va \square}[N] |\widetilde{\psi^\lambda_{\va \square}}\rangle_{(1)}&\approx&\frac{2\epsilon_r\epsilon_\theta\epsilon_\varphi }{\kappa}\frac{N(v) }{\sqrt{{\rm det}(E)}}
\Bigg[\frac{1}{2}E^r_3 \left( E^\theta_1(-A_r^2 A_2+2A_r A'_1+A''_2)+E^\theta_2(A_r^2 A_1 +2A_r A'_2-A''_1)  \right)\epsilon_r\n\\
&-&\frac{\sin\theta}{2}E^r_3 \left(  E^\varphi_2 (A_r^2A_2 -2A_rA'_1-A''_2) +E^\varphi_1(A_r^2 A_1 +2A_r A'_2-A''_1)\right)\epsilon_r\n\\
&-&\frac{\sin(2\theta)}{4} \left(E^r_3A_r(E^\varphi_2 A_2+E^\varphi_1 A_1 )+E^r_3 \left(E^\varphi_1A'_2-E^\varphi_2A'_1\right)\right)\epsilon_\varphi\n\\
&+&\cos\theta (E^\theta_1 E^\varphi_2-E^\theta_2 E^\varphi_1)\left(
A_1^2+A^2_2-1\right)\epsilon_\theta
\Bigg]\n\\
&=&\frac{2\epsilon_r\epsilon_\theta\epsilon_\varphi }{\kappa}\frac{N(v) }{\sqrt{((E^1)^2 +(E^2)^2  )E^r}}
\Bigg[\n\\
&&\sin\theta E^r \left( E^1(-A_r^2 A_2+2A_r A'_1+A''_2)+E^2(A_r^2 A_1 +2A_r A'_2-A''_1)  \right)\epsilon_r\n\\
&+&\cos\theta ((E^1)^2 +(E^2)^2)\left(
A_1^2+A^2_2-1\right)\epsilon_\theta\n\\
&-&\frac{\sin(2\theta)}{4} \left(E^rA_r(E^1 A_2-E^2 A_1 )-E^r \left(E^2A'_2 + E^1A'_1\right)\right)\epsilon_\varphi
\Bigg]\,,
\ea
where in the last equality we have used Eq. \eqref{SE2}. Notice that since the  second and the last correction terms proportional to $\epsilon_\theta$ and $\epsilon_\varphi$ contain an overall $\theta$-dependent part of the form  respectively $\cos\theta$ and $\sin(2\theta)$, in the continuum limit when integrating over $\theta\in[0,\pi]$ they both vanish. However, the first correction term proportional  to $\epsilon_r$  has  an overall $\theta$ dependence  that survives the integral  and it represents the only correction  at first order.

Therefore, after performing the sum over all cuboidal cells of the triangulation  and taking the continuum limit $\epsilon_r, \epsilon_\theta, \epsilon_\varphi \rightarrow 0$, we obtain the effective Euclidean Hamiltonian constraint for a spherically symmetric spacetime given by the quantum corrected expression
\ba
{}^{\va R} H^{\va E}_{\rm eff}[N]
&=&\frac{1}{G}\int_\Sigma dr \frac{ N(r) }{\sqrt{((E^1)^2 +(E^2)^2  )E^r}}
\Bigg[\Bigg(
2E^r A_r\left(E^1 A_1+E^2A_2
\right)
+2E^r (E^1A'_2-E^2A'_1)\n\\
&+&((E^1)^2 +(E^2)^2)\left(
A_1^2+A^2_2-1\right)
\Bigg)\n\\
&+&
\epsilon_r\Bigg(
E^r A_r\left(A_r E^2 A_1-A_r E^1A_2+2E^1 A'_1 +2 E^2 A'_2)+E^r\left(E^1A''_2-E^2A''_1\right)
\right)
\Bigg)
\Bigg]\,.
\ea

\subsection{Lorentzian term}\la{sec:Lor-term}

The expectation value of the Lorentzian Hamiltonian constraint operator can be computed using its expression in terms of the densitized scalar curvature expressed as a function of the fluxes and their derivatives alone, as obtained in Eq. \eqref{RE}. Let us set $\epsilon_\theta=\epsilon_\varphi=\epsilon$ and quantize derivatives of the fluxes again in terms of discrete differences as defined in Eqs. \eqref{partialE} and \eqref{partial2E}. We have

\ba
&&\langle \widetilde{\psi^\lambda_{\va \square}}| {}^{\va R}\hat H^{\va L}_{\va \square}[N] |\widetilde{\psi^\lambda_{\va \square}}\rangle\approx
\frac{1}{\kappa}\left(1+\frac{1}{\gamma^2}\right)\frac{1}{\epsilon_r^3 \epsilon^8}  \frac{N(v)}{2 (E^r _3)^{5/2} \sqrt{\sin{\theta}} \ [(E^{\varphi} _1 )^2+(E^{\varphi} _2 )^2]^{3/2}} \n\\
&\times & \bigg[-2 (\sin {\theta})^2\epsilon^8 \epsilon_r^4 (E^{\varphi} _1)^4  (E^r )^2 \left[ (\sin{(\theta+\epsilon)}-\sin{ \theta})^2
- \sin{\theta} (\sin{(\theta+2\epsilon)}-2\sin{(\theta+\epsilon)}+\sin{\theta})
 \right]\n\\
 & -& 2 (\sin{\theta})^2 \epsilon^8 \epsilon_r^4  (E^{\varphi} _1)^3  E^2(E^r)^2 \left[(\sin{(\theta+\epsilon)}-\sin{ \theta})^2 
 + \sin{\theta} (\sin{(\theta+2\epsilon)}-2\sin{(\theta+\epsilon)}+\sin{\theta})\right] \n\\
&-& 2 (\sin{\theta})^2 \epsilon^8 \epsilon_r^2   E^{\varphi} _1  (E^r )^2 \Big( \epsilon_r^2 E^2 E^{\varphi} _2  \big[( E^{\varphi} _2    - 4   E^1 )    (\sin{(\theta+\epsilon)}-\sin{ \theta}) ^2
+    E^{\varphi} _2   \sin{\theta}  (\sin{(\theta+2\epsilon)}-2\sin{(\theta+\epsilon)}+\sin{\theta})\big] \n\\
&-& 2 \epsilon^2(\sin{\theta})^2 E^r  \left( E^2(r+\epsilon_r)-E^2(r)\right) \left( E^r(r+\epsilon_r) -E^r(r)\right) \Big) \n\\
&+& (E^{\varphi} _1)^2 (\sin{\theta})^2\epsilon^8 \epsilon_r^2 (E^r)^2 \Big( -4  \epsilon_r^2 (E^{\varphi} _2)^2 \big[(\sin{(\theta+\epsilon)}-\sin{ \theta})^2 - \sin{\theta}  (\sin{(\theta+2\epsilon)}-2\sin{(\theta+\epsilon)}+\sin{\theta})\big]\n\\
&+& 2 \epsilon_r^2 E^{\varphi} _2 E^1 \big[(\sin{(\theta+\epsilon)}-\sin{ \theta})^2 + \sin{\theta} (\sin{(\theta+2\epsilon)}-2\sin{(\theta+\epsilon)}+\sin{\theta})\big] \n\\
&+&  \big[4  \epsilon_r^2(E^1)^2(\sin{(\theta+\epsilon)}-\sin{ \theta})^2 + (\sin{\theta})^2  \epsilon^2 (E^r(r+\epsilon_r) -E^r(r))^2 
+  4 (\sin{\theta})^2  \epsilon^2 E^r(r)  \left( E^r(r+2\epsilon_r) -2E^r(r+\epsilon_r)+ E^r(r)\right) \big] \Big) \n\\
&+& \epsilon^8 \epsilon_r^2 E^{\varphi} _2  (\sin{\theta})^2(E^r)^2
\Big( -2  \ (E^{\varphi} _2)^3 \epsilon_r^2 \big[(\sin{(\theta+\epsilon)}-\sin{ \theta})^2 - \sin{\theta} (\sin{(\theta+2\epsilon)}-2\sin{(\theta+\epsilon)}+\sin{\theta})\big] \n\\
&+& 2  (E^{\varphi} _2)^2  \epsilon_r^2  E^1 \big[(\sin{(\theta+\epsilon)}-\sin{ \theta})^2 
+ \sin{\theta} (\sin{(\theta+2\epsilon)}-2\sin{(\theta+\epsilon)}+\sin{\theta})\big] \n\\
&-& 4 \epsilon^2 (\sin{\theta})^2E^r  (E^1(r+\epsilon_r)-E^1(r)) (E^r (r+\epsilon_r)-E^r (r))\n\\
&+&  E^{\varphi} _2 \big[4  \epsilon_r^2(E^2)^2 (\sin{(\theta+\epsilon)}-\sin{ \theta})^2
 +  \epsilon^2  (\sin{\theta})^2
\left[ (E^r (r+\epsilon_r)-E^r (r))^2 + 4 E^r \left( E^r(r+2\epsilon_r) -2E^r(r+\epsilon_r)+ E^r(r)\right)\right]\big]\Big) \bigg] \n\\
&=&\frac{1}{\kappa}\left(1+\frac{1}{\gamma^2}\right) \frac{N(v)}{2 \sqrt{E^r} [(E^1(r))^2+(E^2(r))^2]^{3/2} } \n\\
&\times & \Bigg[
 4  \epsilon_r \bigg[(E^2(r))^2+ (E^1(r))^2 \bigg]^2
     \left(\sin{(\theta+2\epsilon)}-2\sin{(\theta+\epsilon)}+\sin{\theta}\right)\n\\
&+& \frac{\epsilon^2}{\epsilon_r}\sin{\theta} \bigg[(E^1(r))^2+(E^2(r))^2\bigg]\Big(    \left(E^r(r+\epsilon_r) -E^r(r)\right)^2 
+  4  E^r(r)  \left( E^r(r+2\epsilon_r) -2E^r(r+\epsilon_r)+ E^r(r)\right) \Big) \n\\
&-& 4 \frac{\epsilon^2}{\epsilon_r } \sin{\theta} E^r (r)(E^r (r+\epsilon_r)-E^r (r)) \bigg[E^1(r) (E^1(r+\epsilon_r)-E^1(r)) 
+E^2(r) \left( E^2(r+\epsilon_r)-E^2(r)\right)\bigg]
\Bigg]\,.
\ea
By expanding the above expression for the expectation value up to the fourth order in $\epsilon$'s, we get
\ba
\langle \widetilde{\psi^\lambda_{\va \square}}| {}^{\va R}\hat H^{\va L}_{\va \square}[N] |\widetilde{\psi^\lambda_{\va \square}}\rangle
&\approx&\frac{1}{\kappa}\left(1+\frac{1}{\gamma^2}\right) \frac{ \epsilon_r\epsilon^2 N(v)}{2   \sqrt{E^r} [(E^1(r))^2+(E^2(r))^2]^{3/2} } \n\\
&\times & \Bigg[
 -4  \sin{\theta} \bigg[(E^2(r))^2+ (E^1(r))^2 \bigg]^2 \n\\
&+&\sin{\theta} \bigg[(E^1(r))^2+(E^2(r))^2\bigg]\bigg(  (E^{r'}(r))^2+4E^r(r)E^{r''}(r) \bigg) \n\\
&-& 4 \sin{\theta} E^r (r) E^{r'} (r)\bigg[E^1(r) E^{1'}(r)+E^2(r)E^{2'}(r)\bigg]\n\\
&-&4\epsilon  \cos{\theta}  \bigg[(E^2(r))^2+ (E^1(r))^2 \bigg]^2 \n\\
&+&\epsilon_r \sin{\theta} \Bigg[\bigg[(E^1(r))^2+(E^2(r))^2\bigg]\bigg(E^{r'}(r)E^{r''}(r)+4E^{r}(r)E^{r'''}(r)\bigg)\n\\
&-& 2  E^r (r)\bigg[E^1(r) \left(E^{r'}E^{1''}(r)+E^{r''}E^{1'}(r)\right)
+E^2(r) \left(E^{r'}E^{2''}(r)+E^{r''}E^{2'}(r)\right)
\bigg]\Bigg]
\Bigg]\la{HLcor-1}\\
&=&\frac{2}{\kappa}\left(1+\frac{1}{\gamma^2}\right)  \epsilon_r\epsilon^2 N(v)\n\\
&\times&\Bigg\{\sin{\theta} 
\Bigg[\Lambda(r)\left( \frac{(R'(r))^2}{\Lambda^2(r) }-1 \right)
+2   \frac{R(r)}{ \Lambda(r) }\left(  R''(r)-  \frac{\Lambda'(r) R'(r)}{ \Lambda(r) }\right)
 \Bigg]\n\\
 &+&
\epsilon_r  \frac{\sin{\theta}}{\Lambda(r)}   \Bigg[  
2R'(r) R''(r) + 2R(r)R'''(r)
- 3\frac{\Lambda'(r) (R'(r))^2}{\Lambda(r)}- \frac{\Lambda''(r) R(r)R'(r)}{\Lambda(r)} -2  \frac{\Lambda'(r)R(r)R''(r)}{\Lambda(r)}
\Bigg]\n\\
&-&
\epsilon  \cos{\theta} \Lambda(r)
\Bigg\}\,.\la{HLcor-2}
\ea
We see that the leading term in the last equality above reproduces the classical expression given in Eq. \eqref{H_L} for the Lorentzian part of the Hamiltonian constraint. The last two subleading terms correspond to quantum corrections. However, 
notice that the second correction term proportional to $\epsilon$ vanishes when the integral over $\theta$ is performed and only the correction proportional to $\epsilon_r$ remains. Equivalently, the classical and the correction terms expressions in terms of densitized triads can be read off of the first equality in Eq. \eqref{HLcor-1}.

\section{Concluding Remarks}

In this article we laid the foundations for a systematic treatment of spherically symmetric spacetimes in the framework of LQG. Applying the QRLG proposal, we implemented a quantization program that is aimed at identifying a symmetric sector at the quantum level, thus reverting the process of symmetry reduction and quantization that is frequently adopted in all existing treatments of quantum black holes.
The main result of this paper is the construction of an effective Hamiltonian that can now be used to evolve black hole initial data sets while incorporating quantum corrections.  To construct this Hamiltonian, we first built a convenient quantum gauge fixed kinematical Hilbert space that is compatible with a radial gauge even in the absence of symmetry. This was used to define coherent states where a notion of spherical symmetry could be imposed at the level of expectation values of geometrical operators. We then quantized the modified Hamiltonian constraint resulting from the gauge unfixing procedure as explained in Sec. \ref{sec:phase-space}. 
Finally we computed the effective Hamiltonian as the expectation value of the modified Hamiltonian operator on the coherent states that, if sharply peaked,  are the best candidates to describe classical geometries. 

The classical data entering the coherent states can now be seen as the initial data set to be evolved with the effective Hamiltonian.
The importance of our result lies in the fact that it is not tied to a particular choice of foliation, allowing one to treat on equal footing various sets of coordinate systems such as horizon penetrating coordinates or coordinates restricted to the interior or exterior of the event horizon of a black hole. This is a significant addition to the existing literature that mainly deals with either the interior or the exterior of event horizons.  In most of the previous treatments of this problem, one has been forced to use different Hilbert spaces for the interior and the exterior (as a result of the classical symmetry reduction process) which is normally plagued by ambiguities associated with gluing together interior and exterior geometries. \footnote{In the symmetry reduced phase space quantization scheme of  \cite{Gambini:2013ooa,Gambini:2013hna, Gambini:2014qta} one is still able to use the same kinematical Hilbert space for the solutions to the Hamiltonian constraint both in the exterior and the interior, in the sense that they have finite norm with respect to the same inner product. However, one of the quantum number characterizing the solutions changes from real to pure imaginary when going from the exterior to the interior. This implies that, effectively, one ends up treating the two regions separately and the structure of the complete solution at the horizon is not specified, leaving the gluing amibiguity. 
Let us also point out that the Hamiltonian that is quantized in \cite{Gambini:2013ooa,Gambini:2014qta}  corresponds to the correct equation of motion on-shell but it results in an algebra that is not equivalent to Dirac algebra restricted to the symmetric subspace. Here, in contrast, we deal with the original set of constraints. More generally, some issues associated with the covariance of the quantization scheme in the symmetry-reduced phase space approach were raised in \cite{Bojowald:2015zha} and  \cite{BenAchour:2017jof}, and their implications for the effective line element were investigated in \cite{BenAchour:2018khr} and \cite{Bojowald:2018xxu}. The question of covariance within our framework is an important aspect left for future investigations.}

We are now in a position to study the equations of motion generated by the effective Hamiltonian, derive the dressed metric that incorporates the quantum corrections and verify its compatibility with the existing results based on polymerlike quantization that is used in LQC or the proposed Planck star metric. In particular, it was shown in the cosmological context that scenarios different from a symmetric or asymmetric bounce are possible \footnote{Namely, the emergent bouncing Universe \cite{Alesci:2016xqa}.} and it is interesting to explore the consequences of this type of singularity resolution on the black hole-white hole scenario. The first results of this investigation have been presented in \cite{Alesci:2019pbs}, where the effective dynamics generated by the quantum corrected Hamiltonian derived here has been solved for the black interior geometry, by exploiting the simplification of adopting an homogeneous foliation inside. The analysis in \cite{Alesci:2019pbs}  shows how the white hole horizon picture is indeed replaced by an expanding Bianchi I Universe inside, once the corrections deriving from the full theory framework are actually taken into account.
Our final aim in the foreseeable future is to solve the quantum gravitational collapse problem in the presence of matter through the simplifications introduced by the QRLG approach, which facilitates the inclusion of extra fields \cite{Bilski:2015dra, Bilski:2016pib}.  

Another important application of our construction that is related to the issues of singularity resolution and the black hole information loss paradox is to illuminate the quantum nature of black hole entropy. This problem has  quite a long history in the LQG literature. The first ideas on how to microscopically describe the degrees of freedom accounting for the Bekenstein-Hawking entropy formula date back to the works \cite{Smolin:1995vq, Rovelli:1996dv, Krasnov:1996wc}. This approach was then refined within the framework of ``isolated horizons'' in
\cite{Ashtekar:1997yu, Ashtekar:1999wa, Ashtekar:2000eq} and generalized to the full gauge-invariant case in \cite{Engle:2009vc, Engle:2010kt, Perez:2010pq, Engle:2011vf, Frodden:2012en}. Despite the remarkable success of these results in recovering the entropy-area law from a quantum description of the horizon gravitational degrees of freedom, there are two open issues that still affect the LQG derivation of black hole entropy. The first concerns the role of the Barbero-Immirzi parameter in recovering the exact numerical coefficient  1/4 in the Bekenstein-Hawking entropy formula (see \cite{Jacobson:2007uj, Ghosh:2011fc, Frodden:2012dq, Ghosh:2013iwa, Pranzetti:2013lma, Bodendorfer:2013hla, Ghosh:2014rra, Pranzetti:2014tla, Achour:2014eqa, BenAchour:2016mnn} for an extensive debate on this topic). The second somehow unsatisfactory feature of the LQG black hole entropy calculation that is oftentimes simply glossed over is the {\it assumption} of the validity of the ``weak holographic principle'' \cite{Smolin:2000ag} leading to a horizon density matrix in which {\it both} the interior and the exterior of the black hole quantum geometry degrees of freedom are traced over.

Our construction has the potential to solve both issues, or at least to provide important insights about them. In fact, concerning the fixation of the Barbero-Immirzi parameter, it has recently been pointed out in the literature \cite{Ghosh:2013iwa, Oriti:2015rwa, Oriti:2018qty, Freidel:2016bxd, Eder:2018uzm} that new degrees of freedom should be included in the partition function in order to set $\gamma$ free from any numerical constraint. \footnote{A possible source of the ambiguity behind the role of $\gamma$ may also be related to the issue pointed out in \cite{Bodendorfer:2016pxg}, namely the inadequacy of the isolated horizon boundary condition usually implemented in the quantum theory to single out the notion of a horizon. The characterization of an isolated horizon through new degrees of freedom emerging from some particular boundary conditions, as well as through a maximal entropy principle, may settle this  ambiguity.} This is in addition  to the internal gauge degrees of freedom already accounted for in the standard calculation. These new degrees of freedom have been identified with either graph combinatorial structures or inclusion of matter (see, however, \cite{Freidel:2016bxd} for a possible unification of the two). Since our construction of a spherically symmetric  black hole quantum geometry derives from the full theory and does not rely on Chern-Simons techniques to model the horizon as a single intertwiner Hilbert space, new horizon graph degrees of freedom are automatically included in the horizon partition function. At the same time, inclusion of matter  can be implemented in a straightforward manner as pointed out above. This  provides the possibility to investigate the role of the Barbero-Immirzi parameter in the entropy calculation through a physically richer modelization of the horizon quantum geometry.

Concerning the validity of the weak holographic principle, i.e. the idea that the degrees of freedom relevant to the Bekenstein-Hawking entropy formula are only those lying at the horizon and in its vicinity, this is expected to be proven by the implementation of the quantum dynamics. More precisely, it is the solution of the Hamiltonian constraint, as well as the implementation of semiclassical consistency conditions that should introduce correlations between the horizon and the interior degrees of freedom. 
In fact, contrary to the AdS/CFT proposal, we expect the notion of holography to emerge only at the semiclassical level (see, e.g., \cite{Perez:2014xca} for a discussion of this point of view). 
An intriguing scenario would be the possibility to construct  physical solutions from the repeated action of the Hamiltonian constraint operator on a seed state, along the lines of the GFT condensates philosophy \cite{Oriti:2015rwa, Oriti:2018qty} but now with a concrete notion of the dynamics at hand. This could allow for the construction of a physical black hole interior density matrix given by a weighted sum over graphs with weights provided by matrix elements of the Hamiltonian constraint. In this picture then, a concrete notion of holography could be described and tested by understanding how dynamics is implemented as a refinement operation and by going to a continuum limit by means of coarse graining techniques, in the spirit of \cite{Livine:2006xk, Dittrich:2014ala}.
This is clearly a very ambitious and long-term plan that we leave for future investigations. However, all the necessary ingredients and tools are now at our disposal.



\section*{Acknowledgements}
We acknowledge the John Templeton Foundation
for the supporting Grant No.\#51876.
This work was supported in part by the NSF Grant No. PHY-1505411, the
Eberly research funds of Penn State, and the grant of Polish Narodowe Centrum Nauki No. DEC-2011/02/A/ST2/00300.
Research at the Perimeter Institute for Theoretical Physics is supported in part by the Government of Canada through NSERC and by the Province of Ontario through MRI.
\begin{appendix}

\section{Connection coefficients and curvature for geometries in spherical symmetry} \la{sec:A}

The components of an antisymmetric spin connection solution to the torsion-free equation $de^I=-\omega^I\!_J\wedge e^J$ can be written as
\be\la{sp-con}
\omega_{IJ}=\frac{1}{2}(c_{IJK}+c_{IKJ}-c_{JKI})e^K\,,
\ee
where $c_{IJK}$ are the structure functions. Using this one can read off the expression
\be\la{structure-constants}
de^I=-\frac{1}{2}c_{JK}\!^I e^J\wedge e^K\,.
\ee

From Eq. \eqref{tet0} we have
\baa
de^0&=&N' dr\wedge dt=\frac{N' }{N\Lambda}e^3\wedge e^0\,,\\
de^3&=&(\Lambda' N^r +\Lambda {N^r}')dr\wedge dt - \dot{\Lambda} dr \wedge dt
=\frac{(\Lambda' N^r +\Lambda {N^r}'- \dot{\Lambda} )}{N\Lambda}e^3\wedge e^0\,,\\
de^1&=&R'\cos{\alpha}\, dr\wedge d\theta-R'\sin{\theta}\sin{\alpha}\,dr\wedge d\varphi\n\\
&+&\dot R\cos{\alpha}\, dt\wedge d\theta-\dot R\sin{\theta}\sin{\alpha}\,dt\wedge d\varphi\n\\
&-&R\cos{\theta}\sin{\alpha}\,d\theta\wedge d\varphi
\n\\
&=&\frac{R'}{R\Lambda}e^3\wedge e^1+\frac{(\dot R-R' N^r)}{R N}e^0\wedge e^1-\frac{\cot{\theta}\sin{\alpha}}{R}e^1\wedge e^2\,,\\
de^2&=&R'\sin{\alpha} \,dr\wedge d\theta+R'\sin{\theta}\cos{\alpha} dr\wedge d\varphi\n\\
&+&\dot R\sin{\alpha} \,dt\wedge d\theta+\dot R\sin{\theta}\cos{\alpha} dt\wedge d\varphi\n\\
&+&R\cos{\theta}\cos{\alpha}\,d\theta\wedge d\varphi\n\\
&=&\frac{R'}{R\Lambda}e^3\wedge e^2+\frac{ (\dot R-R'N^r)}{R N}e^0\wedge e^2+\frac{\cot{\theta}\cos{\alpha}}{R}e^1\wedge e^2\,.
\eaa
Using Eq. \eqref{structure-constants}, the corresponding nonvanishing structure functions are
\baa
c_{30}\!^0&=&-c_{03}\!^0=-\frac{N' }{N\Lambda}\,\\
c_{30}\!^3&=&-c_{03}\!^3=-\frac{(\Lambda' N^r +\Lambda {N^r}'- \dot{\Lambda} )}{N\Lambda}\,\\
c_{01}\!^1&=&-c_{10}\!^1=\frac{R'N^r-\dot R}{R N}\,\\
c_{02}\!^2&=&-c_{20}\!^2=\frac{R'N^r-\dot R}{R N}\,\\
c_{31}\!^1&=&-c_{13}\!^1=-\frac{R'}{R\Lambda}\,\\
c_{12}\!^1&=&-c_{21}\!^1=\frac{\cot{\theta}\sin{\alpha}}{R}\,\\
c_{32}\!^2&=&-c_{23}\!^2=-\frac{R'}{R\Lambda}\,\\
c_{12}\!^2&=&-c_{21}\!^2=-\frac{\cot{\theta}\cos{\alpha}}{R}\,.
\eaa
We insert these in Eq. \eqref{sp-con} and find
\begin{subequations}
\ba\la{omega1}
\omega_{03}&=&-\frac{N' }{N\Lambda} e^0 +\frac{(\Lambda' N^r +\Lambda {N^r}'- \dot{\Lambda} )}{N\Lambda} e^3\n\\
&=& \Big (\frac{N^r }{N}(\Lambda' N^r +\Lambda {N^r}'- \dot{\Lambda} ) -\frac{N' }{\Lambda} \Big)dt
+\frac{(\Lambda' N^r +\Lambda {N^r}'- \dot{\Lambda} )}{N}dr \,,\la{omega2}\\
\omega^{01}&=&-\omega_{01}=\frac{(R'N^r-\dot R)}{RN} e^1
= \frac{(R'N^r-\dot R)}{N}\cos{\alpha} d\theta-\frac{(R'N^r-\dot R)}{N}\sin{\theta}\sin{\alpha}d\varphi\,,\la{omega3}\\
\omega^{02}&=&-\omega_{02}=\frac{(R'N^r-\dot R)}{RN} e^2
=\frac{(R'N^r-\dot R)}{N} \sin{\alpha} d\theta+\frac{(R'N^r-\dot R)}{N} \sin{\theta}\cos{\alpha} d\varphi\,,\la{omega4}\\
\omega^{12}&=&\frac{\cot{\theta}\sin{\alpha}}{R}e^1-\frac{\cot{\theta}\cos{\alpha}}{R}e^2=-\cos{\theta} d\varphi\,,\la{omega5}\\
\omega^{13}&=&\frac{R'}{R\Lambda}e^1= \frac{R'}{\Lambda}\left( \cos{\alpha} d\theta-\sin{\theta}\sin{\alpha}d\varphi\right)\,,\la{omega6}\\
\omega^{23}&=&\frac{R'}{R\Lambda}e^2= \frac{R'}{\Lambda}\left( \sin{\alpha} d\theta+\sin{\theta}\cos{\alpha} d\varphi\right)\,.\la{omega7}
\ea
\end{subequations}
For the connection coefficients $\Gamma^i_a=-\frac{1}{2}\epsilon^i\!_{jk} \omega^{jk}_a$, from the expressions above we obtain
\begin{subequations}
\ba
\Gamma^1_\varphi&=&-\sin\theta\, \Gamma^2_\theta=-\cos\alpha \sin\theta \frac{R'}{\Lambda}\,,\\
\Gamma^2_\varphi&=&\sin\theta\, \Gamma^1_\theta=-\sin\alpha \sin\theta \frac{R'}{\Lambda}\,,\\
\Gamma^3_\varphi&=&\cos\theta\,,\\
\Gamma^i_r&=&0\,,
\ea
\end{subequations}
from which we compute the intrinsic curvature components $ R^k_{ab}=2\partial_{[a} \Gamma^k_{b]}+\epsilon^k\!_{lm} \Gamma^l_a  \Gamma^m_b $, 
\begin{subequations}
\ba
R^1_{r\varphi}&=&-\sin\theta\, R^2_{r\theta}=\partial_r \Gamma^1_\varphi=-\cos\alpha \sin\theta \left(\frac{R''}{\Lambda}-\frac{R'\Lambda'}{\Lambda^2}\right)\,,\\
R^2_{r\varphi}&=&\sin\theta\, R^1_{r\theta}=\partial_r \Gamma^2_\varphi=-\sin\alpha \sin\theta \left(\frac{R''}{\Lambda}-\frac{R'\Lambda'}{\Lambda^2}\right)\,,\\
R^3_{\theta\varphi}&=&\partial_\theta \Gamma^3_\varphi+2 \ \Gamma^{[1}_\theta\Gamma^{2]}_\varphi=\sin\theta\left(\frac{(R')^2}{\Lambda^2}-1\right)\,\\
R^1_{\theta\varphi}&=&R^2_{\theta\varphi}=0\,,
\ea
\end{subequations}
and
\ba
R&=&-\epsilon_{ijk}R^k_{ab}e^a_i e^b_j=
4e^r_3\left(R^1_{r\theta} e^\theta_2-R^2_{r\theta} e^\theta_1\right)
+2R^3_{\theta\varphi} (e^\theta_2 e^\varphi_1-e^\theta_1 e^\varphi_2)\n\\
&=&\frac{4}{R\Lambda^2}\left(\frac{R'\Lambda'}{\Lambda}-R''\right)
+\frac{2}{R^2}\left(1-\frac{(R')^2}{\Lambda^2}\right)\,.\la{R}
\ea

For the components of the Ashtekar-Barbero connection $A^i_a=\Gamma^i_a+\gamma K^i_a$, namely
$
F^i_{ab}(A)= 2\partial_{[a}A^i_{b]}+\epsilon^i\!_{jk}A^j_{a}A^k_{b}
$,
we use Eq. \eqref{Sconn2} to get
\begin{subequations}\la{FA}
\ba
F^1_{r\theta }(A)&=&\partial_r A^1_\theta-A^2_\theta A^3_r
=A'_1-A_2 A_r\,,\\
F^1_{r\varphi }(A)&=&\partial_r A^1_\varphi-A^2_\varphi A^3_r
=-\sin\theta \left( A'_2+A_1 A_r\right)\,,\\
F^1_{\theta \varphi}(A)&=&\partial_\theta A^1_\varphi + A^2_\theta A^3_\varphi
=0\,,\\
F^2_{r\theta}(A)&=& \partial_r A^2_\theta+ A^1_\theta A^3_r
= A'_2+A_1 A_r\,,\\
F^2_{r\varphi}(A)&=&\partial_r A^2_\varphi+ A^1_\varphi A^3_r
=\sin\theta \left( A'_1-A_2 A_r\right)\,,\\
F^2_{\theta \varphi}(A)&=&\partial_\theta A^2_\varphi -A^1_\theta A^3_\varphi=0\,,\\
F^3_{r\theta}(A)&=&
F^3_{r\varphi}(A)=0\,,\\
F^3_{\theta \varphi}(A)&=&\partial_\theta A^3_\varphi + A^1_\theta A^2_\varphi-A^1_\varphi A^2_\theta=
\sin\theta \left[
A_1^2+A^2_2-1\right]\,.
\ea
\end{subequations}
\subsection{Densitized scalar curvature in terms of fluxes}
Here we express the densitized scalar curvature, $\sqrt{\text{det(E)}} R$, in terms of the fluxes and their
derivatives. The final result of this calculation appears
in Eq. \eqref{H_L}. 

The starting equation is 
\ba\la{denR1} 
\sqrt{\text{det(E)}} R &=& - \sqrt{\text{det}(E)} \epsilon_{ijk} R^k _{ab} e^a _i e^b _j \n\\
& = & - \frac{\epsilon_{ijk}}{\sqrt{\text{det}(E)} } E^a_i E^b_j \Big(2 \partial_{[a} \Gamma^k _{b]} + \epsilon ^k \  _{lm} \Gamma ^l _a \Gamma^m _b \Big).
\ea 
Note that by our gauge condition, the only nonvanishing fluxes
are 
\ba 
E^r _3, E^{\theta} _a, E^{\theta} _2, E^{\varphi} _1, E^{\varphi} _2.
\ea
We are also going to exploit some simplifications resulting from spherical symmetry. Let us emphasize that this is not going to undermine the generality of our quantum result when computing the expectation value of the Lorentzian Hamiltonian constraint operator. This is due to the fact that the coherent states constructed to implement the spherical symmetry enforce these simplifications in the final result for the expectation value. This is just a matter of convenience to avoid even lengthier expressions which in the end yield the same effective result. However, we are not going to use the explicit spherically symmetric expressions for the fluxes within terms containing nonvanishing derivatives (in the spherically symmetric case) since this would yield simplifications  which {\it a priori}, while preserving the semiclassical expression, could remove sources of quantum corrections at higher orders.

So first, as it was also previously shown due to spherical symmetry, the 
only nonzero $\Gamma^i _a$s are 
\ba  
\Gamma^1 _{\varphi}, \Gamma^2 _{\varphi}, \Gamma^3 _{\varphi}, \Gamma^1 _{\theta}, \Gamma^2 _{\theta}.
\ea
Additionally, due to spherical symmetry neither the fluxes
nor the spin connections  depend on the coordinate $\varphi$.
Taking advantage of these simplifications, Eq. \eqref{denR1} becomes
\ba\la{denR2} 
\sqrt{\text{det(E)}} R &=& - \frac{\epsilon_{ijk}}{\sqrt{\text{det}(E)} } E^a_i E^b_j \Big(2 \partial_{[a} \Gamma^k _{b]} + \epsilon ^k \  _{lm} \Gamma ^l _a \Gamma^m _b \Big) \n\\
& =& - \frac{2}{\sqrt{\text{det}(E)}} \bigg[E^a _1 E^b _2 \Big(2 \partial_{[a} \Gamma^3 _{b]} + \epsilon ^3 \  _{lm} \Gamma ^l _a \Gamma^m _b \Big) + E^a _2 E^b _3 \Big(2 \partial_{[a} \Gamma^1 _{b]} + \epsilon ^1 \  _{lm} \Gamma ^l _a \Gamma^m _b \Big) \n\\
& +&  E^a _3 E^b _1 \Big(2 \partial_{[a} \Gamma^2 _{b]} + \epsilon ^2 \  _{lm} \Gamma ^l _a \Gamma^m _b \Big) \bigg] \n\\
& =& - \frac{2}{\sqrt{\text{det}(E)}} \bigg[E^a _1 E^b _2 \Big(2 \partial_{[a} \Gamma^3 _{b]} +  \Gamma ^1 _a \Gamma^2 _b - \Gamma ^2 _a \Gamma^1 _b \Big) + E^a _2 E^b _3 \Big(2 \partial_{[a} \Gamma^1 _{b]} + \Gamma^2 _a \Gamma^3 _b    \n\\
& - &\Gamma^3 _a \Gamma^2 _b \Big) + E^a _3 E^b _1 \Big( 2 \partial_{[a} \Gamma^2 _{b]} + \Gamma^3 _a \Gamma^1 _b - \Gamma^1 _a \Gamma^3 _b \Big) \bigg] \n\\
& =& - \frac{2}{\sqrt{\text{det}(E)}} \bigg[(E^a _1 E^{\varphi} _2 - E^a _2 E^{\varphi} _1) \partial_a \Gamma ^3 _{\varphi} + E^a _1 E^b _2 (\Gamma^1 _a \Gamma^2 _b - \Gamma^2 _a \Gamma^1 _b) - E^a _2 E^r _3 \partial_{r} \Gamma^1 _{a}   \n\\
&  + & E^a _1 E^r _3 \partial_r \Gamma^2 _a \bigg] \n\\
& = & - \frac{2}{\sqrt{\text{det}(E)}} \bigg[2 E^{[\theta} _1 E^{\varphi]} _2  \partial_{\theta} \Gamma ^3 _{\varphi}
+ 4 E^{[\theta}_1 E^{\varphi]} _2 \Gamma^1 _{[\theta} \Gamma^2 _{\varphi]} -E^{\theta}_2 E^r _3 \partial_r \Gamma^1 _{\theta}-E^{\varphi}_2 E^r _3 \partial_r \Gamma^1 _{\varphi}\n\\
& + & E^{\theta}_1 E^r _3 \partial_r \Gamma^2 _{\theta} +  E^{\varphi}_1 E^r _3 \partial_r \Gamma^2 _{\varphi} \bigg].\n\\
\ea
For our purposes, we can further simplify Eq. \eqref{denR2}
by using the following relations which are due to  
spherical symmetry:
\ba 
&& E^{\theta}_1 = \sin{\theta} E^{\varphi} _2, \hspace{0.5cm} E^{\theta}_2 = - \sin{\theta} E^{\varphi} _1, \n\\
&&\Gamma^1 _{\varphi} = - \sin{\theta} \Gamma^2 _{\theta}, \hspace{0.5cm} \Gamma^2 _{\varphi} = \sin{\theta} \Gamma^1 _{\theta}.
\ea
We shall use these equations everywhere except for the fluxes that are acted on by $\partial_{\theta}$. Doing this, we find
\ba \la{denR3}
\sqrt{\text{det}(E)} R &=& - \frac{2}{\sqrt{\sin{\theta} \ E^r _3 \big[(E^{\varphi} _1) ^2  + (E^{\varphi} _2)^2\big]}} \bigg[\sin{\theta} \ \big[(E^{\varphi} _1)^2+(E^{\varphi} _2)^2\big] \partial_{\theta} \Gamma^3 _{\varphi} \n\\
&& + \sin^2{\theta} \ \big[(E^{\varphi} _1)^2+(E^{\varphi} _2)^2\big]
\big[(\Gamma^1 _{\theta})^2+(\Gamma^2 _{\theta} )^2\big] + 2\sin{\theta} E^r _3 E^{\varphi}_1 \partial_r \Gamma^1 _{\theta}\nonumber\\
&& + 2\sin{\theta}  E^{\varphi} _2 E^r _3 \partial_r \Gamma^2 _{\theta} \bigg] \nonumber\\
&&= - \frac{2}{\sqrt{\sin{\theta} \ E^r _3 \big[(E^{\varphi} _1) ^2  + (E^{\varphi} _2)^2\big]}} \bigg[\sin{\theta} \ \big[(E^{\varphi} _1)^2+(E^{\varphi} _2)^2\big] \Big( \partial_{\theta} \Gamma^3 _{\varphi} \nonumber\\
&& + \sin{\theta} \ 
\big[(\Gamma^1 _{\theta})^2+(\Gamma^2 _{\theta})^2\big]\Big) +2\sin{\theta} E^r _3  \big(E^{\varphi} _1 \partial_r \Gamma^1 _{\theta} + E^{\varphi} _2 \partial_r \Gamma^2 _{\theta} \big) \bigg].
\ea

We now eliminate $\Gamma^i_a$ in favor of $E^a_i$ and its derivatives. By definition we have 
\ba \la{gammadef}
\Gamma^{j}_{b} = - \frac{1}{2} \epsilon^{jik} E^{a}_{k} \bigg[E^{i}_{b,a}-E^{i}_{a,b} + E^c_{i} E^{l}_b E^{l}_{c,a} + E^{i}_b \frac{\text{det}(E)_{,a}}{\text{det}(E)} - \frac{1}{2} E^{i}_{a} \frac{\text{det}(E)_{,b}}{\text{det}(E)} \bigg],
\ea
with 
\ba \la{fluxinverse}
 E^i _{a} = \frac{1}{2 \text{det}(E)} \epsilon_{a bc} \epsilon^{i jk} E^b _{j} E^c _{k}. 
\ea
For $\Gamma^i_{\theta}$ and $\Gamma^3_{\varphi}$ we find
\begin{subequations}\la{gammanonvan}
\ba 
\Gamma^3 _{\varphi} &=& - \frac{1}{2} E^{\theta} _2 \bigg[E^1 _{\varphi, \theta}
+ E^{\theta}_1 (E^1 _{\varphi} E^1 _{\theta,\theta} + E^2 _{\varphi} E^2 _{\theta,\theta} ) + E^{\varphi}_1 (E^1 _{\varphi} E^1 _{\varphi,\theta} + E^2 _{\varphi} E^2 _{\varphi,\theta} )\n\\
&& + E^1 _{\varphi} \log{[\text{det}(E)]}_{,\theta} \bigg] + \frac{1}{2} E^{\theta} _1 \bigg[ E^2 _{\varphi, \theta} + E^{\theta} _2 (E^1 _{\varphi} E^1 _{\theta,\theta} + E^2 _{\varphi} E^2 _{\theta,\theta}) + E^{\varphi} _2 (E^1 _{\varphi} E^1 _{\varphi,\theta} \n\\
&&  + E^2 _{\varphi} E^2 _{\varphi,\theta}) +  E^2 _{\varphi} \log{[\text{det}(E)]}_{,\theta} \bigg] \n\\
&& = -\frac{1}{2 E^r _3 (E^{\theta} _2 E^{\varphi} _1 - E^{\theta} _1 E^{\varphi} _2)^2} \bigg[ -(E^{\theta}_1)^3 E^{\varphi} _2 E^r _{3,\theta}
+(E^{\theta} _1)^2 \Big( E^{\theta} _2 E^{\varphi} _1 E^r _{3,\theta}-E^r _3 \n\\
&& \times [E^{\varphi} _2 E^{\theta} _{1,\theta} +E^{\varphi} _1 E^{\theta} _{2, \theta}] \Big) + (E^{\theta}_2)^2 \Big(E^{\theta} _2 E^{\varphi} _1 E^r _{3,\theta} + E^r _3 [E^{\varphi}_2 E^{\theta} _{1,\theta} + E^{\varphi} _1 E^{\theta} _{2,\theta} ] \Big) \n\\
&& - E^{\theta} _1 E^{\theta} _2 \Big(E^{\theta} _2 E^{\varphi} _2 E^{r} _{3,\theta} + 2 E^r _3 [-E^{\varphi} _1 E^{\theta} _{1,\theta} + E^{\varphi} _2 E^{\theta} _{2, \theta}] \Big) \bigg] \n\\
&=& - \frac{1}{2 (E^r _3)^2 [(E^{\varphi} _1)^2 +(E^{\varphi} _2)^2]^2} \bigg[-\frac{2(E^r _3)^2}{\sin{\theta}} \Big(E^{\varphi} _1  E^{\theta} _{1,\theta} + E^{\varphi} _2  E^{\theta} _{2,\theta}\Big)^2 + \sin{\theta} \ [(E^{\varphi} _1)^2 \n\\
&&+(E^{\varphi} _2)^2]^2  \ [( E^r _{3,\theta})^2 - E^r _3 E^r _{3,\theta \theta}] + E^r _3 [(E^{\varphi} _1)^2+(E^{\varphi} _2)^2]  \Big(-E^{\varphi} _2 [ E^r _{3,\theta} E^{\theta} _{1,\theta} + E^r _3  E^{\theta} _{1,\theta \theta}] \n\\ 
&&+ E^{\varphi} _1 [ E^r _{3,\theta}  E^{\theta} _{2,\theta} + E^r _3 E^{\theta} _{2,\theta \theta}] \Big) \bigg], \\
&& \n\\
&&\n\\
\Gamma^1 _{\theta} &=& - \frac{1}{2}E^r _3 \bigg[ E^2 _{\theta,r}  + E^{\theta} _2 (E^1 _{\theta}  E^1 _{\theta,r} + E^2 _{\theta}  E^2 _{\theta,r}) + E^{\varphi} _2 (E^1 _{\theta}  E^1 _{\varphi,r} + E^2 _{\theta}  E^2 _{\varphi,r}) \n\\
&&+ E^2 _{\theta} \log{[\text{det}(E)]_{,r}} \bigg] \n\\
&=&- \frac{1}{2 (E^{\theta} _2 E^{\varphi} _1 - E^{\theta} _1 E^{\varphi} _2)^2} \bigg[E^{\varphi} _1 E^{r} _{3,r} (E^{\theta} _2 E^{\varphi} _1 - E^{\theta} _1 E^{\varphi} _2)  + E^r _3 \Big(E^{\theta} _1 [-E^{\varphi} _2 E^{\varphi} _{1,r} \n\\
&& + E^{\varphi} _{1} E^{\varphi} _{2,r}] + E^{\theta} _2 [E^{\varphi} _1 E^{\varphi} _{1,r} + E^{\varphi} _2 E^{\varphi} _{2,r}] - E^{\theta} _{2,r} [(E^{\varphi} _1 )^2 +(E^{\varphi} _2)^2]\Big) \bigg]\n\\
&=&\frac{E^{\varphi} _1 E^r _{3,r}}{2 \sin{\theta} \big[(E^{\varphi} _1)^2 + (E^{\varphi} _2)^2\big]} \\
&& \n\\
&&\n\\
\Gamma^2 _{\theta} &=&  \frac{1}{2} E^r _3 \bigg[ E^1 _{\theta,r}  + E^{\theta} _1 (E^1 _{\theta}  E^1 _{\theta,r} + E^2 _{\theta}  E^2 _{\theta,r}) + E^{\varphi} _1 (E^1 _{\theta}  E^1 _{\varphi,r} + E^2 _{\theta}  E^2 _{\varphi,r}) \n\\
&& + E^1 _{\theta} \log{[\text{det}(E)]_{,r}} \bigg] \n\\
&&= - \frac{1}{2 (E^{\theta} _2 E^{\varphi} _1 - E^{\theta} _1 E^{\varphi} _2)^2} \bigg[E^{\varphi} _2 E^{r} _{3,r} (E^{\theta} _2 E^{\varphi} _1 - E^{\theta} _1 E^{\varphi} _2)  + E^r _3 \Big(E^{\theta} _2 [-E^{\varphi} _2 E^{\varphi} _{1,r} \n\\
&& + E^{\varphi} _{1} E^{\varphi} _{2,r}] - E^{\theta} _1 [E^{\varphi} _1 E^{\varphi} _{1,r} + E^{\varphi} _2 E^{\varphi} _{2,r}]  + E^{\theta} _{1,r} [(E^{\varphi} _1 )^2 +(E^{\varphi} _2)^2]\Big) \bigg] \n\\
&=& \frac{E^{\varphi} _2 E^r _{3,r}}{2 \sin{\theta} \big[(E^{\varphi} _1)^2 + (E^{\varphi} _2)^2\big]}.
\ea
\end{subequations}
In Eq. \eqref{gammanonvan} we used $\partial_{\theta}E^{\varphi} _i =0$ as suggested by spherical symmetry. Using the above, Eq. \eqref{denR3} reduces to
\ba 
\sqrt{\text{det}(E)} R &=& -\frac{1}{2 (E^r _3)^{5/2} \sin^{1/2}{\theta} \ [(E^{\varphi} _1 )^2+(E^{\varphi} _2 )^2]^{3/2}}  \bigg[-2 \sin^2 {\theta} (E^{\varphi} _1)^4 \Big[(\partial_{\theta} E^r _3)^2 - E^r _3 \partial^2 _{\theta} E^r _3 \Big] \n\\
& -& 2 \sin{\theta} E^r _3 (E^{\varphi} _1)^3 [\partial_{\theta} E^r _3 \partial_{\theta} E^{\theta} _2 + E^r _3 \partial^2 _{\theta} E^{\theta} _2] - 2 E^r _3 E^{\varphi} _1 \Big(E^{\varphi} _2  \big[(\sin{\theta} \ E^{\varphi} _2 \partial_{\theta} E^r _3 - 4 E^r _3 \partial_{\theta} E^{\theta} _1) \partial_{\theta} E^{\theta} _2 \n\\
& +& \sin{\theta} \ E^r _3 E^{\varphi} _2 \partial^2_{\theta} E^{\theta} _2\big] +  2 (E^r _3)^2 \partial_r E^{\varphi} _1 \partial_r E^r _3 \Big) + (E^{\varphi} _1)^2 \Big( -4 \sin^2{\theta} (E^{\varphi} _2)^2 \big[(\partial_{\theta} E^r _3)^2 - E^r _3 \partial^2 _{\theta} E^r _3\big]\n\\
&+& 2\sin{\theta} E^r _3 E^{\varphi} _2 \big[\partial_{\theta} E^r _3 \partial_{\theta} E^{\theta} _1 + E^r _3 \partial^2 _{\theta} E^{\theta} _1\big] + (E^r _3)^2 \big[4 (\partial_{\theta} E^{\theta} _1)^2 + (\partial_{r}E^r _3)^2 +   4 E^r _3 \partial^2 _r E^r _3 \big] \Big) + E^{\varphi} _2 \Big( -2 \sin^2 {\theta} \n\\
& \times& (E^{\varphi} _2)^3 \big[(\partial_{\theta} E^r _3)^2 - E^r _3 \partial^2 _{\theta} E^r _3\big] + 2 \sin{\theta}  E^r _3 (E^{\varphi} _2)^2 \big[\partial_{\theta} E^r _3 \partial_{\theta} E^{\theta} _1 + E^r _3 \partial^2_{\theta} E^{\theta} _1\big] - 4 (E^r _3)^3 \partial_r E^{\varphi} _2 \partial_r E^r _3 + (E^r _3)^2 E^{\varphi} _2 \n\\
& \times &  \big[4 (\partial_{\theta} E^{\theta} _2)^2 + (\partial_r E^r _3)^2 + 4 E^r_3  \partial^2 _r E^r _3\big]\Big) \bigg]. \la{RE}
\ea

\section{Coherent states}\la{sec:B}

 The derivation of Eq. \eqref{psir} is straightforward given that only $\tau_3$ appears. Let us then show how we arrive at Eqs. \eqref{psit}  and \eqref{psiv} by providing a few more details. First of all, given the property $u_\ell^{-1} \tau_\ell u_\ell=\tau_3$ and the relations given in Eq. \eqref{uzxy}, we have
 \baa
 \tau_x&=&\tau_1\cos{\tilde\alpha}+\tau_2\sin{\tilde\alpha}\,,\\
  \tau_y&=&-\tau_1\sin{\tilde\alpha}+\tau_2\cos{\tilde\alpha}\,,\\
 \tau_1&=&\tau_x\cos{\tilde\alpha}-\tau_y\sin{\tilde\alpha}\,,\\
  \tau_2&=&\tau_x\sin{\tilde\alpha}+\tau_y\cos{\tilde\alpha}\,,\\
 \eaa
from which
\ba
E^1\tau_1+E^2\tau_2&=&E\tau_x\,,\\
E^1\tau_2-E^2\tau_1&=&E\tau_y\,,
\ea
with $E=\Lambda R$. Notice that we have expressed the relation between the internal directions (1,2) and $(x,y)$ in terms of the angle $\tilde \alpha$ since we are interested in the classical solution for the triad and connection to define the $SL(2,\C)$ group elements around which the semiclassical states are peaked. 
It follows that for the coherent state along the $\theta$-direction we have 
\ba
 {}^x\!D^{j_x}_{\bar n_x \bar m_x}\left(e^{\epsilon_\theta(A_1\tau_1 + A_2\tau_2)} e^{\frac{\lambda \delta^2_\theta}{\kappa\gamma }(E^1\tau_1+E^2\tau_2)\sin{\theta}}\right)
&=& D^{j_x}_{\bar n_x \bar m_x}\!\left(u_x^{-1}e^{\epsilon_\theta(A_1\tau_1 + A_2\tau_2)} e^{\frac{\lambda \delta^2_\theta}{\kappa\gamma }E\tau_x\sin{\theta}}u_x\right)\n\\
&=&e^{\lambda \bar m_x\frac{  \delta^2_\theta E^x}{\kappa\gamma }} \, {}^x\!D^{j_x}_{\bar n_x \bar m_x}\!\left(e^{\epsilon_\theta(A_1 \tau_1  + A_2\tau_2 )}\right)
\ea
where $E^x=E\sin{\theta}$.

Similarly, for the coherent state along the $\varphi$-direction we have
\ba
&&{}^y\!D^{j_y}_{\bar n_y \bar m_y}\left(e^{\epsilon_\varphi[(A_1\tau_2 -A_2\tau_1)\sin{\theta}+\cos{\theta}\tau_3]} e^{\frac{\lambda \delta^2_\varphi}{\kappa\gamma }(E^1\tau_2-E^2\tau_1)}\right)\n\\
&&=e^{\lambda \bar m_y \frac{ \delta^2_\varphi E^y}{\kappa\gamma }}
\,{}^y\!D^{j_y}_{\bar n_y \bar m_y}\left(e^{\epsilon_\varphi[(A_1\tau_2 -A_2\tau_1)\sin{\theta}+\cos{\theta}\tau_3]} 
 \right)\,,\la{su2y}
\ea
where $E^y=E$.

\section{Approximating the Lorentzian Hamiltonian via techniques of Regge calculus }\la{sec:C}

In this appendix we provide an alternative method for quantizing the Lorentzian part of the Hamiltonian constraint. 

As we noted in Sec. \ref{sec:phase-space}, the Lorentzian term can be written either in terms of the extrinsic curvature or in terms of the 3D Ricci scalar  as we do so in Eq. \eqref{LHR}. In \cite{Alesci:2014aza} a proposal  was introduced aiming at providing an alternative approach to Thiemann's construction \cite{Thiemann:1996aw} for quantizing the Lorentzian term. The advantage of this approach is that it is computationally straightforward. Nonetheless, this approach is inherently ``perturbative'' as will become clear below. In situations where nonperturbative quantum gravity effects are likely to be influential (e.g. black hole singularity resolution), this approach may fall short of providing both the correct qualitative and quantitative pictures. Nevertheless, we now briefly review the regularization scheme of \cite{Alesci:2014aza} and then describe how the Lorentzian part of the Hamiltonian constraint can be quantized in our framework, using this approach.

The main idea behind the construction of \cite{Alesci:2014aza} is to regularize the integral of the Ricci scalar over $\Sigma_t$ by means of Regge calculus \cite{Regge:1961px}, i.e. in terms of lengths and angles of the triangulation. More precisely, assuming that curvature lies only on the hinges $h$ of the simplicial decomposition $\Delta$ of the 3D manifold $\Sigma$, we have the simplicial approximation
\be \la{regge1}
\frac{1}{2}\int_\Delta \sqrt{\text{det}(E)} R=\sum_s\sum_{h\in s} L_h^s\left(\frac{2\pi}{\alpha_h}-\theta_h^s  \right)\,,
\ee
where the first sum is over the simplices $s$ of $\Delta$ and the second one is over the hinges in the given simplex. The geometrical quantity $L_h^s$ represents the length of the hinge $h$ in the simplex $s$, $\theta_h^s$ is the dihedral angle at the hinge $h$, and $\alpha_h$ is the number of simplices sharing the hinge $h$. The continuum limit can be obtained by sending the typical length of the lattice to zero and the construction can be straightforwardly generalized to nonsimplicial decompositions, as long as the hinges are straight lines. \footnote{In particular to the case of a cuboidal triangulation which we are  interested in.}

\begin{figure}[h]
\centerline{ \(
\begin{array}{c}
\includegraphics[height=4cm]{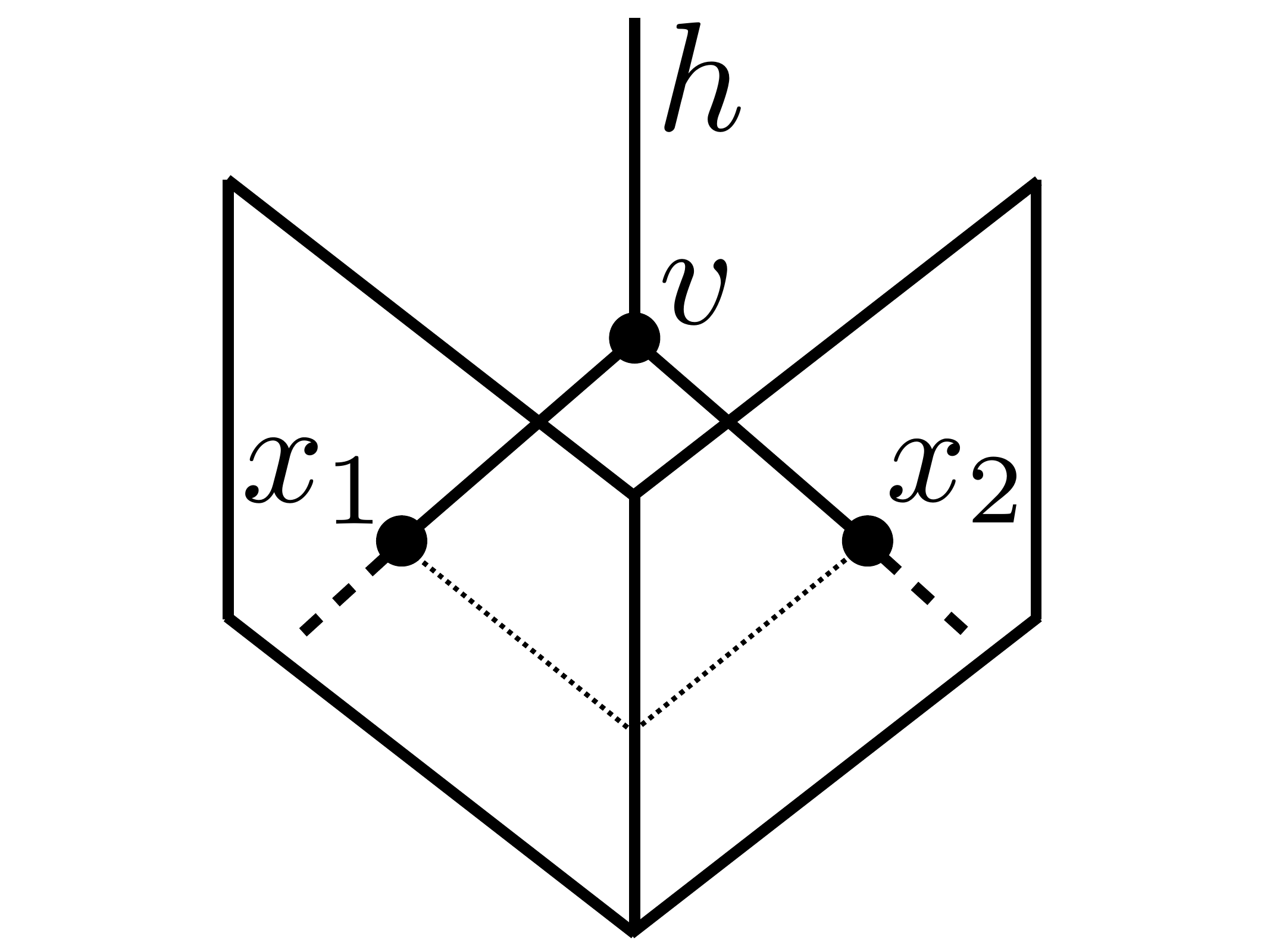}
\end{array}\) } \caption{Intersection of the graph with the dual surfaces  where the fluxes are evaluated in order to compute the dihedral angle around the hinge $h$.}
\label{directions}
\end{figure}

Let us focus on a 3-valent vertex $v$ of our cuboidal decomposition. The three edges in the directions $r,\theta,\varphi$  emanating from the vertex $v$ represent the three hinges on which the curvature is concentrated (see Fig. \ref{directions}). The lengths of these three hinges are given respectively by
\begin{subequations}\la{lengths}
\ba
L_r&=& \frac{\sqrt{
\epsilon_i\!^{jk} E_j(S^\theta) E_k(S^\varphi)
\epsilon^{ilm} E_l(S^\theta) E_m(S^\varphi)
}}{ V(v)}
= \frac{|E(S^\theta)||E(S^\varphi)|}{ V(v)}
\,,\\
L_\theta&=& \frac{\sqrt{
\epsilon_i\!^{jk} E_j(S^r) E_k(S^\varphi)
\epsilon^{ilm} E_l(S^r) E_m(S^\varphi)
}}{ V(v)}
= \frac{|E(S^r)||E(S^\varphi)|}{ V(v)}
\,,\\
L_\varphi&=& \frac{\sqrt{
\epsilon_i\!^{jk} E_j(S^r) E_k(S^\theta)
\epsilon^{ilm} E_l(S^r) E_m(S^\theta)
}}{ V(v)}
= \frac{|E(S^r)||E(S^\theta)|}{ V(v)}
\,.
\ea
\end{subequations}
Here the flux $E_i (S^a)$ is defined in Eq. \eqref{E}. The corresponding dihedral angles are
\begin{subequations}\la{angles}
\ba
\theta_r&=& \pi-\arccos{\left[\frac{\delta^{ij}E_i(S^\theta[x_1])
E_j(S^\varphi[x_2])}{|E(S^\theta)||E(S^\varphi)|}
\right]}
\,,\\
\theta_\theta&=& \pi-\arccos{\left[\frac{\delta^{ij}E_i(S^r[x_1])
E_j(S^\varphi[x_2])}{|E(S^r)||E(S^\varphi)|}
\right]}
\,,\\
\theta_\varphi&=& \pi-\arccos{\left[\frac{\delta^{ij}E_i(S^r[x_1])
E_j(S^\theta[x_2])}{|E(S^r)||E(S^\theta)|}
\right]}
\,,\\ \nonumber
\ea
\end{subequations}
where $|E(S^a)|=\sqrt{\delta^{ij}E_i(S^a)E_j(S^a)}$. We promote the 
right-hand side of Eq. \eqref{regge1} to an operator by replacing the classical length and angle variables by their quantum counterparts. 
Now the expectation value of the quantum version of the right-hand side of Eq. \eqref{regge1} on the semiclassical states is simply
given by the classical expression described below.  

To compute the right-hand side of Eq. \eqref{regge1}, we need the following expansion in terms of the holonomies of spin connections
\be \la{holonomy1}
E_i(S^a[x_1])E_j(S^b[x_2])=E_i(S^a[v])\big[\delta^l_j-\epsilon^a\epsilon^b\epsilon_{jk}\!^lR^k_{ab}(v)\big]E_l(S^b[v]) + o(\epsilon^3)\,.
\ee
On the left-hand side of the expression above, the two fluxes that read the dihedral angle around the hinge $h$ are not computed at the same point. They intersect the dual links (the edges of the graph) away from the vertex $v$, at two points $x_1$ and $ x_2$. On the right-hand side we have expressed their product in terms of the fluxes evaluated at the same point $v$ times the parallel transport through  holonomies of the intrinsic curvature $\Gamma^i_a$ from $x_1$ and $x_2$ to $v$. Such parallel transport can be written as a Wilson loop on the plane dual to the hinge and thus expressed in terms of the curvature of $\Gamma^i_a$.

It is immediate to see that the zeroth order in $\epsilon$ inside the $\arccos$ function vanishes, since the fluxes are orthogonal, and thus the leading order gives a $o(\epsilon^2)$ term. 

The contribution of a single 3-valent vertex to the integral of the Ricci scalar is hence given by (considering that each hinge is shared by four cubes)
\ba
\sum_{a=r,\theta,\varphi}L_a\left(\frac{\pi}{2}-\theta_a\right)&=&
L_r\frac{\left(E_2(S^\theta )E_1(S^\varphi)-E_1(S^\theta )E_2(S^\varphi)\right)R^3_{\theta\varphi}\epsilon_\theta\epsilon_\varphi}{|E(S^\theta)||E(S^\varphi)|}\n\\
&+&L_\theta\frac{E_3(S^r ) \left( E_2(S^\varphi)R^1_{r\varphi}-E_1(S^\varphi) R^2_{r\varphi}\right)\epsilon_r \epsilon_\varphi}{|E(S^r)||E(S^\varphi)|}\n\\
&+&L_\varphi\frac{E_3(S^r ) \left( E_2(S^\theta)R^1_{r\theta}-E_1(S^\theta) R^2_{r\theta}\right)\epsilon_r \epsilon_\theta}{|E(S^r)||E(S^\theta)|}\,.
\ea
By means of Eqs. \eqref{lengths} and \eqref{angles}, and the relation $E^a_i=\sqrt{\text{det}(E)} e^a_i$, it is straightforward to check that
\ba
\sum_{a=r,\theta,\varphi}L_a\left(\frac{\pi}{2}-\theta_a\right)
&=&
\epsilon_r\epsilon_\theta\epsilon_\varphi \sqrt{\text{det}(E)}
\Big[\left(e_2^\theta e_1^\varphi-e_1^\theta e_2^\varphi\right)R^3_{\theta\varphi}
+e_3^r  \left( e_2^\varphi R^1_{r\varphi}-e_1^\varphi R^2_{r\varphi}\right)
+e_3^r  \left( e_2^\theta R^1_{r\theta}-e_1^\theta R^2_{r\theta}\right)
\Big]\n\\
&=&
\epsilon_r\epsilon_\theta\epsilon_\varphi \sqrt{\text{det}(E)}R
\,,
\ea
as expected.

\subsection{Higher order holonomy correction to the triangulation formula for total curvature}

Equation \eqref{holonomy1} contains the first nontrivial correction due to the curvature. Here we derive the next order correction. 

From the derivation of \cite{Vines:2014uoa}, we have 
\ba \label{holonomy}
&& E_i (S^a[x_1]) E_j (S^b[x_2]) = E_i (S^a[v]) \Big[\delta^l \ _{j}  + R^l \ _{j ab} \epsilon^a \epsilon^b + \frac{1}{2} R^l \ _{j ab ; a} \epsilon^{a 2} \epsilon^b \nonumber\\
&&+ \frac{1}{2} R^l \ _{j ab ; b} \epsilon^{a} \epsilon^{b 2} \Big] E_l (S^b[v]) + o(\epsilon^4).
\ea
For the length-angle formula, the first term in the square
bracket does not contribute. The contribution of the second term was worked out above. Here we focus on the third and fourth terms. 

Note that since total curvature is invariant under arbitrary spatial diffeomorphisms, the final result of this calculation cannot depend on the internal angle $\alpha$. Therefore, for simplicity and to reduce the number of expressions we set $\alpha = 0$ at the vertex $v$. This way $e_1 ^{\varphi} = e_2 ^{\theta} = 0$ at $v$. The nonvanishing holonomy corrections are 
\ba \label{holterms3}
&&E_1(S^{\theta}[x_1]) E_1(S^{\varphi}[x_2]) \Big|_{\epsilon ^3} = \frac{\epsilon^{\theta} \epsilon^{\varphi}}{2} E_1(S^{\theta})E_2 (S^{\varphi}) \Big[R^2 \ _{1 \theta \varphi ;\theta} \epsilon^{\theta} +R^2 \ _{1 \theta \varphi ;\varphi} \epsilon^{\varphi}\Big], \nonumber\\
&&E_3(S^{r}[x_1]) E_3(S^{\varphi}[x_2]) \Big|_{\epsilon ^3} =\frac{\epsilon^{r} \epsilon^{\varphi}}{2} E_3(S^{r})E_2 (S^{\varphi}) \Big[R^2 \ _{3 r \varphi ;r} \epsilon^{r} +R^2 \ _{3 r \varphi ;\varphi} \epsilon^{\varphi}\Big], \nonumber\\
&&E_3(S^{r}[x_1]) E_3(S^{\theta}[x_2]) \Big|_{\epsilon ^3} =\frac{\epsilon^{r} \epsilon^{\theta}}{2} E_3(S^{r})E_1 (S^{\theta}) \Big[R^1 \ _{3 r \theta ;r} \epsilon^{r} +R^1 \ _{3 r \theta ;\theta} \epsilon^{\theta}\Big]. \nonumber\\
\ea 
The correction to the length-angle term becomes
\ba \label{defangcor}
\sum_{a = r, \theta, \varphi} L_{a} \Big(\frac{\pi}{2} - \theta_a \Big) \Big|_{\epsilon^4} &&= - \frac{\epsilon^r \epsilon^{\theta} \epsilon^{\varphi}}{2} \sqrt{\text{det}(E)} \bigg[ e_1 ^{\theta} e_2 ^{\varphi}(R^2 \ _{1 \theta \varphi ;\theta} \epsilon^{\theta} +R^2 \ _{1 \theta \varphi ;\varphi} \epsilon^{\varphi}) \nonumber\\
&& + e_3 ^r e_2 ^{\varphi} (R^2 \ _{3 r \varphi ;r} \epsilon^{r} +R^2 \ _{3 r \varphi ;\varphi} \epsilon^{\varphi}) + e_3 ^r e_1 ^{\theta} (R^1 \ _{3 r \theta ;r} \epsilon^{r} +R^1 \ _{3 r \theta ;\theta} \epsilon^{\theta}) \bigg] \nonumber\\
&& = - \frac{\epsilon^r \epsilon^{\theta} \epsilon^{\varphi}}{2} \sqrt{\text{det}(E)} \bigg[ \epsilon^{\theta} \Big (e_1 ^{\theta} e_2 ^{\varphi} R^2 \ _{1 \theta \varphi ;\theta} + e_1 ^{\theta} e_3 ^r R^1 \ _{3 r \theta ;\theta} \Big) \nonumber\\
&&+ \epsilon^{\varphi} \Big (e_2 ^{\varphi} e_1 ^{\theta} R^2 \ _{1 \theta \varphi ;\varphi} + e_2 ^{\varphi} e_3 ^r R^2 \ _{3 r \varphi ;\varphi} \Big)
+ \epsilon^r \Big(e_3 ^r e_1 ^{\theta} R^1 \ _{3 r \theta ;r} + e_3 ^r e_2 ^{\varphi} R^2 \ _{3 r \varphi ;r}\Big) \bigg] \nonumber\\
&&  = - \frac{\epsilon^r \epsilon^{\theta} \epsilon^{\varphi}}{2} \sqrt{\text{det}(E)} \bigg[ \epsilon^{\theta} \Big (R^2 \ _{112 ; \theta} + R^1 \ _{331;\theta} - R^2 \ _{11\varphi} e_{2;\theta} ^{\varphi}
-R^2 \ _{1 \theta 2} e_{1;\theta} ^{\theta} \nonumber\\
&&-R^1 \ _{3r1} e_{3;\theta}^r - R^1 \ _{33\theta} e_{1;\theta}^{\theta}\Big)+ \epsilon^{\varphi} \Big(R^2 \ _{112;\varphi} + R^2 \ _{332;\varphi} - R^2 \ _{11\varphi} e_{2;\varphi}^{\varphi} \nonumber\\
&& - R^2 \ _{1 \theta 2} e_{1;\varphi} ^{\theta} - R^2 \ _{33\varphi} e_{2;\varphi}^{\varphi} - R^2 \ _{3r2} e_{3;\varphi}^{r}\Big)
+ \epsilon^r \Big(R^1 \ _{331;r} + R^{2} \ _{332;r} \nonumber\\
&& - R^1 \ _{3r1}e_{3;r}^r - R^1 \ _{33\theta}e_{1;r}^{\theta}
- R^2 \ _{3r2} e_{3;r}^r - R^2 \ _{33\varphi} e_{2;r}^{\varphi}\Big)\bigg]. \nonumber\\
\ea
We can simplify the above expression by noting that 
\ba 
&& e_{2;\theta}^{\varphi} = e_{2,\theta}^{\varphi} + \Gamma^{\varphi}_{\varphi \theta} e_2 ^{\varphi} = -\frac{1}{R \sin{\theta}} \cot{\theta} + \frac{1}{R \sin{\theta}} \cot{\theta} = 0, \nonumber\\
&& e_{1;\theta}^{\theta} = e_{3;\theta}^r = e_{2;\varphi}^{\varphi}
= e_{1;\varphi}^{\theta} = e_{3;\varphi}^r = 0, \nonumber\\
&&e_{3;r}^r = e_{3,r}^r + \Gamma^{r} _{rr} e_3 ^r = -\frac{\Lambda'}{\Lambda 2} + \frac{\Lambda'}{\Lambda 2} = 0, \nonumber\\
&& e_{1;r}^{\theta} = e_{1,r}^{\theta} + \Gamma^{\theta} _{\theta r} e_1 ^{\theta } = - \frac{R'}{R^2} + \frac{R'}{R^2}  = 0, \nonumber\\
&&e_{2;r}^{\varphi} = e_{2,r}^{\varphi} + \Gamma^{\varphi}_{\varphi r} e^{\varphi}_2 = - \frac{R'}{R^2 \sin{\theta}} + \frac{R'}{R^2 \sin{\theta}} = 0.
\ea
Thus, \eqref{defangcor} becomes 
\ba \label{defangcor1}
\sum_{a = r, \theta, \varphi} L_{a} \Big(\frac{\pi}{2} - \theta_a \Big) \Big|_{\epsilon^4} &&= - \frac{\epsilon^r \epsilon^{\theta} \epsilon^{\varphi}}{2} \sqrt{\text{det}(E)} \bigg[\epsilon^{\theta}\Big(R^2 \ _{112;\theta} + R^1 \ _{331;\theta}\Big) \nonumber\\
&& + \epsilon^{\varphi} \Big(R^2 \ _{112;\varphi} + R^2 \ _{332;\varphi} \Big)
+ \epsilon^r \Big(R^1 \ _{331;r} + R^2 \ _{332;r} \Big) \bigg].
\ea
It follows from symmetries of the Riemann tensor and the definition of the internal metric that 
\ba 
&&R^2 \ _{112} = R_{2112} = - R_{1212}, \ \ R^1 \ _{331} = R_{1331} = - R_{1313}, \nonumber\\
&&R^2 \ _{332} = R_{2332} = -R_{2323}, \nonumber\\
&& R_{1212} + R_{1313} = R_{11}, \ \ R_{1212} + R_{3232} = R_{22}, \nonumber\\
&& R_{1313} + R_{2323} = R_{33}.
\ea
Therefore, \eqref{defangcor1}  reduces to 
\ba \label{defangcor2} 
&&\sum_{a = r, \theta, \varphi} L_{a} \Big(\frac{\pi}{2} - \theta_a \Big) \Big|_{\epsilon^4} =\frac{\epsilon^r \epsilon^{\theta} \epsilon^{\varphi}}{2} \sqrt{\text{det}(E)} \bigg[\epsilon^{\theta} R_{11;\theta} + \epsilon^{\varphi} R_{22;\varphi} +\epsilon^r R_{33;r}\bigg].
\ea
Due to spherical symmetry we have
\ba \label{classical}
&& R_{11; \theta} = e_1 ^{\theta 2} R_{\theta \theta ;\theta}
 = e_1 ^{\theta 2} R_{\theta \theta,\theta} = 0, \nonumber\\
&&R_{22;\varphi} = e_{2}^{\varphi 2} R_{\varphi \varphi;\varphi} = e_{2}^{\varphi 2} R_{\varphi \varphi,\varphi} = 0, \nonumber\\
&&R_{33;r} = e_3 ^{r 2} R_{rr;r} = e_3 ^{r 2} \Big[R_{rr,r} - 2 \Gamma^r _{rr} R_{rr}\Big] = \frac{2}{R^2 \Lambda^4} \bigg[-3 R R' \Lambda^{\prime 2} \nonumber\\
&& + \Lambda \Big(-\Lambda'[R^{\prime 2} - 3 R R''] + R R' \Lambda'' \Big) + \Lambda^2 \Big(R' R'' - R R''' \Big) \bigg].
\ea
Putting everything together, \eqref{defangcor2} reduces to
\ba \label{defangcor3} 
\sum_{a = r, \theta, \varphi} L_{a} \Big(\frac{\pi}{2} - \theta_a \Big) \Big|_{\epsilon^4} &=&\frac{(\epsilon^{r})^2 \epsilon^{\theta} \epsilon^{\varphi}}{2} \sqrt{\text{det}(E)}  R_{33;r} \n\\
&=& (\epsilon^{r})^2 \epsilon^{\theta} \epsilon^{\varphi} \frac{\sin{\theta}}{\Lambda^3} \bigg[-3 R R' \Lambda^{\prime 2}+ \Lambda \Big(-\Lambda'[R^{\prime 2} - 3 R R''] + R R' \Lambda'' \Big)
+ \Lambda^2 \Big(R' R'' - R R''' \Big) \bigg]. \n\\
\ea
That this result is somewhat different from what was obtained in 
Eq. \eqref{HLcor-2} is not all that surprising. Indeed the 
method here and the one is Sec. \ref{sec:Lor-term} correspond to two different regularization schemes for the spin connections.

\end{appendix}
%

\end{document}